\documentclass[conference]{IEEEtran}
\IEEEoverridecommandlockouts
\usepackage{cite}
\usepackage{amsmath,amssymb,amsfonts}
\usepackage{algorithmic}
\usepackage{graphicx}
\usepackage{textcomp}
\usepackage{xcolor}
\def\BibTeX{{\rm B\kern-.05em{\sc i\kern-.025em b}\kern-.08em
    T\kern-.1667em\lower.7ex\hbox{E}\kern-.125emX}}

\usepackage{graphicx}
\usepackage{balance}  
\usepackage{xcolor}
\usepackage{url}
\usepackage{enumitem} 
\usepackage{pifont}

\usepackage{marginnote} %

\usepackage{multirow}
\usepackage[normalem]{ulem}
\usepackage{amsmath}
\usepackage{caption}
\usepackage{subcaption}

\usepackage{listings}
\usepackage{bm}
\usepackage[linesnumbered,ruled,vlined]{algorithm2e}
\usepackage{graphicx}
\usepackage{hyperref}

\usepackage{makecell}
\usepackage{fontawesome}

\usepackage{multicol}

\usepackage{amsmath} 
\usepackage{balance} 
\usepackage{booktabs} 
\usepackage{caption}
\usepackage{comment}
\usepackage{enumitem}
\usepackage{url}
\setlist[itemize]{leftmargin=2.2em}

\usepackage{fancyvrb} 
\usepackage[symbol]{footmisc}
\usepackage{graphicx} 
\usepackage{makecell} 
\usepackage{multirow} 
\usepackage{soul}
\usepackage{subcaption}
\usepackage{xcolor}
\usepackage{xspace} 
\usepackage{bm}
\usepackage{utfsym}
\usepackage{cancel}

\usepackage{graphicx}
\usepackage{balance}  
\usepackage{xcolor}
\usepackage{url}
\usepackage{enumitem} 
\usepackage{pifont}

\usepackage{multirow}
\usepackage[normalem]{ulem}
\usepackage{amsmath}
\usepackage{caption}
\usepackage{subcaption}

\usepackage{listings}
\usepackage{bm}
\usepackage[linesnumbered,ruled,vlined]{algorithm2e}
\usepackage{graphicx}

\usepackage{makecell}
\usepackage{fontawesome}

\usepackage{multicol}

\usepackage{amsmath} 
\usepackage{balance} 
\usepackage{booktabs} 
\usepackage{caption}
\usepackage{comment}
\usepackage{enumitem}
\setlist[itemize]{leftmargin=2.2em}

\usepackage{fancyvrb} 
\usepackage[symbol]{footmisc}
\usepackage{graphicx} 
\usepackage{makecell} 
\usepackage{multirow} 
\usepackage{soul}
\usepackage{subcaption}
\usepackage{xcolor}
\usepackage{xspace} 
\usepackage{bm}
\usepackage{utfsym}
\usepackage{cancel}
\usepackage{authblk}

\SetKwInOut{Input}{input}\SetKwInOut{Output}{output}

\newtheorem{example}{\indent Example}

\newcommand\lion{\ensuremath{\texttt{Lion}}\xspace}

\newif\ifextended\extendedfalse
\newcommand{\maintext}[1]{\ifextended\relax\else#1\fi} 
\newcommand{\extended}[1]{\ifextended#1\else\relax\fi}

\begin{document}

\extendedtrue

\maintext{
\newcommand{\comments}[2]{\noindent{\textbf{Comment #1}}: {#2}}
\newcommand{\response}[2]{\vspace{0mm}\noindent\textbf{Response #1}: {#2} \vspace{0.5mm}}
\newcommand{\revhighlight}[1]{\emph{\textbf{\uuline{#1}}}}
\pagestyle{empty}
\thispagestyle{empty}

\newcommand{\question}[1]{{\textcolor{blue}{(#1)}}}

\onecolumn

\centerline{\textbf{\Large Response to Reviewers' Comments}}
\vspace{1mm}

\begin{multicols}{2}

\setcounter{page}{1}

\vspace{-2mm}
\noindent We would like to thank the reviewers and meta reviewer for their constructive comments.
We have addressed all the comments in the revised paper. We highlight the major changes made to the manuscript in
{\color{blue} blue}.

\vspace{-2mm}
\section*{Summary of Changes}

\begin{itemize}[leftmargin=*]

\item We have improved Section~\ref{sec:overview} by introducing the remastering mechanism and providing a comprehensive discussion on its correctness. 

\item We have enhanced the presentations to detail the system design in Section~\ref{sec:design} as suggested.
We have revised Section~\ref{sec.design.prediction.timeseries} to outline the rationale for our selection of LSTM, its training process, and modeling the predicted co-access partitions into a graph. 

\item We have clarified in Section~\ref{sec:implementation} that \lion employs remastering which can be automatically or manually invoked in existing production databases.  

\item We have substantially improved the experiments in Section~\ref{sec:evaluation} according to the reviewers' constructive comments.
First, we have refined the experiment setup in Section~\ref{sec:exp.setup}, explaining the upper limit for replicas and settings of the LSTM. 
Second, we have conducted the ablation studies in Section~\ref{sec.exp.optimization.analysis} and show the results in Table~\ref{tab:ablation} and Figure~\ref{Fig.experiment.ablation} to demonstrate the individual effectiveness of our proposed optimizations, where we also compare the partitioning strategy of \lion to Schism. 
Third, we have included additional experiments to compare \lion with Lotus in Figures~\ref{fig.non-deterministic.experiment.throughput}--\ref{Fig.exp.scale}, \ref{Fig.latency.experiment.timeline}.
Then we have updated Figure~\ref{fig.non-deterministic.experiment.throughput} and Figure~\ref{Fig.dm.experiment.throughput} to show the performance with a more realistic remastering delay.
Finally, we have evaluated the system performance under two dynamic workloads in Figure~\ref{Fig.experiment.nondmtimeline} and Figure~\ref{Fig.experiment.timeline}. 


\item To make space for the above changes, we have moved the evaluation on Calvin previously in Section~\ref{sec:evaluation} to the extended version~\cite{lionExtended}. Due to space limitations, we have also condensed Section~\ref{sec:pre}, Section ~\ref{sec:overview}, Section~\ref{sec:exp.setup}, Section~\ref{exp.overhead}, Section~\ref{sec:relatedwork}. We have tried our best to make sure that these sections remain concise, and their fuller versions can be found in the extended manuscript~\cite{lionExtended}.
\end{itemize}

\vspace{-2.5mm}
\section*{Response to Meta-Reviewer}

We have addressed all the required changes accordingly. 
Please kindly refer to our responses to R1, R2, and R3 below for details.


\vspace{-2mm}
\section{Response to Reviewer \#1}


\response{O1}{
We have conducted additional experiments to compare the proposed partitioning strategy with Schism.
We implement an alternation of \lion using Schism as the partitioning strategy, denoted as \ensuremath{\texttt{Lion(S)}\xspace}.
As depicted in Figure~\ref{Fig.experiment.ablation}, \lion outperforms \ensuremath{\texttt{Lion(S)}\xspace} by up to 1.7$\times$. This is attributed to the fact that, unlike Schism, our partitioning strategy additionally considers replications and takes future transactions into account. 
We further compare \ensuremath{\texttt{Lion(S)}\xspace} with \ensuremath{\texttt{Lion(R)}\xspace}, where \ensuremath{\texttt{Lion(R)}\xspace} represents \lion with only our proposed replica rearrangement strategy, to evaluate the partitioning strategy effectiveness exclusively.
As observed in Figure~\ref{Fig.experiment.ablation}, the replica rearrangement strategy outperforms Schism by up to 31.1\%, primarily because Schism does not account for the placement of secondary replicas, leading to unnecessary migrations. 
We also examine the effectiveness of the prediction mechanism by integrating it with Schism, denoted as \ensuremath{\texttt{Lion(SW)}\xspace}.
As shown in Figure~\ref{Fig.experiment.ablation}, \ensuremath{\texttt{Lion(SW)}\xspace} outperforms \ensuremath{\texttt{Lion(S)}\xspace} by up to 29.4\%, because of the reduced migration cost facilitated by predictions.
}
\response{O2}{
We have conducted additional ablation experiments to exclusively evaluate the effectiveness of each optimization in \lion.
We outline all the variants of \lion with different optimizations in Table~\ref{tab:ablation}, and plot the throughput in Figure~\ref{Fig.experiment.ablation}. 
As observed, each optimization improves transaction performance. 
First, compared to \ensuremath{\texttt{2PC}\xspace}, \ensuremath{\texttt{Lion(R)}\xspace} which represents \lion with only our proposed replica rearrangement strategy, demonstrates up to 3.5$\times$ performance improvement.
Second, by additionally leveraging workload prediction in \ensuremath{\texttt{Lion(R)}\xspace}, \ensuremath{\texttt{Lion(RW)}\xspace} shows a performance increase of up to 52.3\% over \ensuremath{\texttt{Lion(R)}\xspace}. 
Third, with the further employment of batch optimization, \ensuremath{\texttt{Lion}\xspace} achieves up to 20\% higher throughput than \ensuremath{\texttt{Lion(RW)}\xspace}. 

}

\section{Response to Reviewer \#2}


\response{O1}{
We have revised Section~\ref{sec:prediction_optimization} and Section~\ref{sec:exp.setup} to explore the system-level aspects of maintaining an LSTM model within a database. 
In our implementation, we train an LSTM model using the CPU and encode this model as an execution binary. This allows the database to directly invoke the model for inference tasks, i.e., workload prediction.
As discussed in Section~\ref{sec:exp.setup}, we utilize a lightweight LSTM model in \lion, which does not necessarily require a GPU for training because the training latency is acceptable even on a CPU.
We train the model periodically based on the logs that record the partitions accessed by transactions in a given time period. 
When the mean squared error (MSE) between predicted and actual results falls below a predefined threshold, we retrain the model to maintain the model accuracy.
We acknowledge GPU can be used to expedite LSTM model training~\cite{DBLP:journals/pvldb/LeeZLHTLZ21}.
However, as we primarily focus on leveraging LSTM for workload prediction to improve transaction performance, rather than optimizing the LSTM, we consider the GPU-based optimization to be orthogonal to \lion, and we defer this optimization to our future work.


}


\response{O2}{
We have conducted additional experiments by comparing \lion with Lotus over YCSB and TPC-C, as detailed in Section~\ref{sec:evaluation}. 
The results depicted in Figure~\ref{Fig.dm.ycsb.skew} and Figure~\ref{Fig.dm.tpcc.skew} demonstrate that Lotus performs well in low cross-ratio scenarios due to its specific epoch-based transaction processing.
However, Lotus's performance decreases significantly as the cross-ratio increases, because Lotus requires a costly commit protocol for distributed transactions and lacks optimizations for load balancing.
Further, we evaluate the performance of Lotus under dynamic changing workloads.
As observed in Figure~\ref{Fig.experiment.timeline}, \lion outperforms Lotus significantly under dynamic workloads.

}


\response{O3}{
Following your suggestions, we have conducted additional experiments with distribution shift, as detailed in Section~\ref{sec.exp.comparison.non-deter} and Section~\ref{sec.exp.comparison.deter}. 
We would like to emphasize that \lion, similar to other protocols, does not assume the key distribution to remain unchanged.
Apart from the workload with varying hotspot intervals used in our initial submission, we introduce a new workload with varying hotspot positions as recommended to assess the overall adaptability of \lion further. 
The results, shown in Figure~\ref{Fig.shift.timeline} and Figure~\ref{Fig.ycsb.shift.timeline.dm}, demonstrate that \lion can adapt to more complex workloads due to its proposed partitioning strategy and prediction mechanism, delivering up to 1.7$\times$ higher throughput than the next-best approach.
Further, we have refined the experiment setup in Section~\ref{sec.exp.comparison.non-deter.cross-ratio} and Section~\ref{sec.exp.comparison.non-deter.dynamicworkload} to clarify that we chose YCSB and TPC-C to evaluate the overall performance of \lion while using the dynamic workloads to evaluate the adaptability of \lion with workload shift.
}


\response{D1}{
We have enhanced Section~\ref{sec.design.predict} to explain why our proposed workload prediction algorithm is based on the LSTM model.
We recognize the existence of various models for time-series prediction, such as linear regression, kernel regression, and traditional recurrent neural networks.
However, compared to LSTM, traditional methods struggle to effectively capture long-term dependencies and handle non-linear dynamics within sequences, thus exhibiting limitations in handling complex time series patterns~\cite{DBLP:conf/uai/BootsGG13, DBLP:conf/sigmod/MaAHMPG18, DBLP:conf/icde/GaoHZ00C23}, such as the co-accessed partitions focused in our paper.
Further, we would like to clarify that we impose no restrictions on the choice of prediction model.
As we primarily focus on leveraging replicas to minimize distributed transactions, we directly use a standard LSTM for workload prediction without specific improvement or fine-tuning.
We consider conducting future work to incorporate more superior time-series prediction methods, such as attention-based models and transformer architectures~\cite{vaswani2017attention, devlin2018bert}, to further improve the prediction accuracy in \lion.

}

\section{Response to Reviewer \#3}



\response{D1}{
As suggested, we have enhanced Section~\ref{sec:overview} to describe the process of remastering, which is widely used in~\cite{DBLP:conf/sigmod/TaftSMVLGNWBPBR20, DBLP:journals/pvldb/HuangLCFMXSTZHW20, DBLP:journals/pvldb/YangYHZYYCZSXYL22, DBLP:journals/corr/abs-2206-07278}.
First, a secondary replica is selected as the candidate based on our proposed replica rearrangement algorithm and new read/write operations will get blocked on the primary replica.
Then the lagging logs will be synchronized from the leader to the target secondary replica, ensuring its state is consistent with that of the primary replica.
After that, a leader election starts to prompt the candidate to be the new primary replica, who then continues to perform operations.
}


\response{D2}{
We have improved Section~\ref{sec:overview} to clarify how we guarantee the correctness of remastering in \lion. 
We understand that during remastering, there are potential risks of data inconsistency and split-brain problems~\cite{DBLP:phd/us/Ongaro14}. 
We address this following the TiDB's approach~\cite{DBLP:journals/pvldb/HuangLCFMXSTZHW20}.
This approach utilizes log synchronization to maintain consistency between the new primary replica and the old one during remastering.
Further, to prevent split-brain issues, it blocks new operations during remastering to ensure that only one primary replica provides service at any given time.
}

\response{D3}{
We have updated Section~\ref{sec.design.arrangement.cost} to clarify that \lion does not inherently impose a theoretical upper limit on the number of replicas.
This limitation was set in our initial submission due to the constraints on the number of nodes used in our experiments.
As we use 4 nodes by default, we set the max replica number to 4.
We have improved Section~\ref{sec:exp.setup} to clarify this experimental setting as well.
}


\response{D4}{
We would like to state that \lion employs a replica rearrangement algorithm that creates a unified plan based on both historical and predicted workloads, rather than producing separate adjustment plans for each type of workload.
We model historical and predicted workloads as a single graph for that purpose. 
To fine-tune the influence of predicted workloads on our planning process, we employ a parameter $w_p$ that determines their weight within the graph. 
We have revised Section~\ref{sec.design.prediction.timeseries} to better clarify this statement.

}


\response{D5}{
We would like to clarify that distributed databases, such as TiDB~\cite{TidbDocs} and Oceanbase~\cite{oceanbaseDocs} typically offer APIs for replica remastering and replica management, e.g., adding a replica.
For example, \texttt{transfer-leader} API provided in TiDB can enable replica remastering without the necessity of a leader failure.
While these databases are indeed built upon consensus protocols, their remastering processes generally follow the approach we described in our response to D1.
Therefore, \lion can be integrated into these real-world distributed databases by calling the APIs they provide.
To make this clearer, we have improved Section~\ref{sec:overview} accordingly.
}



\response{D6}{
In our initial submission, we conducted experiments on remastering latency and plotted the results in Figure 12, which we have moved to the extended manuscript~\cite{lionExtended} (Figure 13) due to space constraints.
As observed, the transaction performance of batch processing surpasses that of standard processing with increased remastering latency. 
Following your suggestion, we have adjusted the default remastering latency to 3000 microseconds (previously we set it to 500 microseconds) to better simulate the latency in real-world scenarios, and rerun experiments in Section~\ref{sec.exp.comparison.non-deter} and Section~\ref{sec.exp.comparison.deter}.
As depicted in Figure~\ref{fig.non-deterministic.experiment.throughput} and Figure~\ref{Fig.dm.experiment.throughput}, \lion achieves up to 21.2\% better throughput in the batch mode compared to the standard mode.
This observation is consistent with the results of our ablation study shown in Figure~\ref{Fig.experiment.ablation}. 
}


\response{D7}{
We have enhanced Section~\ref{sec.exp.comparison.non-deter.dynamicworkload} and Section~\ref{sec.exp.comparison.deter.dynamicworkload} by conducting experiments using a combined workload,  
where we simultaneously change the cross-partition ratio and the \texttt{skew\_factor} to evaluate the comprehensive adaptability of \lion.
The results, shown in Figure~\ref{Fig.shift.timeline} and Figure~\ref{Fig.ycsb.shift.timeline.dm}, demonstrate that \lion can adapt quickly with new workloads, while delivering up to 1.7$\times$ higher throughput than the next-best approach, due to our proposed partitioning strategy and prediction mechanism.


}
\end{multicols}

\twocolumn

\clearpage
}

\maintext{\title{Lion: Minimizing Distributed Transactions through Adaptive Replica Provision}}

\extended{\title{Lion: Minimizing Distributed Transactions through Adaptive Replica Provision (Extended Version)}}

\author{
Qiushi Zheng$^\dagger$, Zhanhao Zhao$^\dagger$, Wei Lu$^\dagger$, Chang Yao$^\ddagger$, Yuxing Chen$^\S$, Anqun Pan$^\S$, Xiaoyong Du$^\dagger$ \\
\fontsize{10}{10}\textit{$^\dagger$ Renmin University of China} 
\qquad\fontsize{10}{10}\textit{$^\ddagger$ Zhejiang University} 
\qquad\fontsize{10}{10}\textit{$^\S$ Tencent Inc.} \\
\fontsize{9}{9}\texttt{$^\dagger$\{zhengqiushi, zhanhaozhao, lu-wei, duyong\}@ruc.edu.cn} \\
\fontsize{9}{9}\texttt{$^\ddagger$changy@zju.edu.cn} \qquad 
\fontsize{9}{9}\texttt{$^\S$\{axingguchen, aaronpan\}@tencent.com}  
}

\maketitle

\thispagestyle{plain}
\pagestyle{plain}

\begin{abstract}
Distributed transaction processing often involves multiple rounds of cross-node communications, and therefore, tends to be slow. 
To improve performance, existing approaches convert distributed transactions into single-node transactions by either migrating co-accessed partitions onto the same nodes or establishing a super node housing replicas of the entire database.
However, migration-based methods might cause transactions to be blocked due to waiting for data migration, while the super node can become a bottleneck.


In this paper, we present \lion, 
a novel transaction processing protocol that utilizes partition-based replication to reduce the occurrence of distributed transactions.
Inspired by the fact that modern distributed databases horizontally partition data, with each partition having multiple replicas, 
\lion aims to assign a node with one replica from each partition involved in a given transaction's read or write operations.
To ensure such a node is available, we propose an adaptive replica provision mechanism, enhanced with an LSTM-based workload prediction algorithm, to determine the appropriate node for locating replicas of co-accessed partitions.
The adaptation of replica placement is conducted preemptively and asynchronously, 
thereby minimizing its impact on performance.
By employing this adaptive replica placement strategy, we ensure that the majority of transactions can be efficiently processed on a single node without additional overhead. Only a small fraction of transactions will need to be treated as regular distributed transactions when such a node is unavailable.
%
Consequently, \lion effectively minimizes distributed transactions, while avoiding any disruption caused by data migration or the creation of a super node.
We conduct extensive experiments to compare {\lion} against various transaction processing protocols.
The results show that {\lion} achieves up to 2.7x higher throughput and 76.4\% better scalability against these state-of-the-art approaches.

\end{abstract}

\begin{IEEEkeywords}
Transaction Processing, Replica Provision
\end{IEEEkeywords}
\setcounter{page}{1}
\setcounter{section}{0} 
\section{Introduction}
\label{text:intro}

Modern distributed databases, with representative examples including Spanner~\cite{DBLP:conf/osdi/CorbettDEFFFGGHHHKKLLMMNQRRSSTWW12}, CockroachDB~\cite{DBLP:conf/sigmod/TaftSMVLGNWBPBR20}, TiDB~\cite{DBLP:journals/pvldb/HuangLCFMXSTZHW20}, are essential to today's large-scale online applications.
These databases horizontally partition data across several nodes to improve scalability.
However, data partition comes at the cost of requiring distributed transaction processing, where a transaction must be executed and committed through rounds of communication with multiple nodes.
As the involved network overhead is non-negligible, it is commonly accepted that distributed transactions are slow~\cite{DBLP:journals/pvldb/HardingAPS17}.

To boost performance, it is prevalent to execute transactions on a single node as many as possible to avoid distributed transaction processing.
Thus far, existing approaches, either based on data migration~\cite{DBLP:journals/pvldb/CurinoZJM10, DBLP:conf/edbt/QuamarKD13, DBLP:journals/pvldb/SerafiniTEPAS16, DBLP:conf/sigmod/LinC0OTW16, DBLP:journals/pvldb/AbebeGD20} or full-replication~\cite{DBLP:journals/pvldb/LuYM19, DBLP:conf/icde/AbebeGD20}, are sub-optimal.
Data migration improves the partition quality by transferring the required data for a transaction to a specific node. 
This synchronous migration may lead to substantial performance degradation, as transactions relying on those partitions will get blocked until the migration completes.
Worse still, when distinct transactions on different nodes require the same data, a ``ping-pong''~\cite{DBLP:conf/sigmod/LinTLCW21} problem can emerge, with the data continuously migrating between nodes back and forth to satisfy transaction demands.
Full-replication-based methods, on the other hand, establish an additional node containing replicas of the entire database.
Instead of involving data migration, distributed transactions can be converted into single-node transactions by executing on this ``super node''. 
However, this node can potentially become a performance bottleneck.
More importantly, due to the large data volume, establishing such a node may be infeasible. 
Therefore, designing a transaction processing protocol to efficiently minimize distributed transactions still remains an open problem.

In real-world distributed databases~\cite{DBLP:conf/osdi/CorbettDEFFFGGHHHKKLLMMNQRRSSTWW12,DBLP:conf/sigmod/TaftSMVLGNWBPBR20,DBLP:journals/pvldb/HuangLCFMXSTZHW20}, it is fundamental that each partition has multiple replicas for high availability.
Given the fact that a node can host several replicas of different partitions, it is possible that one node houses all the replicas needed for a given transaction's read/write operations.
Therefore, we consider reducing distributed transactions by executing them as single-node transactions on such nodes.
However, using this idea to improve transaction performance requires addressing two major challenges:
First, it is not trivial to ensure a node meeting the transaction's specific requirements always exists.
A sophisticated replica placement mechanism is needed to strategically locate replicas on appropriate nodes, thus avoiding the creation of a super node.
Second, optimizing replica placement without causing transactions to be delayed or blocked is not straightforward.
We must prepare a node hosting all the necessary replicas in advance of the transaction's execution to avoid data migration overheads.

In this paper, we present {\lion}, an efficient transaction processing protocol that employs partition-based replication to minimize distributed transactions.
We propose an adaptive replica provision mechanism to optimize the replica placement, ensuring that transactions can find a node with all the required replicas.
Specifically, we first employ a graph-based workload analysis algorithm to identify partitions that are frequently accessed together and then adjust the replica placement to ensure the co-accessed partitions are allocated onto an appropriate node.
The replica placement adaptation is guided by a tailor-designed cost model for minimized distributed transactions and load balancing.
Further, we introduce a workload prediction algorithm based on Long Short Term Memory (LSTM) to facilitate non-intrusive replica adjustment.
By analyzing these predicated workloads, we anticipate partitions that are likely to be co-accessed in the future and pre-allocate their replicas to the appropriate nodes in advance.
Therefore, we guarantee the replica adjustment is performed asynchronously with transaction processing.
During the adjustment, we add a secondary replica to the corresponding node in the background, without interrupting the execution of transactions on the primary replica. 
We utilize the remastering technique~\cite{DBLP:conf/usenix/WeiSCC17} to prompt the secondary replica to the primary replica when necessary.
Because replication protocols~\cite{DBLP:conf/usenix/OngaroO14,DBLP:conf/opodis/Lamport02} are adopted to guarantee the majority (or even all) of the replicas for a partition are consistent with each other, this remastering process is generally lightweight and free of data migration overheads.



Based on the replica placement, we execute transactions on a single node as many as possible.
Due to the constraint that each partition typically has one primary replica and several secondary replicas, but only the primary replica handles write requests, we process transactions efficiently and correctly as follows.
1) If a transaction finds a node housing primary replicas of all relevant data, it can directly execute on that node as a single-node transaction.
2) In situations where the node lacks the necessary primary replicas but contains secondary ones, the transaction can still execute on that node after remastering these replicas to be primary.
3) Otherwise, transactions that fail to find such a node are treated as regular distributed transactions.
We utilize the proposed cost model to ensure most transactions can be directly executed without remastering and distributed processing.
Further, {\lion} supports two kinds of transaction processing schemes, namely standard (ad-hoc) execution and batch execution, ensuring its general applicability to real-world distributed databases.



In summary, we make the following contributions.

\begin{itemize}[leftmargin=*]

\item We introduce {\lion}, a new transaction processing protocol that minimizes distributed transactions.
{\lion} is generally applicable to modern distributed databases that leverage partition-based replication.

\item We propose an adaptive replica provision mechanism that 
uses a graph-based workload analysis algorithm to determine a proper replica placement and a cost-model-based strategy to minimize distributed transactions as well as achieve load balance.

\item We design an LSTM-based workload prediction algorithm to ensure the replica placement adjustment is non-intrusive while accommodating workload changes with sustainable performance.

\item We conduct extensive evaluations on two popular benchmarks, namely YCSB and TPC-C, and compare {\lion} against various existing works for optimizing distributed transactions.
The results show that {\lion} is efficient, and outperforms the state-of-the-art approaches by up to 2.7x.

\end{itemize}

The remainder of the paper is structured as follows.
The next section provides relevant background and presents the problem statement.
Section~\ref{sec:overview} overviews the architecture of {\lion}.
Section~\ref{sec:design} details the adaptive replica provision mechanism, and elaborates on the prediction-based optimization.
Section~\ref{sec:implementation} describes the system implementation, and Section~\ref{sec:evaluation} presents the experimental results.
Section~\ref{sec:relatedwork} discusses the related work, and Section~\ref{sec:conclusion} concludes.






\section{PRELIMINARIES}
\label{sec:blg}

In this section, we first describe distributed transaction processing and existing approaches to minimize distributed transactions.
We then present the problem statement.


\begin{figure}
\centering
\includegraphics[width=0.44\textwidth]{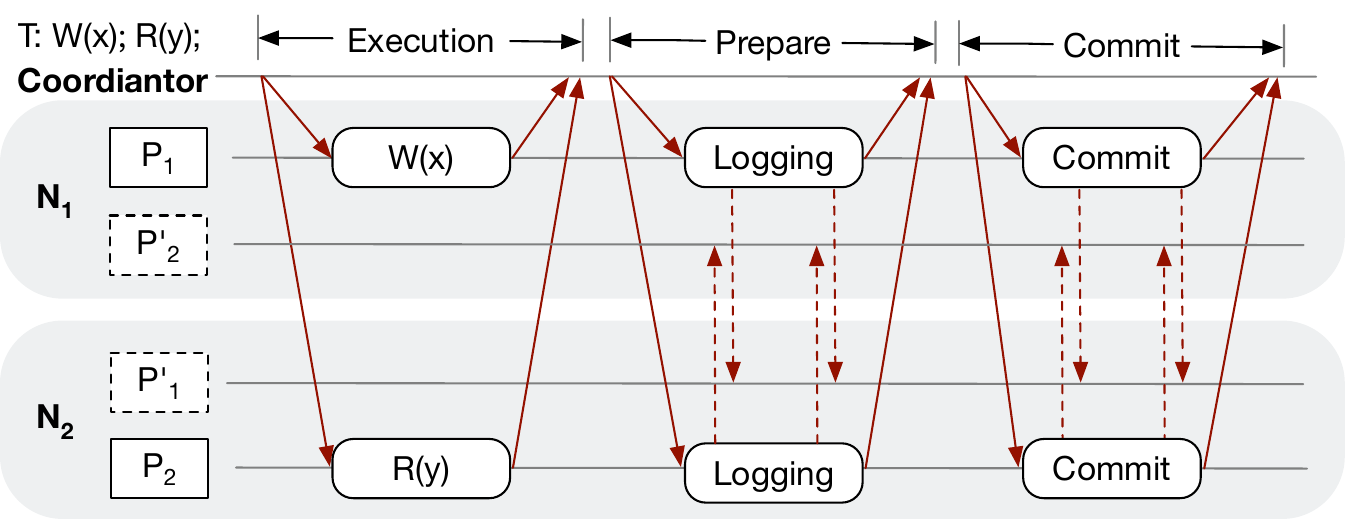}
\caption{Standard distributed transaction processing}
\label{fig:execution}
\vspace{-4mm}
\end{figure}

\subsection{Distributed Transaction Processing}

In modern distributed database systems, data is typically divided into several partitions.
Each of these partitions has one primary replica and may have multiple secondary replicas.
For example, as shown in Figure~\ref{fig:execution}, there are two nodes $N_1$ and $N_2$ and two partitions $P_1$ and $P_2$. 
The primary replica of $P_1$ is located on $N_1$ while its secondary replica is on $N_2$. Conversely, $P_2$ has its primary replica on $N_2$ and its secondary on $N_1$. 
When the context is clear, we use $P_1$ to denote the primary replica of partition $P_1$, and use $P'_1$ to represent the secondary replica of partition $P_1$.


\extended{
Distributed transaction processing is essential to manipulate data on multiple nodes while ensuring ACID properties and availability.
The standard approach~\cite{DBLP:conf/osdi/CorbettDEFFFGGHHHKKLLMMNQRRSSTWW12} for processing a distributed transaction $T$ consists of three phases: 1) execution phase, 2) prepare phase, and 3) commit phase.
As shown in Figure~\ref{fig:execution}, suppose the transaction $T$ involves a write operation on the data item $x$, denoted as $W(x)$, and a read operation on the data item $y$, denoted as $R(y)$. 
In the execution phase, the coordinator of transaction $T$ first distributes these read/write requests to the corresponding primary replicas for local execution.
For example, $T$ sends the write operation $W(x)$ to $P_1$, while sending the read operation $R(y)$ to $P_2$, respectively. 
These primary replicas ($P_1$ and $P_2$) are then regarded as participants. 
}

\maintext{
Distributed transaction processing is essential to manipulate data on multiple nodes while ensuring ACID properties and availability.
The standard approach~\cite{DBLP:conf/osdi/CorbettDEFFFGGHHHKKLLMMNQRRSSTWW12} for processing a distributed transaction $T$ consists of three phases: 1) execution phase, 2) prepare phase, and 3) commit phase.
As shown in Figure~\ref{fig:execution}, suppose $T$ involves a write operation on the data item $x$, denoted as $W(x)$, and a read operation on $y$, denoted as $R(y)$. 
In the execution phase, the coordinator of $T$ first distributes these requests to the corresponding primary replicas for local execution.
For example, $T$ sends $W(x)$ to $P_1$, while sending $R(y)$ to $P_2$, respectively. 
These primary replicas ($P_1$ and $P_2$) are then regarded as participants. 
}
Once all read/write operations of $T$ are successfully executed, the coordinator employs the two-phase commit (2PC) to commit/abort $T$.
First, the prepare phase begins, during which the coordinator sends a prepare message to all participants. 
Upon receiving this message, each participant votes for whether to commit $T$ and replicates its prepare log, containing this vote and the data item to be written (if applicable), to the corresponding secondary replicas.
For example, supposing $P_1$ decides to commit $T$, it replicates the commit vote with the data item $x$ to $P'_1$.
If all participants agree to commit $T$, the coordinator then moves $T$ to the commit phase.
During this phase, the commit decision is sent to these participants and replicated to all involved secondary replicas.

According to this process, executing and committing a distributed transaction requires multiple ($\geq 5$) round trips of blocking communication, which limits transaction throughput~\cite{DBLP:journals/pvldb/HardingAPS17}.
In this paper, we therefore focus on enabling most transactions can be executed as single-node transactions to avoid the costly network overheads.

\subsection{Distributed Transactions Minimization}
\label{sec:pre}
We now analyze existing approaches that minimize distributed transactions to clarify the design space of \lion.

\subsubsection{The data migration technique}

\extended{
Data migration is commonly used to co-locate correlated partitions, through either a ``pull'' or ``push'' method~\cite{DBLP:conf/sigmod/ElmoreATPAA15, DBLP:conf/sigmod/ElmoreDAA11, DBLP:conf/sigmod/CubukcuEPSS21, DBLP:journals/pvldb/DasNAA11}. However, during the migration, transactions accessing these partitions may abort or get blocked, resulting in service downtime and performance degradation.

Several earlier works are designed as offline tools for repartitioning the physical layout of databases. For example, Schism~\cite{DBLP:journals/pvldb/CurinoZJM10} and Sword~\cite{DBLP:conf/edbt/QuamarKD13} identify co-accessed partitions based on historical data and initiate the partition replacement before the database restarts. 
However, the downtime is unbearable and the layout becomes stale when the workload changes. Plus, they do not account for the placement of secondary replicas, leading to unnecessary migrations. 

Leap~\cite{DBLP:conf/sigmod/LinC0OTW16} pursues adaptivity to dynamical workload through an aggressive migration strategy at the transaction level. This approach allows each node to migrate remote data to the local node for each operation before execution. 
However, this approach may lead to two potential issues: the ``ping-pong'' problem and the load imbalance problem since all partitions will be migrated to the same node facing the skewed workload.

In contrast, Clay~\cite{DBLP:journals/pvldb/SerafiniTEPAS16} and Hermes~\cite{DBLP:conf/sigmod/LinTLCW21} adopt a sophisticated migration strategy by analyzing more transactions. 
Clay periodically monitors transaction execution on each node and devises a migration plan upon detecting load imbalances. Its approach involves transferring partitions from overloaded nodes to others to resolve the imbalance. However, as the primary focus is load balancing through repartitioning, Clay can not eliminate all distributed transactions. Because it sometimes misinterprets the overloaded node, running single-node transactions, as having a similar load to nodes with fewer distributed transactions and fails to start the repartitioning.
Hermes combines migration with deterministic protocols. Similar to Leap, it migrates data on the fly with distributed transaction processing. But it collects transactions in batches and reorders the batch sequence to keep transactions accessing the same partitions together. This approach reduces the ``ping-pong'' effect since the following transactions may reuse the migration caused by earlier ones. 
However, migration is still inevitable and severely interferes with transaction processing.
}
\maintext{
Data migration can be used to move partitions to a new target node through either a ``pull'' ~\cite{DBLP:conf/sigmod/ElmoreATPAA15, DBLP:conf/sigmod/ElmoreDAA11} or a ``push'' method~\cite{DBLP:conf/sigmod/CubukcuEPSS21, DBLP:journals/pvldb/DasNAA11}. 
Note that migration may result in service downtime and performance degradation since transactions that access these partitions will abort or get blocked during the process.
Earlier works focus on repartitioning databases, with tools like Schism~\cite{DBLP:journals/pvldb/CurinoZJM10} and Sword~\cite{DBLP:conf/edbt/QuamarKD13} identifying co-accessed partitions based on historical data and initiating partition replacement before database restarts. However, they suffer from unbearable downtime and stale layouts during workload changes, and neglect the placement of secondary replicas, leading to unnecessary migrations.
Leap~\cite{DBLP:conf/sigmod/LinC0OTW16} aims for adaptivity by employing an aggressive transaction-level migration strategy. Each node migrates remote data to the local node before execution, yet faces potential "ping-pong" and load imbalance problems due to migrating all partitions to a single node under skewed workloads.
Clay~\cite{DBLP:journals/pvldb/SerafiniTEPAS16} monitors transaction execution and transfers partitions from overloaded nodes to resolve imbalances but cannot eliminate all distributed transactions due to misinterpretations of node loads. 
Hermes~\cite{DBLP:conf/sigmod/LinTLCW21} integrates migration with deterministic protocols and reorders transaction batches to mitigate the ``ping-pong'' effect, yet migration still disrupts transaction processing.
}

\subsubsection{The full-replication technique}
\label{sec.preliminary.remaster}

\extended{
The full-replication technique ensures that at least one node contains complete replicas of all partitions. It is usually equipped with data remastering techniques~\cite{DBLP:journals/tpds/ShenWCCZ22, DBLP:conf/usenix/WeiSCC17} to eliminate distributed transactions. 
Compared with data migration, remastering is a lightweight technique since it only transfers from the primary replica to another secondary one, instead of the entire partition copy. 
Star~\cite{DBLP:journals/pvldb/LuYM19} introduces a ``super node'' with full replication, routing all distributed transactions to that node. The system periodically remasters primary replicas to the ``super node'' to process them as single-node ones. However, as the cross-partition ratio increases, the ``super node'' becomes overloaded and causes the bottleneck.
DynaMast~\cite{DBLP:conf/icde/AbebeGD20} assumes that all nodes have full replicas and uses dynamic mastering to ensure that each node has an appropriate proportion of mastership, which is not within the scope of partition-based replication methods. However, it is unfeasible to have certain nodes store all data replicas. As the number of nodes increases, the synchronization costs between all replicas can become unsustainable.
}

\maintext{
The full-replication technique ensures that at least one node contains complete replicas of all partitions. It is usually equipped with data remastering techniques~\cite{DBLP:journals/tpds/ShenWCCZ22, DBLP:conf/usenix/WeiSCC17} to eliminate distributed transactions. 
Unlike data migration, remastering is lightweight, involving the leadership transfer, rather than the entire physical partition 
Star~\cite{DBLP:journals/pvldb/LuYM19} introduces a ``super node'' with full replication, where all distributed transactions are routed to that node as single-node ones. However, this strategy leads to bottleneck issues, especially as the distributed transactions increase.
DynaMast~\cite{DBLP:conf/icde/AbebeGD20} suggests dynamic mastering to ensure an appropriate distribution of mastership across nodes, assuming all nodes have full replicas. However, this approach is impractical for partition-based replication methods and becomes unsustainable as the number of nodes increases due to synchronization costs between all replicas.
}

\subsubsection{Summary}

\extended{
In Table~\ref{tab:security}, we compare existing approaches with \lion across four dimensions. 
All methods, excluding 2PC, aim to enhance transaction processing efficiency through data migration and full replication. 
However, Schism and Star lack adaptability to dynamic workloads. 
Schism relies on an offline strategy, while Star cannot dynamically modify its replica placement. 
Leap, Clay, and Hermes suffer performance degradation during migration. 
Leap and Star experience bottlenecks due to inadequate load balancing considerations. 
Additionally, Star and Hermes may impose batch or deterministic constraints on transaction execution. 
In contrast, \lion considers improving across all dimensions by its adaptive replica provision. 
} 

\maintext{
In Table~\ref{tab:security}, we compare existing approaches with \lion across four dimensions. 
All methods aim to enhance transaction processing efficiency through data migration and full replication. 
However, Schism and Star lack adaptability to dynamic workloads. 
Leap, Clay, and Hermes suffer performance degradation during migration. 
Additionally, Star and Hermes may impose batch or deterministic constraints on transaction execution. 
In contrast, \lion considers improving across all dimensions by its adaptive replica provision. 
Due to space constraints, a more detailed analysis of related techniques can be found in the extended version~\cite{lionExtended}.
} 


\extended{
\begin{table}[t]
\centering
\renewcommand\arraystretch{1.1}
\caption{Comparison of {\lion} with existing approaches}
\label{tab:security}
\vspace{1mm}
\resizebox{0.98\columnwidth}{!}{%
\begin{tabular}{|c|c|c|c|c|c|c|}
\hline
         & Key Designs  & \makecell{Dynamic\\Adaptivity} & \makecell{Migration\\Efficiency} & \makecell{Load\\Balancing} & \makecell{Execution\\Constraints} \\
         \hline \hline 
2PC                                     & \makecell{Distributed Transactions}      & N/A & N/A         & \usym{2717} & N/A \\ \hline
Schism~\cite{DBLP:journals/pvldb/CurinoZJM10} & \makecell{Offline Repartitioning}  & \usym{2717} & \usym{2717} & \usym{2717} & N/A \\ \hline
Leap~\cite{DBLP:conf/sigmod/LinC0OTW16} & \makecell{Aggressive Migration}          & \usym{2713} & \usym{2717} & \usym{2717} & N/A \\ \hline
Clay~\cite{DBLP:journals/pvldb/SerafiniTEPAS16}& \makecell{Periodical Migration}   & \usym{2713} & \usym{2717} & \usym{2713} & N/A \\ \hline
Hermes~\cite{DBLP:conf/sigmod/LinTLCW21}& \makecell{Deterministic Migration}        & \usym{2713} & \usym{2717} & \usym{2713} & In batches \\ \hline
Star~\cite{DBLP:journals/pvldb/LuYM19}  & \makecell{Full Replication}              & N/A & \usym{2713} & \usym{2717} & In batches \\ \hline
\lion & \makecell{Adaptive Replication}                                            & \usym{2713} & \usym{2713} & \usym{2713} & N/A \\ \hline
\end{tabular}%
}
\vspace{-3mm}
\end{table}
}

\maintext{
\begin{table}[t]
\centering
\renewcommand\arraystretch{1.1}
\caption{Comparison of {\lion} with existing approaches}
\label{tab:security}
\vspace{1mm}
\resizebox{0.98\columnwidth}{!}{%
\begin{tabular}{|c|c|c|c|c|c|c|}
\hline
         & Key Designs  & \makecell{Dynamic\\Adaptivity} & \makecell{Migration\\Efficiency} & \makecell{Load\\Balancing} & \makecell{Execution\\Constraints} \\
         \hline \hline 
Schism~\cite{DBLP:journals/pvldb/CurinoZJM10} & \makecell{Offline Repartitioning}  & \usym{2717} & \usym{2717} & \usym{2717} & N/A \\ \hline
Leap~\cite{DBLP:conf/sigmod/LinC0OTW16} & \makecell{Aggressive Migration}          & \usym{2713} & \usym{2717} & \usym{2717} & N/A \\ \hline
Clay~\cite{DBLP:journals/pvldb/SerafiniTEPAS16}& \makecell{Periodical Migration}   & \usym{2713} & \usym{2717} & \usym{2713} & N/A \\ \hline
Hermes~\cite{DBLP:conf/sigmod/LinTLCW21}& \makecell{Deterministic Migration}        & \usym{2713} & \usym{2717} & \usym{2713} & In batches \\ \hline
Star~\cite{DBLP:journals/pvldb/LuYM19}  & \makecell{Full Replication}              & N/A & \usym{2713} & \usym{2717} & In batches \\ \hline
\lion & \makecell{Adaptive Replication}                                            & \usym{2713} & \usym{2713} & \usym{2713} & N/A \\ \hline
\end{tabular}%
}
\vspace{-3mm}
\end{table}
}


\subsection{Problem Definition}
\label{sec.pre.objective}
\extended{
We now present a formal definition of the problem addressed by \lion. 
The database comprises multiple nodes denoted as $N = \{N_1, \ldots, N_n\}$.
Each node's storage includes a collection of partitions represented as $P = \{P_1, \ldots, P_m\}$.
To ensure high availability, each partition must contain a minimum of $k$ replicas, distributed in a default round-robin fashion.
Given a batch of transactions $B = \{T_1, \ldots, T_b\}$, our objective is to determine a new replica placement $P'$. 
Let $\epsilon$ represent the percentage of load imbalance permissible within the system, and let $\theta$ denote the average load across all nodes in $P'$ multiplied by $1+\epsilon$.
Further, we outline three requirements in the following formula.
First, we aim to minimize distributed transactions. 
Second, the replica rearrangement cost from $P$ to $P'$ should be minimized. 
Third, the load should be balanced under $P'$.
}
\maintext{
We now present a formal definition of the problem addressed by \lion. 
The database comprises multiple nodes denoted as $N = \{N_1, \ldots, N_n\}$.
Each node includes multiple partitions represented as $P = \{P_1, \ldots, P_m\}$.
For high availability, each partition has a minimum of $k$ replicas.
Given a batch of transactions $B = \{T_1, \ldots, T_b\}$, our objective is to determine a new replica placement $P'$. 
Let $\epsilon$ represent the percentage of load imbalance permissible within the system, and let $\theta$ denote the average load across all nodes in $P'$ multiplied by $1+\epsilon$.
We outline three goals in the following formula:
1) minimize distributed transactions, 
2) minimize the replica rearrangement cost from $P$ to $P'$, 
3) ensure the load balance under $P'$.
}

\vspace{-4mm}
\begin{equation}
\label{eq1}
\begin{split}
minimize\  C_{e}(B, P') = \sum\nolimits_{i=1}^B f_c(n_i, T_i), C_{p}(P, P') \\
s.t.\ \forall x \in N ,\ f_b(x) < \theta 
\end{split}
\end{equation}
Given a transaction $T_i$ routed to the node $n_i$, $f_c(n_i,T_i)$ represents the execution cost of $T_i$ on $n_i$, mainly including the remote access and commit cost. 
$C_{e}(B, P')$ represents the cost summation for $B$ under $P'$.
$C_{p}(P, P')$ signifies the replica rearrangement cost, including migration and remastering costs. 
We impose an upper limit of $\theta$ on the load $f_b(x)$ for each node to maintain load balance.

\section{System Overview}
\label{sec:overview}

\begin{figure}
\centering
\includegraphics[width=0.9\columnwidth]{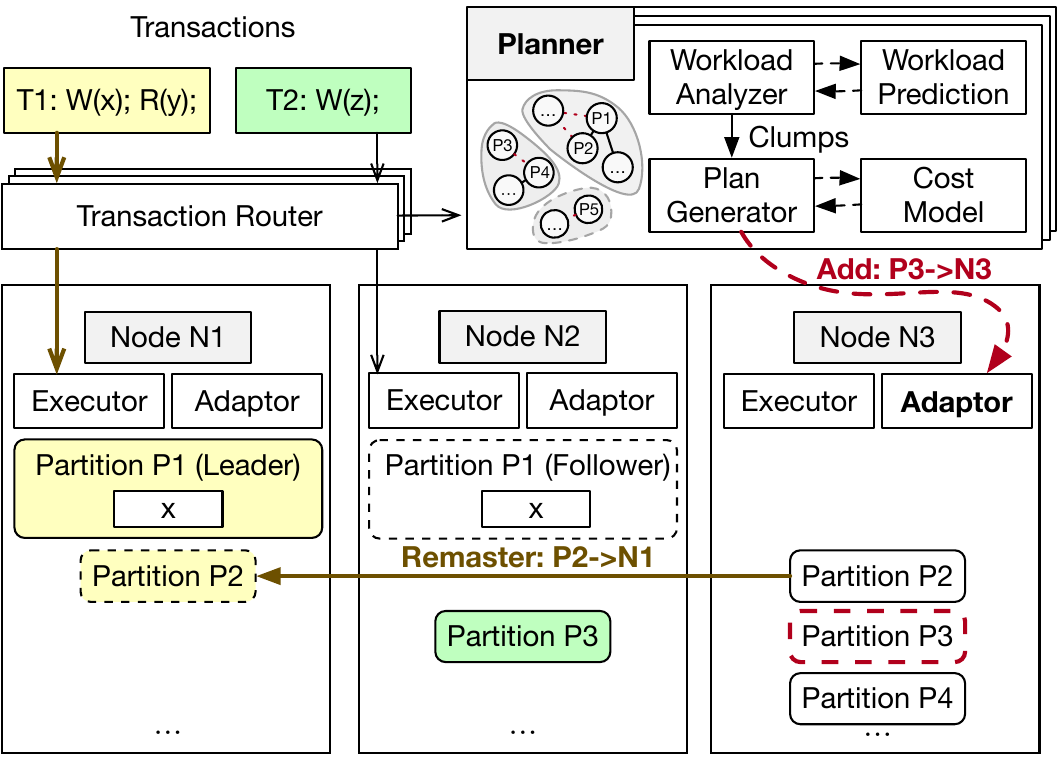}
\caption{System overview of {\lion}}
\label{fig:arch}
\vspace{-4mm}
\end{figure}

\lion is applicable for modern distributed databases~\cite{DBLP:conf/osdi/CorbettDEFFFGGHHHKKLLMMNQRRSSTWW12,DBLP:journals/pvldb/HuangLCFMXSTZHW20,DBLP:conf/sigmod/TaftSMVLGNWBPBR20} that maintain multiple replicas of data partitions.
For illustration purposes, we overview \lion based on the share-nothing architecture consisting of multiple monolithic nodes.
Each node is responsible for storing data replicas and processing incoming transactions.
Figure~\ref{fig:arch} shows the system overview of \lion.
To minimize distributed transactions, \lion introduces two system components: 
1) \textbf{planner}, a specific kind of node, including a workload analyzer and a plan generator, to determine an optimal replica placement plan;
2) \textbf{adaptor}, a component within each node to adjust the replica placement based on the generated plan.

We dynamically adjust replica provision according to the workload through the following steps: 
\begin{itemize} [leftmargin=*]

\item \textsl{Workload analysis}: 
After collecting $B$ recently received transactions, we invoke the workload analyzer to identify co-accessed partitions.
In addition to analyzing the past $B$ transactions, our approach includes $K$ predicted future transactions, which are generated by the proposed workload prediction technique.
We construct a graph based on the read/write sets of these $B+K$ transactions.
In the graph, we represent partitions as vertices and transactions as edges connecting partitions that are co-accessed.
We utilize a clustering algorithm to precisely group these partitions into clumps.
Each clump consists of partitions and indicates that the partitions within it should be placed at the same node.
This method represents both current and anticipated future access patterns of transactions. 
The detailed workload analysis and prediction techniques will be presented in Section~\ref{design:workload} and Section~\ref{sec:prediction_optimization}.

\item \textsl{Plan generation}: 
We then employ the plan generator to determine the optimal replica placement plan.
We design a cost model to assign each clump to a specific node, taking into account factors such as replica rearrangement cost and load balancing.
By considering multiple replicas, this model ensures less adjustment overhead and better load balancing than the prior methods designed for the single replica setting. 
We shall detail the plan generation in Section~\ref{sec:design_replacement}.

\item \textsl{Asynchronous adjustment}: 
Lastly, the adaptor asynchronously adjusts the replicas according to the established plan.
In particular, the adaptor will add or remove secondary replicas by invoking the replica manipulation functions~\cite{DBLP:conf/sigmod/TaftSMVLGNWBPBR20, DBLP:journals/pvldb/HuangLCFMXSTZHW20} inherent in the replication-based database. 



\end{itemize}

\extended{
After pre-allocating replicas according to the plan, \lion dynamically adjusts the position of primary replicas and converts the distributed transactions into single-node ones. 
This is achieved by the following two steps. 
Firstly, we dispatch the transaction $T$ to the node with well-prepared replicas. To accomplish this, we introduce a set of transaction routers, each of which is equipped with a cost model identical to the planner's. 
The router will dispatch $T$ to a node with maximum requisite replicas, where the execution cost is the lowest. 
Secondly, we utilize the remastering technique to finalize the conversion.
On the executor, $T$ is processed through the execution, prepare, and commit phases.
During the execution phase, operations are directly executed on the node if it possesses the necessary primary replicas.
\lion remasters before executing that operation if it has a secondary replica. 
If all operations can be executed on a single node, the transaction can be directly committed, omitting the prepare phase. 
Otherwise, \lion processes $T$ as a standard distributed transaction with 2PC. 
}
\maintext{
After the pre-replication, \lion dynamically adjusts primary replica positions and converts distributed transactions into single-node ones. 
The router first dispatches the transaction $T$ to a node with well-prepared replicas, based on identical strategies of the planner.
Then the executor tries to turn $T$ into a single-node transaction through remastering.
During execution phases, operations are performed directly on nodes with necessary primary replicas or remastered if secondary replicas exist. 
If all operations can be executed on a single node, the transaction is directly committed; otherwise, \lion processes $T$ with standard 2PC.
\textcolor{blue}{
The \marginpar[\textcolor{blue}{R3.D1}]{\textcolor{blue}{R3.D1}}process of remastering is performed as follows, which is widely used in~\cite{DBLP:conf/sigmod/TaftSMVLGNWBPBR20, DBLP:journals/pvldb/HuangLCFMXSTZHW20, DBLP:journals/pvldb/YangYHZYYCZSXYL22, DBLP:journals/corr/abs-2206-07278}.
First, a secondary replica is selected as the candidate based on our proposed replica rearrangement algorithm and new read/write operations will get blocked on the primary replica.
Then the lagging logs will be synchronized from the leader to the target secondary replica, ensuring its state is consistent with that of the primary replica.
After that, a leader election starts to prompt the candidate to be the new primary replica, who then continues to perform operations.
}
\textcolor{blue}{
During 
\marginpar[\textcolor{blue}{R3.D2}]{\textcolor{blue}{R3.D2}}
remastering, there are potential risks of data inconsistency and split-brain problems~\cite{DBLP:phd/us/Ongaro14}. 
We address these following the TiDB's approach~\cite{DBLP:journals/pvldb/HuangLCFMXSTZHW20}.
This approach utilizes log synchronization to maintain consistency between the new primary replica and the old one during remastering.
Further, to prevent split-brain issues, it blocks new operations during remastering to ensure that only one primary replica provides service at any given time.
}
}
\extended{
The process of remastering is performed as follows, which is widely used in~\cite{DBLP:conf/sigmod/TaftSMVLGNWBPBR20, DBLP:journals/pvldb/HuangLCFMXSTZHW20, DBLP:journals/pvldb/YangYHZYYCZSXYL22, DBLP:journals/corr/abs-2206-07278}.
First, a secondary replica is selected as the candidate based on our proposed replica rearrangement algorithm and new read/write operations will get blocked on the primary replica.
Then the lagging logs will be synchronized from the leader to the target secondary replica, ensuring its state is consistent with that of the primary replica.
After that, a leader election starts to prompt the candidate to be the new primary replica, who then continues to perform operations.
During remastering, there are potential risks of data inconsistency and split-brain problems~\cite{DBLP:phd/us/Ongaro14}. 
We address these following the TiDB's approach~\cite{DBLP:journals/pvldb/HuangLCFMXSTZHW20}.
This approach utilizes log synchronization to maintain consistency between the new primary replica and the old one during remastering.
Further, to prevent split-brain issues, it blocks new operations during remastering to ensure that only one primary replica provides service at any given time.
}

The replica provisioning ensures that co-access partitions will be placed on one node and transactions accessing the same partitions are deliberately routed to the same node, which reduces ``ping-pong'' remastering across the nodes. 
In scenarios where a remastering conflict emerges, one transaction completes the remastering successfully while others resort to committing as distributed transactions. 
For example, transactions $T_1$ and $T_2$ are routed to nodes $N_1$ and $N_2$, respectively, and both attempt to remaster the overlapping replicas simultaneously. 
Assuming the success of $T_1$ and the failure of $T_2$, subsequent transactions resembling $T_2$, which access the same partitions, will endeavor to route to $N_1$ whenever possible. Otherwise, they will execute through 2PC.

\begin{example}
\label{eg1}
As shown in Figure~\ref{fig:arch}, there are three nodes $N_1$, $N_2$, and $N_3$, and three partitions $P_1$, $P_2$, and $P_3$. 
Suppose $x$, $y$, and $z$ is stored in $P_1$, $P_2$, and $P_3$.
The primary replica of $P_1$, $P_2$, and $P_3$ is located on $N_1$, $N_3$, and $N_2$, respectively.
Let us consider two transactions, $T_1$ and $T_2$. 
$T_1$ contains a write $W(x)$ and a read $R(y)$, while $T_2$ involves a write $W(z)$.
For $T_1$, the router sends $T_1$ to $N_1$ as it has the primary replica of $P_1$.
However, the replica of $P_2$ is on $N_3$.
Therefore, after $T_1$ executes $W(x)$, its read is executed after the adaptor on $N_1$ remasters $P_2$ to $N_1$.
In contrast, $T_2$ is routed and executed as a single-node transaction on $N_2$ without remastering since all the required primary replica exists in $N_2$.
Consider a transaction $T_3$ writes to partitions $P_3$ and $P_4$.
Suppose the primary replicas of $P_3$ and $P_4$ are on $N_2$ and $N_3$, respectively. 
\extended{
$T_3$ can not be converted into a single-node transaction through remastering since neither of the nodes has all replicas for it. In this case, $T_3$ would be executed as a distributed transaction.
After collecting and analyzing transactions periodically, \lion decides to add a secondary replica of $P_3$ on $N_3$ to co-locate partitions $P_3$ and $P_4$. 
As a result, when a subsequent transaction $T_4$ arrives that reads $P_3$ and $P_4$, it can be processed as a single-node transaction on $N_3$.
The details of that adaptive replica mechanism are further explained in Section~\ref{sec:design_replacement} (Example~\ref{eg2}) and Section~\ref{sec:prediction_optimization} (Example~\ref{eg3}).
}
\maintext{
$T_3$ would be executed as a distributed transaction since neither of the nodes has all replicas for it.
After periodic transaction analysis, \lion adapts to add a secondary replica of $P_3$ on $N_3$ to co-locate partitions $P_3$ and $P_4$. 
As a result, when $T_4$ arrives that also reads $P_3$ and $P_4$, it can be processed as a single-node transaction on $N_3$ through remastering.
The details of that adaptive replica mechanism are further explained in Section~\ref{sec:design_replacement} (Example~\ref{eg2}) and Section~\ref{sec:prediction_optimization} (Example~\ref{eg3}).
}

\end{example}

\section{The Design of {\lion}}
\label{sec:design}


In this section, we present the design of \lion in detail. 


\subsection{Workload Analysis Technique}
\label{design:workload}

\begin{figure}[t]
    \centering
    \begin{subfigure}[b]{0.23\textwidth}
        \centering
        \includegraphics[width=\textwidth]{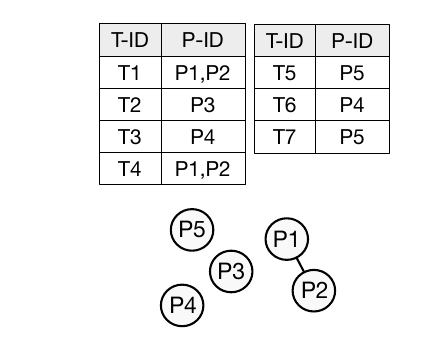}
        \caption{Graph construction}    
        \label{fig:cluster_a}
    \end{subfigure} 
    \hfill
    \begin{subfigure}[b]{0.23\textwidth}
        \centering
        \includegraphics[width=\textwidth]{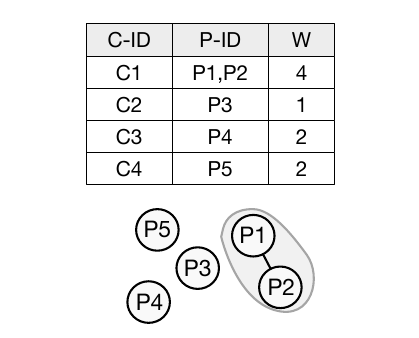}
        \caption{Clumps generation}
        \label{fig:cluster_b}
    \end{subfigure}
       \caption{Example for workload analysis technique}
       \label{fig:cluster}
\vspace{-4mm}
\end{figure}

To determine which partitions should be placed on the same node, we employ a graph-based algorithm within the workload analyzer. This algorithm treats a batch of transactions as a graph and produces a set of clumps, where each clump represents co-accessed partitions. The process of this algorithm is depicted in Figure~\ref{fig:cluster}, comprising two stages: 1) graph construction, defining transactions as a graph, and 2) clump generation, extracting co-access information from the graph.

\textbf{Graph Construction.} 
Within \lion, we retain the partition IDs accessed by each transaction, alongside its $TxnMeta$ information, such as $TxnID$ in the transaction context. The involved partitions of a given query are determined after the SQL parsing procedure and will be further pruned by the query optimization. We record these results as a new variable $TxnParts$ in the $TxnMeta$.

Based on that, we begin by modeling the workload's access patterns into a heat graph, denoted as an indirect weighted graph $G(V, E)$, with vertices in set $V$ and edges in set $E$. We accumulate the weight of both vertices and edges to represent the access frequency of the partition and the co-access possibility between partitions.
Each partition accessed by a transaction is treated as a vertex, labeled as $v$, with its weight denoted as $w(v)$. 
Edges, referred to as $e$, connect pairs of vertices representing partitions accessed by the same transaction.
Notably, the weights of edges connecting multiple nodes denoted as $e_c$ ($e_c = {(u, v) | u \in N_i , v \in N_j, i \neq j}$), are considerably greater than the weights of edges within a single node, termed $e_s$ ($e_s = {(u, v) | u \in N_i , v \in N_j, i = j}$). This emphasizes the higher priority given to $e_c$ when considering edge weights.
Moreover, we employ a priority queue $hVertices$ to record the most frequently accessed vertices, aiming to expedite the subsequent clustering process. 
In Figure~\ref{fig:cluster_a}, we provide an illustrative example illustrating the transactions batch collected by the planner and the resultant construction of the $G(V, E)$ graph.

\textbf{Clump Generation.} 
After the graph $G(V, E)$ is constructed, a clustering algorithm will be triggered to identify partitions that are frequently accessed by the same transactions as clumps. 
Each clump $c$ contains a set of involved vertices ($c.pids$), the weighted sum of the vertices ($c.w$), the destination to place the clump ($c.n$), etc.  

The clustering algorithm selects a vertex and expands on its neighbors until all co-accessed vertices are included. It starts with the hottest unused vertex $v$ from $hVertices$ as a seed and evaluates neighboring vertices $v_{adj}$ based on their connection weight $w(e)$ against the threshold $\alpha$. 
If this weight surpasses $\alpha$, indicating high co-access or different nodes, these vertices are grouped in a clump. Conversely, vertices with weaker correlations or independent access are placed into separate clumps. 
As the clump expands, $c.w$ will get updated with each inclusion of new vertices, which facilitates load balancing for later clump reallocation.
Once all related neighbors have been explored, the algorithm concludes the search for the current clump, selects a new seed and proceeds to the next clump. This process continues until all vertices in $hVertices$ have been visited.
The generated clumps serve as input for the replica rearrangement strategy algorithm (in Section~\ref{sec:design_replacement}). 
Figure~\ref{fig:cluster_b} illustrates the creation of four clumps, representing co-located partition sets. $C_1$ comprises partitions $P_1$ and $P_2$, with a weight of 4 (each vertex weighs 2). Meanwhile, $C_2$, $C_3$, and $C_4$ contain one partition each.

\subsection{Replica Rearrangement Strategy}
\label{sec:design_replacement}
Once the clumps are identified, \lion assigns them across nodes to co-locate replicas for the partitions within a clump.
Nevertheless, an imprudent clump placement can lead to a significant amount of partition movement, consuming system resources and potentially interfering with current transaction execution. 
To mitigate these issues, we introduce a replica rearrangement strategy. This strategy aims to distribute clumps with minimal movement cost while ensuring load balancing.

\subsubsection{Rearrangement Objective}
We first define the problem addressed by this algorithm and refine the objective as discussed in Section~\ref{sec.pre.objective}. 
Given a collection of clumps $C$ and the current replica placement $P$, this algorithm is designed to determine a reconfiguration plan $RP$ for the clump placement. 
To facilitate this, we employ a \texttt{ReconfigurationPlan} structure. The $RP$ constitutes a mapping between clumps and nodes, where each entry $\left \langle c,n \right \rangle $ denotes that clump $c$ will be assigned to the node $n$.
Applying the $RP$ will adjust the replica placement and result in a new placement $P'$.
During this process, the assignment of $C$ will incur additional operational costs. 
Each partition $v$ within clump $c$ should be relocated to node $n$ through a series of operations, involving data migration and remastering.  
The cost fluctuates across different scenarios, as outlined below:

\begin{itemize}[leftmargin=*]
\item \textbf{Case 1}. If node $n$ hosts the primary replica of $v$, i.e. $N_p(v,p) = n$, no additional cost is incurred. $N_p(v,p)$ represents the node that hosts the primary replica of the given partition $v$.
\item  \textbf{Case 2}. If node $n$ has a secondary replica of $v$, i.e. $n \in N_s(v,p)$, remastering can be used to effect the conversion at a cost of $w_r$. $N_s(v,p)$ represents the set of nodes with the secondary replica $v$.
\item \textbf{Case 3}. Otherwise, if node $n$ lacks any replica of partition $v$, data migration becomes inevitable, incurring a cost of $w_m$. 
\end{itemize}


The $RP$ should adhere to the objectives in Equation~\ref{eq1}. 
More specifically, the second objective is extended by Equation~\ref{eq2}. Here, $f_{o}(n_i, c_i)$ represents the cost for assigning $c_i$ to $n_i$.
\begin{equation} \label{eq2}
\begin{array}{c}
minimize\  C_{p}(P, P')=\sum_{\{n_i, c_i\} \in RP_i }{f_{o}(n_i, c_i)}
\end{array}
\end{equation}

\subsubsection{Cost Evaluation}
\label{sec.design.arrangement.cost}
Subsequently, we delineate the process of evaluating cost and identifying the destination for each clump. 
Our approach involves a cost model to assess the cost for each clump $c$ across all nodes and heuristically selects the node $n$ with the lowest cost as the destination. 
Let us start with the following three scenarios for cost evaluation:
If $n$ hosts all primary partitions, the cost model prefers to place $c$ on it with no additional cost.
In case $n$ lacks all primary replicas but possesses all secondary ones, it is still an ideal choice since expensive migration can be avoided. The placement only introduces the lightweight remastering cost for the secondary replicas in the later process. 
However, if $n$ doesn't possess all required replicas, data migration becomes necessary to add replication for the lacking partitions. 


We summarize these scenarios in Equation~\ref{eq3} to calculate the cost of placing $c$ on $n$.

\vspace{-4mm}
\begin{equation} \label{eq3}
\begin{array}{l}
\begin{aligned}
f_{o}(n, c) = w_{r} * \sum_{v \in V_{c}
                } cnt_{r}(v, n)
                +
                w_{m} * \sum_{v \in V_{c}} cnt_{m}(v, n) 
\end{aligned}
\end{array}
\end{equation}

where $V_c$ represents the partitions that $c$ possesses. $\sum_{v \in V_c}{cnt_{r}(v, n)}$ and $\sum_{v \in V_c}{cnt_{m}(v, n)}$ stands for the number of secondary and the lacking partitions on $n$ respectively. The calculation of these two kinds of partition is defined in Equation~\ref{eq4}. 
Note that we also track the normalized access frequency of replicas for each partition as $f(v, n)$. A higher $f(v, N_p(v,p))$ indicates that the remastering cost would be more substantial. When the current primary replica is actively accessed, it might disrupt the ongoing transactions or encounter blocking until the execution is completed. 

\vspace{-4mm}
\begin{equation} \label{eq4}
\begin{aligned}
cnt_{r\ }(v, n) & =\left\{\begin{array}{l}
1 + \log_{2}{(f(v, N_p(v,p)) + 1)}, \ n \in N_s(v, p) \\
0, \text { else }
\end{array}\right. \\
cnt_{m}(v, n) & =\left\{\begin{array}{l}
1, \ n \notin N_p(v, p) \cup N_s(v, p) \\
0, \text { else }
\end{array}\right.
\end{aligned}
\end{equation}


\extended{
We allow users to set a maximum replica limit for each partition according to their requirements.
Note that this replica number is not a theoretical limit of \lion.
}
\maintext{
\textcolor{blue}{
We allow users \marginpar[\textcolor{blue}{R3.D3}]{\textcolor{blue}{R3.D3}} to set a maximum replica number for each partition according to their requirements.
Note that this maximum replica number is not a theoretical limit of \lion.}}
Upon exceeding this limit after adding new replicas, we remove one replica from the replica group. 
We opt to remove the secondary replica with the lowest $f(v, n)$ by designating it with a $delete\_flag$. 
Subsequently, replica synchronization ceases to update the data in the flagged one.
Different from Clay and Hermes, \lion benefits two-fold by considering the placement of multiple replicas: first, it minimizes migration expenses, thereby minimizing disruption to the current system execution; second, it avoids full replication, thereby lessening the synchronization overhead for consistency.

\begin{algorithm}[t]
    \small
    \DontPrintSemicolon
    \caption{Replica Rearrangement Algorithm}
	\label{alg:rearrange}
	\SetKwFunction{FMain}{Rearrangement}
	\SetKwProg{Fn}{Function}{:}{}
	\Fn{\FMain{$C$, $P$}}{
            $N \leftarrow $ the number of nodes in $P$. ; $rp\leftarrow$[];  $m\_c\leftarrow$[][] \;
            $b_i \leftarrow 0\ \forall\ i = 1, ... , N$; $q_i \leftarrow []\ \forall\ i = 1, ... , N$\;
		\For{$c_i \in C$}{
		    $c_i.n\leftarrow FindDstNode(c_i, P, m\_c)$  \;
                $RP$.push($c_i$); $q[c_i.n]$.push($c_i$)\;
                $UpdateBalance(c_i, b)$ \;
		}
            $avg \leftarrow (\frac{LoadSum(C)}{N})$;
            $is\_done \leftarrow false$ \;
            \While{$!CheckBalance(avg, b) $ \textbf{and} $ !is\_done$ } { 
                $step$ $\leftarrow$ A \;
                $oN, iN$ $\leftarrow$ $FindOINodes(b, avg)$ \;
                \If{$|oN| = 0 \ \textbf{or} \ |iN| = 0$}{
                    \textbf{break} \;
                }
                \While {$!CheckBalance(avg, b)$ \textbf{and}  $step > 0$}{
                    $idx, i_n, is\_find$ $\leftarrow$ $PickClump(q, oN, iN, m\_c, b)$\; 
                    \If{$!is\_find$}{
                        \textbf{break} \;
                    }
                    $RP[idx].n \leftarrow i_n$ \;
                    $UpdateBalance(RP[idx], b)$ \;
                    \If{$|oN| = 0 \ \textbf{or} \ |iN| = 0$}{
                        $step  \leftarrow  0 $\;
                    } \Else {
                        $step  \leftarrow  step - 1 $\;
                    }
                }
                \If{$step = A$}{
                    $is\_done \leftarrow true$ \;
                }
            }
            \Return{$RP$}
	}
\end{algorithm}

\begin{figure*}[t]
    \centering
    \begin{subfigure}[b]{0.17\textwidth}
        \centering
        \includegraphics[width=\textwidth]{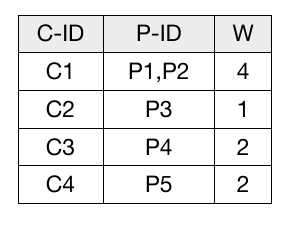}
        \caption{The input clumps}    
        \label{fig:router_a}
    \end{subfigure}
    \hfill
    \begin{subfigure}[b]{0.24\textwidth}
        \centering
        \includegraphics[width=\textwidth]{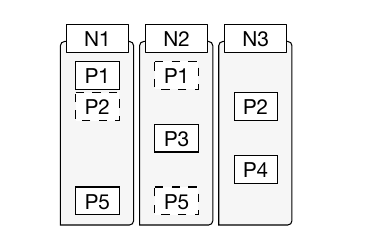}
        \caption{The original replica layout}
        \label{fig:router_b}
    \end{subfigure}
    \hfill
    \begin{subfigure}[b]{0.24\textwidth}
        \centering
        \includegraphics[width=\textwidth]{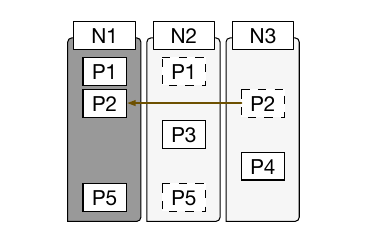}
        \caption{The clump dispatching phase}    
        \label{fig:router_c}
    \end{subfigure}
    \hfill
    \begin{subfigure}[b]{0.24\textwidth}
        \centering
        \includegraphics[width=\textwidth]{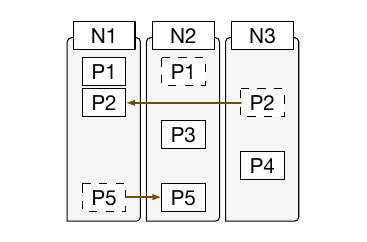}
        \caption{The load fine-tuning phase}    
        \label{fig:router_d}
    \end{subfigure}     
       \caption{Example for the replica rearrangement algorithm}
       \label{fig:router}
\vspace{-4mm}
\end{figure*}

\subsubsection{Rearrangement Algorithm}
\label{design.rearrange}
We finally deploy the replica rearrangement algorithm within the plan generator to achieve the aforementioned objective. This algorithm, operating on a set of clumps $C$ and the existing replica placement $P$, facilitates the determination of the $RP$. The algorithm operates in two distinct steps:

\textbf{Clump dispatching. }  
The plan generator initially assigns the clump to a destination node with a minimal cost. 
It iterates through every clump $c_i$, and employs the function \texttt{FindDstNode()} to select the suitable target node $n$. 
This function utilizes the cost model, as expressed in Equation~\ref{eq3}, to evaluate costs across all nodes, recording the interim costs in $m_c$. 
Subsequently, it identifies the destination with the lowest cost for $c_i$ (lines 5), and the updated $c_i$ is added to the reconfiguration plan $RP$.
To monitor load balance, the balance factor $b_i$ is updated once the plan generator determines the destination for each clump. This factor incorporates the weight $c_w$ of each clump (lines 7). Additionally, a priority queue $q$ is employed to log the clumps assigned to individual nodes, sorted by their respective weights in ascending order (lines 6). These data structures facilitate efficient fine-tuning in subsequent steps if necessary.

\textbf{Load fine-tuning. } 
Subsequently, the plan generator adjusts the current $RP$ to maintain load balance among nodes if necessary. 
The function \texttt{CheckBalance()} calculates the balance variance and compares it against a predefined threshold $\theta$ (line 9). If the variance is below $\theta$, the algorithm concludes.
Otherwise, suggesting a potential load imbalance caused by $RP$, a fine-tuning mechanism is activated to address this issue (lines 9-25). 
The basic idea is to reassign clumps from overloaded nodes to idle ones to bridge the load disparity with additional operation costs. 
Firstly, the function \texttt{FindOINodes()} identifies overloaded nodes $oN$ and idle nodes $iN$ based on the $b_i$ (line 11). 
If no overloaded nodes are detected, the loop terminates. 
Otherwise, the function \texttt{PickOneClump()} is employed to transfer a clump from $oN$ to $iN$ (line 15).
This function initially selects an overloaded node $o_n$ from $oN$ and calculates the required clump size $sz$ based on the load gap between $o_n$ and $iN$. Then it searches the $q[o_n]$ to find a clump less than $sz$. If succeeds, it returns the idle node $i_n$ from $iN$ with the lowest cost. Otherwise, the function retries on other overloaded nodes. 
If all retries fail, the loop exits (line 25). 
Upon successfully identifying a qualified clump, it updates its destination, along with the balance factors and $oN$ and $iN$ (lines 18-19). A variable $step$ is defined to save recalculations of \texttt{FindOINodes()}.
The replica rearrangement algorithm concludes its iterations when the load balance falls within $\theta$.

\begin{example}
\label{eg2}
To illustrate, consider the clumps and replica placement in Figure~\ref{fig:router_a} and~\ref{fig:router_b} following the example discussed in Section~\ref{sec:overview}. For the sake of simplicity in the following examples, we assume that all replicas have approximately the same access frequency.
Upon executing the first step outlined in Algorithm~\ref{alg:rearrange}, all clumps are initially dispatched to nodes with minimized costs, as shown in Figure~\ref{fig:router_c}.
For instance, $C_1$ opts for $N_1$ as its destination. Since the costs for it to $N_1$, $N_2$, and $N_3$ are $w_r$, $w_m + w_r$, and $w_m$, the cost at $N_1$ is lowest. 
Similarly, $C_2$, $C_3$, and $C_4$ choose $N_2$, $N_3$, and $N_1$ without costs.
However, this allocation leads to load imbalance, where $N_1$ is overloaded with a weight of 6, while other nodes maintain weights of 1 and 2. 
To rectify this, the second step is triggered to transfer a clump from $N_1$ to others.
The algorithm selects $C_4$ and designates $N_2$ as the new destination. Given that $N_2$ is an idle node and has a secondary replica, the reassignment of $C_4$ to $N_2$ incurs an additional cost of $w_r$. 
The final replica placement depicted in Figure~\ref{fig:router_d} demonstrates a balanced load with an operation cost of $2*w_r$.

\end{example}


\subsection{Workload Predication} 
\label{sec:prediction_optimization}

While \lion adapts to dynamic workloads through the planner, the time-consuming process of replica rearrangement still poses a formidable challenge. 
As the workload changes, all transactions bear the burden of 2PC overhead until replication becomes fully prepared. 
To expedite this process, we combine \lion with a workload prediction mechanism. 
It is aimed at forecasting workload patterns and proactively adding replication for co-accessed partitions, termed pre-replication.
Unlike Hermes \cite{DBLP:conf/sigmod/LinTLCW21}, which relies on foreknowledge of future read/write operations and suits specific deterministic database scenarios, our prediction mechanism imposes no such constraints and is suitable for non-deterministic systems.


\subsubsection{How to predict co-accessed partitions} 
\label{sec.design.predict}
The basic idea is to convert discrete transaction access into a time-series analysis problem. The prediction technique can be divided into the following three phases:

\textbf{Template Identification.} 
The arrival rate history~\cite{DBLP:conf/sigmod/MaAHMPG18, DBLP:conf/icde/GaoHZ00C23} is a crucial metric used to characterize the access pattern of transaction queries, formulated in Equation~\ref{eq5}. The $ar$ of a query signifies its access frequency changes over time. It can be visualized as a curve with a sampling interval $i$, where the x-axis represents the sampling time point $t$ and the y-axis denotes the sum of query frequency $f(n)$ within $i$.
However, maintaining the $ar$ information for every query can be costly. To mitigate this, we establish a partition-based rule for labeling transactions based on their access patterns. Transactions accessing the same partitions receive the same label, forming identical templates. Once these templates are identified, we track the arrival rate history of each template instead of individual queries. 
In Figure~\ref{fig:prediction_a}, we identify five templates under the partition-based rule, corresponding to the example in Section~\ref{sec:overview}. Of these, four templates ($P_1P_2$, $P_3$, $P_4$, and $P_5$) were active before the timestamp $t_1$, while the other two templates ($P_3P_4$ and $P_5P_6$) became dominant thereafter.
\begin{equation}
\label{eq5}
ar(t, i) = \sum\nolimits_{n = t}^{t + i}  {f(n)} 
\end{equation}






\textbf{Workload Classification.} 
Two templates are deemed similar if their arrival rates increase and decrease simultaneously, a similarity evaluated by computing the cosine distance between their $ar$ values.
To improve prediction efficiency, templates with a calculated distance below a predefined threshold $\beta$ are merged into the same workload class. 
Consequently, subsequent predictions will be performed for the merged workloads rather than individual templates.
Each workload comprehensively stores information for all its constituent templates, including partition IDs and their associated access frequencies. During pre-replication initiation, reservoir sampling \cite{DBLP:journals/toms/Vitter85} assists in identifying partitions from the workload that are highly likely to appear soon.
In Figure~\ref{fig:prediction_b}, we demonstrate the consolidation of five templates into two distinct workloads. Specifically, $P_1P_2$, $P_3$, $P_4$, and $P_5$ constitute workload $W_1$, while $W_2$ encompasses $P_3P_4$ and $P_5P_6$. Timestamp $t_1$ serves as the boundary between these two workloads.

\textbf{Time-series Prediction.} 
\label{sec.design.prediction.timeseries}
We utilize an LSTM model to forecast future workload trends based on the historical $ar$ of each workload. 
\extended{
%
Compared to LSTM, traditional methods like linear regression and traditional RNNs struggle to effectively capture long-term dependencies and handle non-linear dynamics within sequences, thus exhibiting limitations in handling complex time series patterns~\cite{DBLP:conf/uai/BootsGG13, DBLP:conf/sigmod/MaAHMPG18, DBLP:conf/icde/GaoHZ00C23}, such as the co-accessed partitions focused in our paper.
We utilize a lightweight LSTM model in \lion, which does not necessarily require a GPU for training because the training latency is acceptable even on a CPU.
We train the model periodically based on the logs that record the partitions accessed by transactions in a given time period. 
In particular, we build the query arrival rates (as formulated in Equation~\ref{eq5}) based on the raw log.
When the mean squared error (MSE) between predicted and actual results falls below a predefined threshold, we retrain the model to maintain the model accuracy.
%
We acknowledge that more orthogonal optimizations can be explored such as using GPU to expedite model training~\cite{DBLP:journals/pvldb/LeeZLHTLZ21} and incorporating superior algorithms~\cite{vaswani2017attention, devlin2018bert} to improve the prediction accuracy.
As we primarily focus on leveraging replicas to minimize distributed transactions, we directly use a standard LSTM for workload prediction without specific improvement or fine-tuning. 
We defer these optimizations to our future work.
}
\maintext{
\textcolor{blue}{
%
Compared 
\marginpar[\textcolor{blue}{R2.D1, R2.O1}]{\textcolor{blue}{R2.D1, R2.O1}}
to LSTM, traditional methods like linear regression and traditional RNNs struggle to effectively capture long-term dependencies and handle non-linear dynamics within sequences, thus exhibiting limitations in handling complex time series patterns~\cite{DBLP:conf/uai/BootsGG13, DBLP:conf/sigmod/MaAHMPG18, DBLP:conf/icde/GaoHZ00C23}, such as the co-accessed partitions focused in our paper.
We utilize a lightweight LSTM model in \lion, which does not necessarily require a GPU for training because the training latency is acceptable even on a CPU.
We train the model periodically based on the logs that record the partitions accessed by transactions in a given time period. 
In particular, we build the query arrival rates (as formulated in Equation~\ref{eq5}) based on the raw log.
When the mean squared error (MSE) between predicted and actual results falls below a predefined threshold, we retrain the model to maintain the model accuracy.
%
We acknowledge that more orthogonal optimizations can be explored such as using GPU to expedite model training~\cite{DBLP:journals/pvldb/LeeZLHTLZ21} and incorporating superior algorithms~\cite{vaswani2017attention, devlin2018bert} to improve the prediction accuracy.
As we primarily focus on leveraging replicas to minimize distributed transactions, we directly use a standard LSTM for workload prediction without specific improvement or fine-tuning. 
We defer these optimizations to our future work.
}
}

For the upcoming workload with high anticipated arrival rates, we select templates based on their access frequencies. 
\extended{
Subsequently, the co-accessed partitions within the template are integrated into the graph $G(V, E)$. 
\lion employs a replica rearrangement algorithm that creates a unified plan based on both historical and predicted workloads, rather than producing separate adjustment plans for each type of workload.
We model historical and predicted workloads as a single graph for that purpose. 
To fine-tune the influence of predicted workloads on our planning process, we employ a parameter $w_p$ that determines their weight within the graph. 
This parameter serves as a weighted coefficient for incorporating predicted information into the graph. A weight of 0 signifies that the prediction algorithm is inactive. By default, $w_p$ is set to 1.
}
\maintext{
\textcolor{blue}{
Subsequently,
\marginpar[\textcolor{blue}{R3.D4}]{\textcolor{blue}{R3.D4}}
the co-accessed partitions within the template are integrated into the graph $G(V, E)$. 
\lion employs a replica rearrangement algorithm that creates a unified plan based on both historical and predicted workloads, rather than producing separate adjustment plans for each type of workload.
We model historical and predicted workloads as a single graph for that purpose. 
To fine-tune the influence of predicted workloads on our planning process, we employ a parameter $w_p$ that determines their weight within the graph. 
This parameter serves as a weighted coefficient for incorporating predicted information into the graph. A weight of 0 signifies that the prediction algorithm is inactive. By default, $w_p$ is set to 1.
}
}
For instance, in Figure~\ref{fig:prediction_b}, assuming the current timestamp is $t_2$ and the active workload is $W_1$. After forecasting, it indicates that the arrival rate of $W_2$ will surpass $W_1$. Consequently, we sample the template $P_3P_4$ from $W_2$ and incorporate it into $G$. The predicted co-accessed partitions are included as additional edge weights, visualized by the red dashed line connecting $P_3$ and $P_4$ in Figure~\ref{fig:prediction_c}.



\subsubsection{When to trigger pre-replication} 
We've devised a workload variation metric $wv$ to assess the variance between the present workload and its anticipated future state, as defined in Equation~\ref{eq6}. 
After periodic evaluations, the pre-replication will be triggered when $wv$ surpasses a predefined threshold $\gamma$.

\begin{equation} \label{eq6}
   wv(t, h)=\sqrt{\frac{1}{n}{\sum_{k=1}^n(a_k(t + h,  \delta )-a_k(t,  \delta ))^2}}
\end{equation}

\begin{figure}[t]
    \centering
    \begin{minipage}[b]{0.25\textwidth}
        \centering
        \begin{subfigure}[b]{\textwidth}
            \centering
            \includegraphics[width=\textwidth]{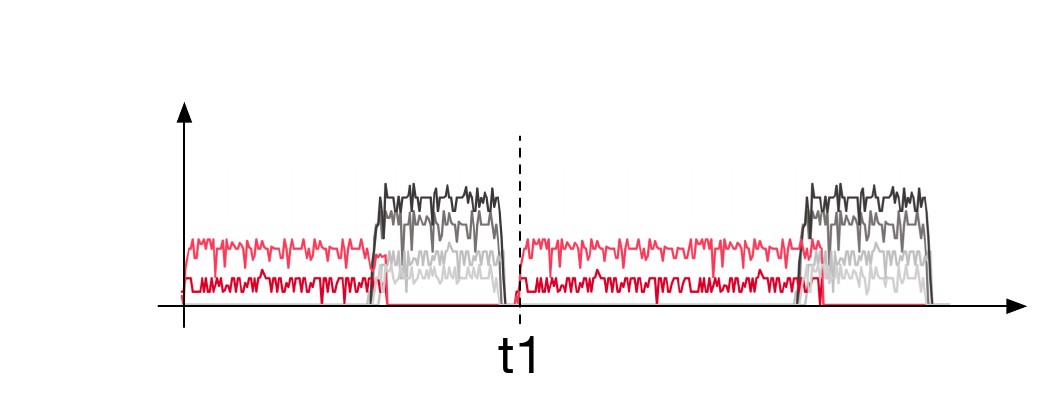}
            \caption{The $ar$ of templates}    
            \label{fig:prediction_a}
        \end{subfigure}
        
        
        \begin{subfigure}[b]{\textwidth}
            \centering
            \includegraphics[width=\textwidth]{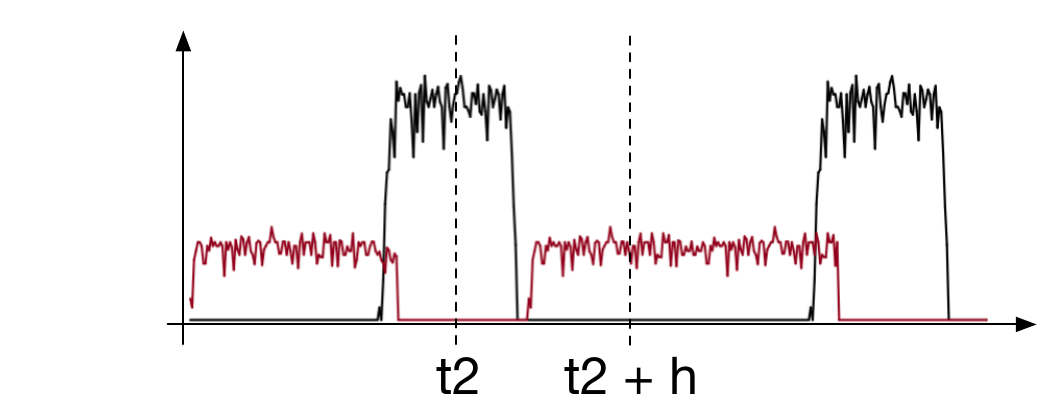}
            \caption{The $ar$ of workloads}
            \label{fig:prediction_b}
        \end{subfigure}
    \end{minipage}%
    \hfill
    \begin{minipage}[b]{0.2\textwidth}
        \centering
        \begin{subfigure}[b]{\textwidth}
            \centering
            \includegraphics[width=\textwidth]{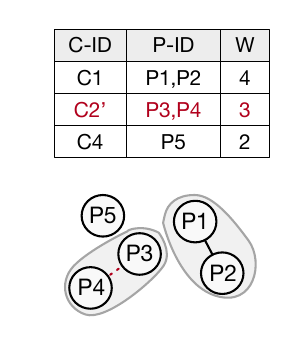}
            \caption{The impact of prediction}
            \label{fig:prediction_c}
        \end{subfigure}
    \end{minipage}
    
    \caption{Example for workload prediction}
    \label{fig:prediction_comparison}
\vspace{-4mm}
\end{figure}

Here, $h$ delineates the prediction horizon, depicting how far into the future a model can forecast.
$ar_k(t,i)$ represents the arrival rate of workload $W_k$ at timestamp $t$ within sampling interval $i$. 
The divergence from timestamp $t$ to $t + h$ is expressed as $wv(t, h)$ across all potential workloads.
When $wv(t, h) > \gamma $, a significant impending workload shift will occur soon, prompting the initiation of pre-replication for forthcoming transactions. 
For example, in Figure~\ref{fig:prediction_b}, the $wv$ reaches its peak at timestamp $t_2$, signifying a drastic alteration in workload expected at the future timestamp $t_2+h$. This indicates an imminent and significant shift in workload dynamics.

\begin{example}
\label{eg3}
Finally, we recap the Example~\ref{eg1} and explain the impact of the prediction mechanism on replica placement rearrangement in Section~\ref{sec:overview}.
In the context of the aforementioned transaction batch and replica placement, the prediction mechanism anticipates that $P_3$ and $P_4$ will be co-accessed soon, corresponding to the transaction $T_3$ that writes to these two partitions.
Consequently, it merges $C_2$ and $C_3$ with a collective weight of 3, as shown in Figure~\ref{fig:prediction_c}. 
During the replica arrangement process, the new $C_2'$ is relocated to $N_3$ to maintain load balance, incurring an additional replication cost of $w_m$. That's why the plan generator instructs $N_3$ to execute $Add:P3 \rightarrow N3$ instead of $N_2$ in Section~\ref{sec:overview}.

\end{example}

\subsection{Batch Optimization}
\label{sec.batch.optimiaztion}
To further reduce the cost of remastering, we've introduced an asynchronous remastering optimization for the batch version of \lion.
The idea is to remaster the partition asynchronously for the whole batch before starts transaction processing, which allows overlapping of network delays and reduces the remastering overhead significantly.

In batch processing, transactions routed to the executors are buffered as a batch and wait to get executed until the global batch epoch is incremented, either by reaching a specified batch size (default 10k) or after a set time window. 
This optimization adds a remastering phase before execution. Upon dispatching a transaction to the executor's buffer queue on the local node, the system checks if it's feasible to convert the transaction into a single-node one, as discussed in Section~\ref{sec:overview}. If viable, the executor sends a remastering request asynchronously; otherwise, it bypasses it directly. 
Unlike the standard execution mode, the executor doesn't stall for the remastering response but proceeds to subsequent incoming transactions. The transaction index within the batch will be piggybacked along with the remastering messages to help the executor locate the transaction context in the buffer.
Once the batch epoch is incremented, the executor proceeds to the next execution phase only after ensuring acknowledgment of all remastering requests. 
To achieve this, barriers are implemented using network round trips between the remastering and execution phases. 
As the execution begins, the executor can locally commit those transactions that have successfully undergone conversion. 
For instance, let's consider the scenario where transactions $T_1$ and $T_2$ are received successively within a time window. Suppose both $T_1$ and $T_2$ meet the conversion conditions. When $T_1$ arrives, the executor initiates the remastering messages. If $T_2$ arrives while $T_1$'s remastering is ongoing, the executor asynchronously triggers the remastering for $T_2$, without waiting for the acknowledgment of $T_1$. 

\section{Implementation}
\label{sec:implementation}

\extended{
Our implementation, built upon the C++ codebase of Star \cite{DBLP:journals/pvldb/LuYM19}, comprises two node types: the distributor node and the executor node. 
Every node is equipped with a global router table, which stores the locations of master and secondary replicas through a hash map.
The distributor node will generate transactions and route the transaction to the appropriate executor node. 
The executor nodes are responsible for partitioned data storage and transaction processing. 
We run multiple threads of three types on these nodes: messenger, worker, and adaptor. 
A messenger thread sends and receives network messages through TCP sockets.
A worker thread is for handling read/write requests. 
We've implemented an adaptor thread that manages data migration and replication with several \texttt{MHandler} functions that can possibly be invoked. 
We can use the \texttt{MigReqHandler()} to migrate partitions from one node to another and use \texttt{AddRepReqHandler()} to add a replication for the given partition on a certain destination node.
\lion operates independently as threads on distributor nodes. It collects transactions and adds them into a \texttt{Clumps} structure to get coarse-grained clumps. 
After that, \texttt{RearrangeFunc()} will be invoked to generate the rearrangement plan ($RP$), which is a list and each entry stands for the target partition ID and its destination.
The $RP$ will be routed to the adaptor on the executor nodes and adjust the replica layout by calling the corresponding \texttt{MHandler} functions. 
\lion adopts an epoch-based group commit mechanism~\cite{DBLP:journals/pvldb/0010Y0M21} to reduce the cost of replication synchronization. 
A global epoch is incremented at 10 millisecond intervals or when reaching a 10k batch size. The committed transactions within an epoch are buffered and asynchronously dispatched to all replicas through \texttt{AsyncReplicationReqHandler()}. Each transaction's outcome remains invisible until the epoch ends and all nodes collectively agree to commit transactions from that specific epoch.
}
\maintext{
Our implementation, built upon the C++ codebase of Star \cite{DBLP:journals/pvldb/LuYM19}, comprises two node types: the distributor node and the executor node. 
The former is for transaction generating and dispatching. The latter is for partitioned data storage and transaction processing. 
We run multiple threads of three types on these nodes: messenger, worker, and adaptor. 
A messenger thread sends and receives network messages through TCP sockets.
A worker thread is for handling read/write requests. 
We've implemented an adaptor thread that manages data migration and replication with several \texttt{MHandler} functions. 
We use \texttt{MigReqHandler()} for partition migration and use \texttt{AddRepReqHandler()} for replication addition, etc.
\lion operates independently as threads on distributor nodes. It collects transactions and generates \texttt{Clumps} structure through workload analysis. 
Then it utilizes \texttt{RearrangeFunc()} to generate a rearrangement plan ($RP$), which is a list and each entry stands for the partition with its new destination.
The $RP$ will be routed to target nodes and the replica layout will be adjusted through the adaptor by calling the corresponding \texttt{MHandler} functions. 
\lion adopts an epoch-based group commit mechanism~\cite{DBLP:journals/pvldb/0010Y0M21} to reduce the cost of replication synchronization. 
A global epoch is incremented at 10 millisecond intervals or when reaching a 10k batch size. The committed transactions within an epoch are buffered and asynchronously dispatched to all replicas through \texttt{AsyncReplicationReqHandler()}. 
}
\extended{
Note that distributed databases, such as TiDB~\cite{TidbDocs} and Oceanbase~\cite{oceanbaseDocs} typically offer APIs for replica remastering and replica management, e.g., adding a replica.
For example, \texttt{transfer-leader} API provided in TiDB can enable replica remastering without the necessity of a leader failure.
While these databases are indeed built upon consensus protocols, their remastering processes generally follow the approach we described in Section~\ref{sec:overview}.
Therefore, \lion can be integrated into these real-world distributed databases by calling the APIs they provide.
}
\maintext{
\textcolor{blue}{
Note 
\marginpar[\textcolor{blue}{R3.D5}]{\textcolor{blue}{R3.D5}}
that distributed databases, such as TiDB~\cite{TidbDocs} and Oceanbase~\cite{oceanbaseDocs} typically offer APIs for replica remastering and replica management, e.g., adding a replica.
For example, \texttt{transfer-leader} API provided in TiDB can enable replica remastering without the necessity of a leader failure.
While these databases are indeed built upon consensus protocols, their remastering processes generally follow the approach we described in Section~\ref{sec:overview}.
Therefore, \lion can be integrated into these real-world distributed databases by calling the APIs they provide.
}
}

\section{Performance Evaluation}
\label{sec:evaluation}

In this section, we evaluate the performance of {\lion} and compare it against state-of-the-art systems.
After introducing the experimental setup, we evaluate {\lion} in a range of settings to demonstrate its superior performance.

\subsection{Experiment Setup}
\label{sec:exp.setup}
We run our experiments on a cluster architecture of multiple nodes. The setup comprises 1 distributor node equipped with 36 Intel(R) Xeon(R) Gold 5220 CPUs and 196 GB of DRAM. Additionally, we employ 10 executor nodes, each featuring 8 Intel Core Processor (Skylake) CPUs with 32 GB of DRAM.
Experiments are conducted using 4 executor nodes by default. Each partition is initially configured to have 2 replicas.
\extended{
We set the max replica number to 4 due to the constraints on the number of nodes used in default experiments.
We use a lightweight LSTM encoder with 2 layers and 20 hidden units for time series prediction and training forecasting models based on the preceding ten-period historical data logs.
}
\maintext{
\textcolor{blue}{
We \marginpar[\textcolor{blue}{R3.D4, R2.O1, R2.D1}]{\textcolor{blue}{R3.D4, R2.O1, R2.D1}}set the max replica number to 4 due to the constraints on the number of nodes used in default experiments.
We use a lightweight LSTM encoder with 2 layers and 20 hidden units for time series prediction and training forecasting models based on the preceding ten-period historical data logs.
}
}
Iperf3 indicates a network throughput of approximately 937 Mbits/sec between each node.
We run 8 worker threads on each executor node, yielding a total of 64 threads. Each node has 2 threads for network communication. 

\subsubsection{Benchmarks}

All experiments are conducted over the following two benchmarks. 


\noindent \textbf{YCSB}.
The Yahoo! Cloud Serving Benchmark (YCSB) is a simplified transactional workload specifically designed to facilitate performance comparisons across various database and key-value systems \cite{DBLP:conf/cloud/CooperSTRS10}. 
In our evaluation, each data node maintains 24 million data items, resulting in a storage space of 200MB per data node. A parameter called \texttt{skew\_factor} is used to control the distribution of the accessed data items. The \texttt{skew\_factor} is set to 0.8 under a skewed workload, resulting in a high load imbalance where 80\% transaction tends to access the partitions in the one node. Under the uniform workload, the \texttt{skew\_factor} is set to 0.  The cross-partitioned transactions always access two partitions.

\noindent \textbf{TPC-C}.
The TPC-C benchmark~\cite{DBLP:conf/tpctc/NambiarWMTLCM11} stands as the industry standard for evaluating OLTP databases. Its dataset comprises 9 relations, and each warehouse is equipped with 100MB of data. By default, we allocate 24 warehouses per node in our experiments. Specifically focusing on NewOrder transactions, the benchmark emulates customers submitting orders to their local district within a warehouse. We simulate scenarios where the same customer makes purchases from different warehouses over time. 

\maintext{

\begin{table}[t]

\vspace{-3mm}
\centering
\renewcommand\arraystretch{1.1}
\caption{\textcolor{blue}{Settings of ablation experiments }} 
\label{tab:ablation}
\vspace{-1mm}
\resizebox{0.60\columnwidth}{!}{%
\begin{tabular}{|c|c|c|c|}
\hline
        Variant  & \makecell{Partitioning\\Strategy}  & \makecell{Workload\\Prediction} & \makecell{Batch\\Optimization}  \\
         \hline \hline 
\ensuremath{\texttt{2PC}\xspace}     & \usym{2717}      & \usym{2717} & \usym{2717} \\ \hline
\ensuremath{\texttt{Lion(S)}\xspace}    & Schism      & \usym{2717} & \usym{2717} \\ \hline
\ensuremath{\texttt{Lion(R)}\xspace}    & Replica Rearrangement        & \usym{2717} & \usym{2717} \\ \hline
\ensuremath{\texttt{Lion(SW)}\xspace}   & Schism      & \usym{2713} & \usym{2717} \\ \hline
\ensuremath{\texttt{Lion(RW)}\xspace}   & Replica Rearrangement        & \usym{2713} & \usym{2717} \\ \hline
\ensuremath{\texttt{Lion(RB)}\xspace}   & Replica Rearrangement        & \usym{2717} & \usym{2713} \\ \hline
\lion      & Replica Rearrangement        & \usym{2713} & \usym{2713} \\ \hline
\end{tabular}%
}
\vspace{-3mm}
\end{table}

\begin{figure}[t]
    \centering
 \begin{subfigure}{1\linewidth}
    \includegraphics[width=\linewidth]{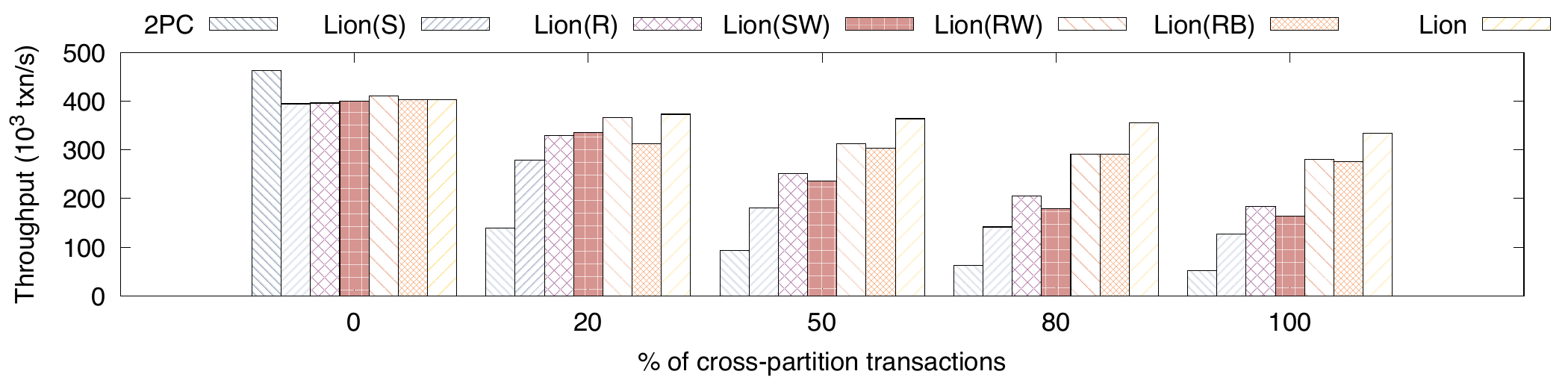}
\end{subfigure}
  \caption{\textcolor{blue}{Performance of ablation experiments}}
  \label{Fig.experiment.ablation}
  \vspace{-6mm}
\end{figure}
}

\extended{

\begin{table}[t]

\vspace{-3mm}
\centering
\renewcommand\arraystretch{1.1}
\caption{Settings of ablation experiments }
\label{tab:ablation}
\vspace{-1mm}
\resizebox{0.60\columnwidth}{!}{%
\begin{tabular}{|c|c|c|c|}
\hline
        Variant  & \makecell{Partitioning\\Strategy}  & \makecell{Workload\\Prediction} & \makecell{Batch\\Optimization}  \\
         \hline \hline 
\ensuremath{\texttt{2PC}\xspace}     & \usym{2717}      & \usym{2717} & \usym{2717} \\ \hline
\ensuremath{\texttt{Lion(S)}\xspace}    & Schism      & \usym{2717} & \usym{2717} \\ \hline
\ensuremath{\texttt{Lion(R)}\xspace}    & Replica Rearrangement        & \usym{2717} & \usym{2717} \\ \hline
\ensuremath{\texttt{Lion(SW)}\xspace}   & Schism      & \usym{2713} & \usym{2717} \\ \hline
\ensuremath{\texttt{Lion(RW)}\xspace}   & Replica Rearrangement        & \usym{2713} & \usym{2717} \\ \hline
\ensuremath{\texttt{Lion(RB)}\xspace}   & Replica Rearrangement        & \usym{2717} & \usym{2713} \\ \hline
\lion      & Replica Rearrangement        & \usym{2713} & \usym{2713} \\ \hline
\end{tabular}%
}
\vspace{-3mm}
\end{table}

\begin{figure}[t]
    \centering
 \begin{subfigure}{1\linewidth}
    \includegraphics[width=\linewidth]{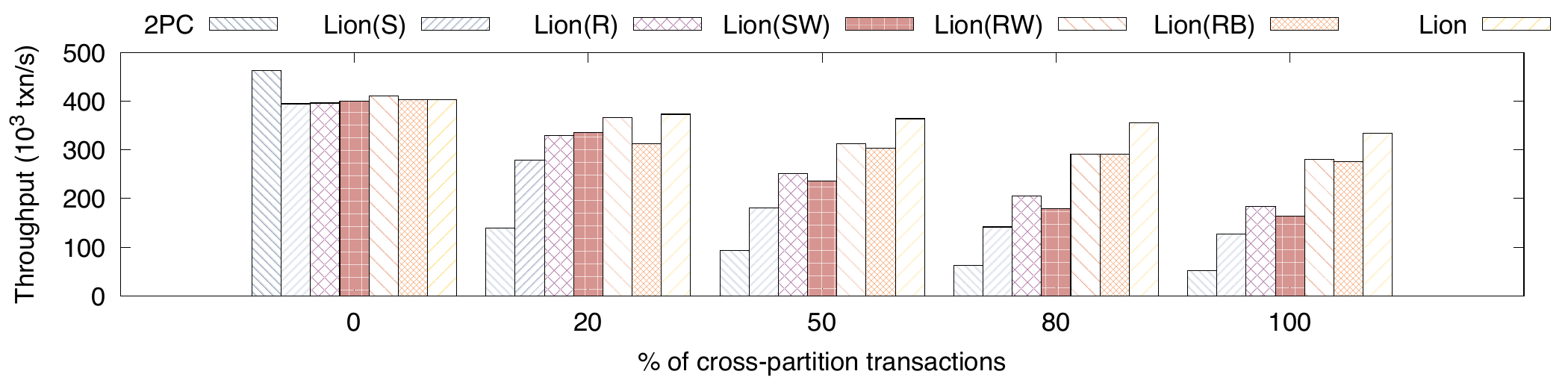}
\end{subfigure}
  \caption{Performance of ablation experiments}
  \label{Fig.experiment.ablation}
  \vspace{-6mm}
\end{figure}

}

\subsubsection{Baselines}
To ensure an apples-to-apples comparison, we implemented existing approaches in the same framework as {\lion}. In our experiments, we use the OCC as the concurrency control algorithm, and compare \lion with both standard execution and batch execution approaches.

\vspace{0.5mm}
\noindent \textit{a. Standard execution approaches.}

\noindent \textbf{2PC}. A classic distributed protocol based on OCC \cite{DBLP:conf/sosp/VandiverBLM07}. The processing of the distributed transaction always undergoes the execute, prepare, and commit phases. 

\noindent \textbf{Leap}. An aggressive transaction management approach. 
Before executing the operations, it always migrates the master replica from the remote node to the local. When all operations are executable, it commits directly and skips the prepare phase.

\maintext{\noindent \textbf{Clay}. An online partitioning approach. It monitors transaction execution and transfers partitions from overloaded nodes to resolve the imbalance. To better compare the partitioning strategy, we implement the asynchronous replication and remastering for Clay as \lion.}

\extended{\noindent \textbf{Clay}. An online partitioning approach. The repartitioning starts when it detects the load imbalance among nodes. Then it generates a partition reconfiguration based on the co-access frequency and adjusts the partitions through data migration. To better compare the cleverness of the reconfiguration, we implement the asynchronous replication and remastering for Clay as \lion.}

\vspace{0.5mm}
\noindent \textit{b. Batch execution approaches.}

\extended{\noindent \textbf{Star}. An asymmetric replication approach with a two-phase switching algorithm. It ensures one node has all the partitions. The transactions will be collected in batches. The distributed transactions in the batch will be routed to that node as the single-node one and get committed without 2PC. }

\maintext{\noindent \textbf{Star}. An asymmetric replication approach. It ensures one ``super node'' to have full replication, where distributed transactions will be executed as the single-node ones without 2PC. }

\extended{\noindent \textbf{Calvin}. A classic distributed deterministic approach. It executes the same transaction batch on each replica to avoid 2PC. It requires the declaration of the read/write set before transaction execution. It uses a lock manager to obtain locks for each transaction in the fixed order and the transaction will not be executed until all locks are acquired.}

\noindent \textbf{Hermes}. A deterministic approach equipped with data migration. It migrates the partition in demand before the lock manager starts to get the locks. It utilizes a prescient transaction routing algorithm to mitigate the ``ping-pong'' effect while achieving load balance.

\extended{\noindent \textbf{Aria}. A distributed deterministic approach. It introduces an optimistic write reservation technique to execute the transactions without coordination and without prior knowledge of a transaction's read/write set.}
\maintext{\noindent \textbf{Aria}. A distributed deterministic approach. It introduces a reservation technique to execute the transactions without coordination and prior knowledge of read/write sets.}

\noindent \textbf{Lotus\cite{DBLP:journals/pvldb/ZhouYGS22}}. 
\extended{
Another distributed epoch-based approach. It is implemented with granule locks to enhance concurrency and introduces batch execution/commit for overlapping computation, communication, and asynchronous replication. }
\maintext{
\textcolor{blue}{
\marginpar[\textcolor{blue}{R2.O2}]{\textcolor{blue}{R2.O2}}
Another distributed epoch-based approach. It is implemented with granule locks to enhance concurrency and introduces batch execution/commit for overlapping computation, communication, and asynchronous replication. }
}

\subsection{Optimization Analysis}
\label{sec.exp.optimization.analysis}

\maintext{
\textcolor{blue}{
We 
\marginpar[\textcolor{blue}{R1.O2}]{\textcolor{blue}{R1.O2}}
first evaluate the effectiveness of three optimizations in \lion: the replica rearrangement algorithm in Section~\ref{sec:design_replacement}, the workload prediction mechanism in Section~\ref{sec:prediction_optimization}, and the batch optimization in Section~\ref{sec.batch.optimiaztion}. 
We conduct comprehensive ablation studies to evaluate the individual contributions of these optimizations in \lion. 
The default workload is uniformed YCSB with 100\% distributed transactions.
We outline all the variants of \lion with different optimizations in Table~\ref{tab:ablation}, and plot the throughput in Figure~\ref{Fig.experiment.ablation}. 
As observed, each optimization improves transaction performance. 
First, compared to \ensuremath{\texttt{2PC}\xspace}, \ensuremath{\texttt{Lion(R)}\xspace} which represents \lion with only our proposed replica rearrangement strategy, demonstrates up to 3.5$\times$ performance improvement.
Second, by additionally leveraging workload prediction in \ensuremath{\texttt{Lion(R)}\xspace}, \ensuremath{\texttt{Lion(RW)}\xspace} shows a performance increase of up to 52.3\% over \ensuremath{\texttt{Lion(R)}\xspace}. 
Third, with the further employment of batch optimization, \ensuremath{\texttt{Lion}\xspace} achieves up to 20\% higher throughput than \ensuremath{\texttt{Lion(RW)}\xspace}. 
}

\textcolor{blue}{
We \marginpar[\textcolor{blue}{R1.O1}]{\textcolor{blue}{R1.O1}} further compare the proposed partitioning strategy with Schism~\cite{DBLP:journals/pvldb/CurinoZJM10}.
We implement an alternation of \lion using Schism as the partitioning strategy, denoted as \ensuremath{\texttt{Lion(S)}\xspace}.
As depicted in Figure~\ref{Fig.experiment.ablation}, \lion outperforms \ensuremath{\texttt{Lion(S)}\xspace} by up to 1.7$\times$. This can be attributed to the fact that, unlike Schism, our partitioning strategy additionally considers replications and takes future transactions into account. 
We further compare \ensuremath{\texttt{Lion(S)}\xspace} with \ensuremath{\texttt{Lion(R)}\xspace}, where \ensuremath{\texttt{Lion(R)}\xspace} represents \lion with only our proposed replica rearrangement strategy, to evaluate the partitioning strategy effectiveness exclusively.
As observed in Figure~\ref{Fig.experiment.ablation}, the replica rearrangement strategy outperforms Schism by up to 31.1\%, primarily because Schism does not account for the placement of secondary replicas, leading to unnecessary migrations. 
We also examine the effectiveness of the prediction mechanism by integrating it with Schism, denoted as \ensuremath{\texttt{Lion(SW)}\xspace}.
As shown in Figure~\ref{Fig.experiment.ablation}, \ensuremath{\texttt{Lion(SW)}\xspace} outperforms \ensuremath{\texttt{Lion(S)}\xspace} by up to 29.4\%, because of the reduced migration cost facilitated by predictions.
}
}
\extended{
We first evaluate the effectiveness of three optimizations in \lion: the replica rearrangement algorithm in Section~\ref{sec:design_replacement}, the workload prediction mechanism in Section~\ref{sec:prediction_optimization}, and the batch optimization in Section~\ref{sec.batch.optimiaztion}. 
We conduct comprehensive ablation studies to evaluate the individual contributions of these optimizations in \lion. 
The default workload is uniformed YCSB with 100\% distributed transactions.
We outline all the variants of \lion with different optimizations in Table~\ref{tab:ablation}, and plot the throughput in Figure~\ref{Fig.experiment.ablation}. 
As observed, each optimization improves transaction performance. 
First, compared to \ensuremath{\texttt{2PC}\xspace}, \ensuremath{\texttt{Lion(R)}\xspace} which represents \lion with only our proposed replica rearrangement strategy, demonstrates up to 3.5$\times$ performance improvement.
Second, by additionally leveraging workload prediction in \ensuremath{\texttt{Lion(R)}\xspace}, \ensuremath{\texttt{Lion(RW)}\xspace} shows a performance increase of up to 52.3\% over \ensuremath{\texttt{Lion(R)}\xspace}. 
Third, with the further employment of batch optimization, \ensuremath{\texttt{Lion}\xspace} achieves up to 20\% higher throughput than \ensuremath{\texttt{Lion(RW)}\xspace}. 

We further compare the proposed partitioning strategy with Schism~\cite{DBLP:journals/pvldb/CurinoZJM10}.
We implement an alternation of \lion using Schism as the partitioning strategy, denoted as \ensuremath{\texttt{Lion(S)}\xspace}.
As depicted in Figure~\ref{Fig.experiment.ablation}, \lion outperforms \ensuremath{\texttt{Lion(S)}\xspace} by up to 1.7$\times$. This can be attributed to the fact that, unlike Schism, our partitioning strategy additionally considers replications and takes future transactions into account. 
We further compare \ensuremath{\texttt{Lion(S)}\xspace} with \ensuremath{\texttt{Lion(R)}\xspace}, where \ensuremath{\texttt{Lion(R)}\xspace} represents \lion with only our proposed replica rearrangement strategy, to evaluate the partitioning strategy effectiveness exclusively.
As observed in Figure~\ref{Fig.experiment.ablation}, the replica rearrangement strategy outperforms Schism by up to 31.1\%, primarily because Schism does not account for the placement of secondary replicas, leading to unnecessary migrations. 
We also examine the effectiveness of the prediction mechanism by integrating it with Schism, denoted as \ensuremath{\texttt{Lion(SW)}\xspace}.
As shown in Figure~\ref{Fig.experiment.ablation}, \ensuremath{\texttt{Lion(SW)}\xspace} outperforms \ensuremath{\texttt{Lion(S)}\xspace} by up to 29.4\%, because of the reduced migration cost facilitated by predictions.
}

\extended{
\begin{figure}[t]
\vspace{-3.5mm}
    \centering
 \begin{subfigure}{0.66\linewidth}
    \includegraphics[width=\linewidth]{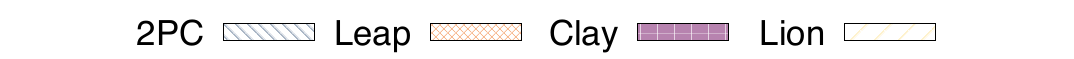}
\end{subfigure}

 \hfill
 \begin{subfigure}{0.48\linewidth}
    \includegraphics[width=\linewidth]{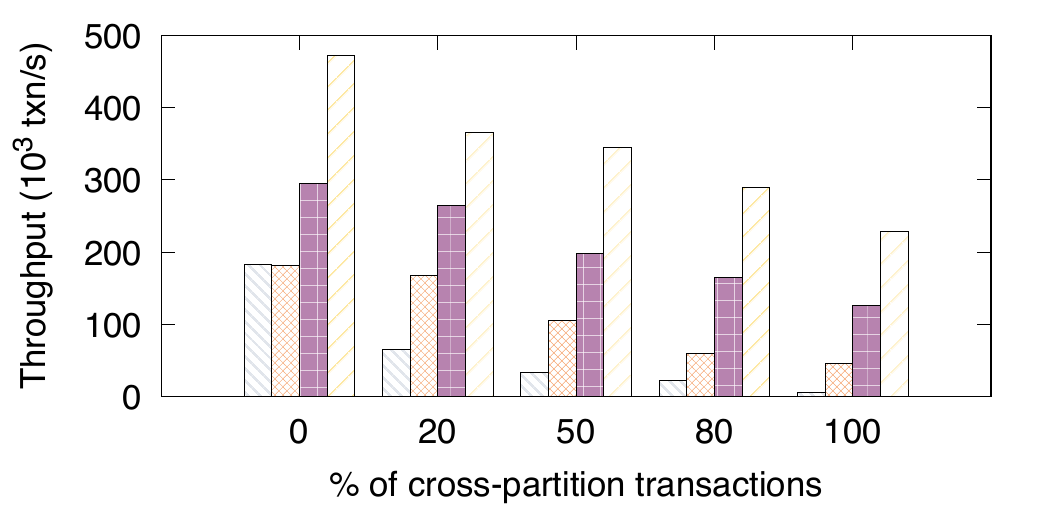}
    \vspace{-6mm}
    \caption{ Skewed workload on YCSB}
    \label{Fig.non.ycsb.skew}
\end{subfigure}
\begin{subfigure}{0.48\linewidth}
    \includegraphics[width=\linewidth]{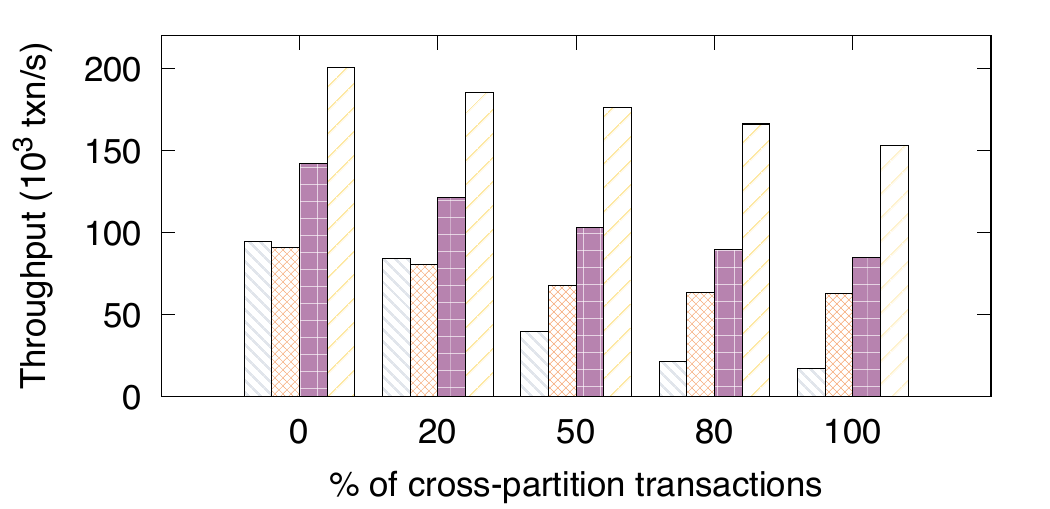}
    \vspace{-6mm}
    \caption{ Skewed workload on TPC-C}
    \label{Fig.non.tpcc.skew}
\end{subfigure}
\hfill
\vspace{-1mm}
  \caption{Impact of varying cross-partition ratios (non-batch)}
  \label{fig.non-deterministic.experiment.throughput}
\vspace{-3mm}
\end{figure}

\begin{figure}[t]
    \centering
\begin{subfigure}{0.58\linewidth}
    \includegraphics[width=\linewidth]{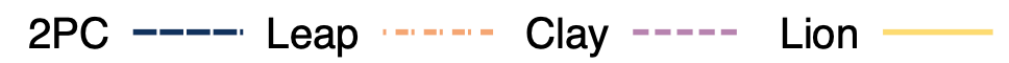}
\end{subfigure}
\begin{subfigure}{0.48\linewidth}
    \includegraphics[width=\linewidth]{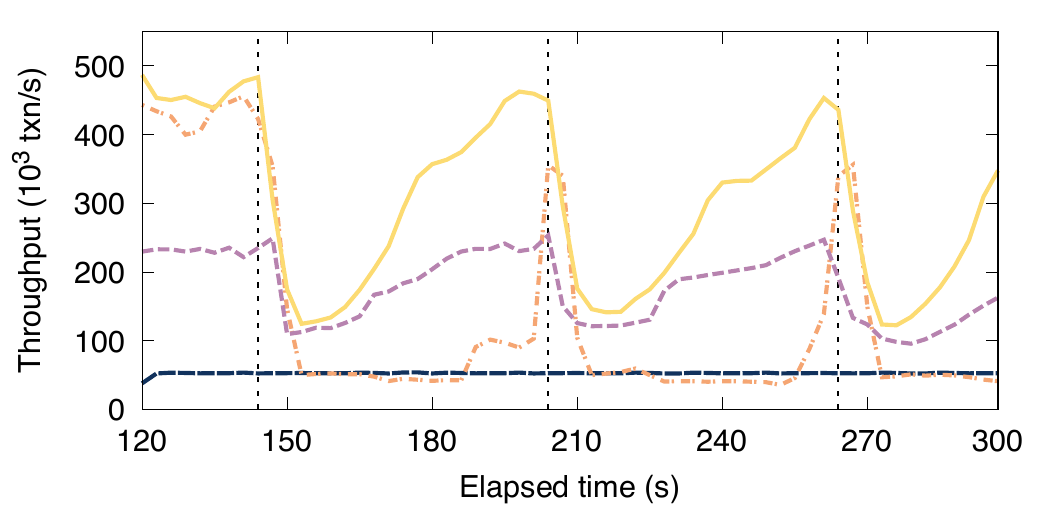}
    \vspace{-6mm}
    \caption{Varying hotspot interval}
    \label{Fig.uniform.timeline}
\end{subfigure}
\begin{subfigure}{0.48\linewidth}
    \includegraphics[width=\linewidth]{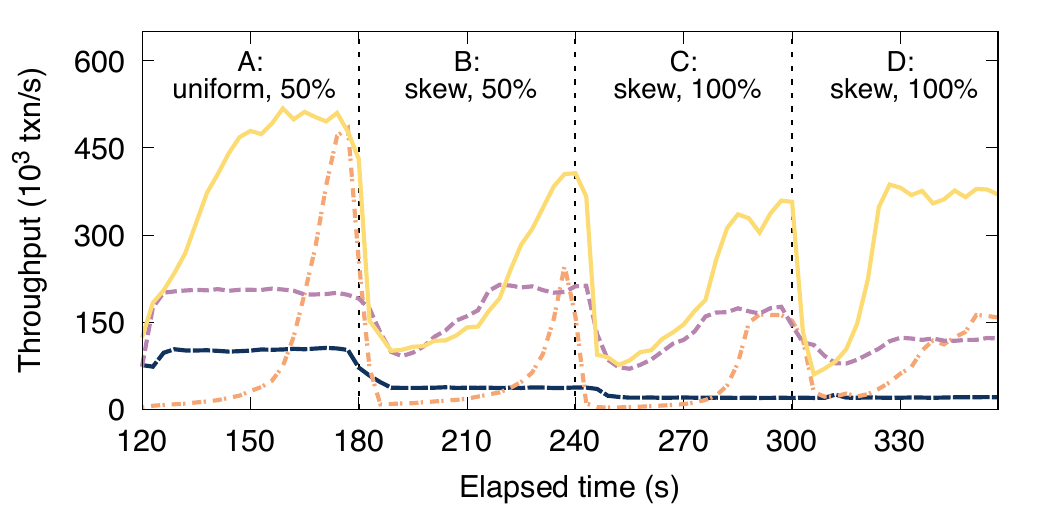}
    \vspace{-6mm}
    \caption{Varying hotspot position}
    \label{Fig.shift.timeline}
\end{subfigure}
\vspace{-1mm}    
  \caption{Impact of dynamic workloads (non-batch)}
  \label{Fig.experiment.nondmtimeline}
\vspace{-6mm}
\end{figure}

}

\subsection{Overall Performance}
\label{sec.exp.comparison.non-deter}
We now compare \lion with other standard execution approaches using both YCSB and TPC-C workloads.

\extended{
\subsubsection{Workloads with varying the cross-partition ratio}
}
\maintext{
\subsubsection{\textcolor{blue}{Workloads with varying the cross-partition ratio}}
}
\label{sec.exp.comparison.non-deter.cross-ratio}
\extended{
We first measure the throughput by increasing the percentage of cross-partition transactions under skewed workloads.
The experiments are conducted on YCSB and TPC-C with an 80\% \texttt{skew\_factor}. We set the default remastering delay to 3000 microseconds to simulate the remastering overhead in real-world scenarios.
The results in Figure~\ref{fig.non-deterministic.experiment.throughput} demonstrate that \lion achieves up to 1.9$\times$ higher throughput than other approaches. 
\lion is designed to eliminate distributed transactions and uses the replica rearrangement algorithm to dissipate imbalances. 
2PC costs expensive network coordination to process distributed transactions. 
Leap and Clay exhibit better performance since they can adapt to the workload by migration. 
However, Leap's aggressive strategy does not consider load balance.
Clay struggles to eliminate all distributed transactions since it perceives the overloaded node running single-node transactions as having an equal load to nodes with fewer distributed transactions.
}
\maintext{
\textcolor{blue}{
We \marginpar[\textcolor{blue}{R3.D6}]{\textcolor{blue}{R3.D6}} first measure the throughput by increasing the percentage of cross-partition transactions under skewed workloads.
The experiments are conducted on YCSB and TPC-C with an 80\% \texttt{skew\_factor}. We set the default remastering delay to 3000 microseconds to simulate the remastering overhead in real-world scenarios.
The results in Figure~\ref{fig.non-deterministic.experiment.throughput} demonstrate that \lion achieves up to 1.9$\times$ higher throughput than other approaches. 
\lion is designed to eliminate distributed transactions and uses the replica rearrangement algorithm to dissipate imbalances. 
2PC costs expensive network coordination to process distributed transactions. 
Leap and Clay exhibit better performance since they can adapt to the workload by migration. 
However, Leap's aggressive strategy does not consider load balance.
Clay struggles to eliminate all distributed transactions since it perceives the overloaded node running single-node transactions as having an equal load to nodes with fewer distributed transactions.
}

}

\maintext{
\subsubsection{\textcolor{blue}{Dynamic workloads with a changing hotspot}}
}
\extended{
\subsubsection{Dynamic workloads with a changing hotspot}
}
\label{sec.exp.comparison.non-deter.dynamicworkload}
\extended{
We evaluate \lion's performance under two dynamically changing workload scenarios. In each scenario, the workload cycles through multiple periods, alternating every 60 seconds. Each period features distinct access patterns, with transactions accessing non-overlapping partitions to create unique hotspots.
In the varying hotspot interval scenario, we create three custom queries with a uniform access pattern. The partition ID intervals within each query are fixed in one period but shift among different periods.
In the varying hotspot position scenario, we design a combined workload to mimic changes in the most frequently accessed keys in the Zipfian distribution. This workload consists of four periods (A, B, C, D), encompassing uniform access with a 50\% cross-ratio, skew with a 50\% cross-ratio, skew with a 100\% cross-ratio, and skew with a 100\% cross-ratio with distribution shift via partition ID offsets.

We then evaluate \lion's throughput fluctuates over time within dynamic workloads. 
As shown in Figure~\ref{Fig.experiment.nondmtimeline}, \lion can adapt to new workloads faster as well as maintain higher stable throughput, due to its prediction mechanism and wise partitioning strategy.
In contrast, 2PC's performance remains consistently low since it fails to adapt to workloads in Figure~\ref{Fig.uniform.timeline}. 
Leap's aggressive migration disrupts transactions and prolongs jitter, evident in skew scenarios B, C, and D (as shown in Figure~\ref{Fig.shift.timeline}).
Clay can not eradicate all distributed transactions when facing cross-partitioned and skewed workloads (as shown in Figure~\ref{Fig.uniform.timeline} and C, D in Figure~\ref{Fig.shift.timeline}).
}

\maintext{
\textcolor{blue}{
We 
\marginpar[\textcolor{blue}{R2.O3}, \textcolor{blue}{R3.D7}]{\textcolor{blue}{R2.O3}, \textcolor{blue}{R3.D7}}
evaluate \lion's performance under two dynamically changing workload scenarios. In each scenario, the workload cycles through multiple periods, alternating every 60 seconds. Each period features distinct access patterns, with transactions accessing non-overlapping partitions to create unique hotspots.
In the varying hotspot interval scenario, we create three custom queries with a uniform access pattern. The partition ID intervals within each query are fixed in one period but shift among different periods.
In the varying hotspot position scenario, we design a combined workload to mimic changes in the most frequently accessed keys in the Zipfian distribution. This workload consists of four periods (A, B, C, D), encompassing uniform access with a 50\% cross-ratio, skew with a 50\% cross-ratio, skew with a 100\% cross-ratio, and skew with a 100\% cross-ratio with distribution shift via partition ID offsets.
}

\textcolor{blue}{
We then evaluate \lion's throughput fluctuates over time within dynamic workloads. 
As shown in Figure~\ref{Fig.experiment.nondmtimeline}, \lion can adapt to new workloads faster as well as maintain higher stable throughput, due to its prediction mechanism and wise partitioning strategy.
In contrast, 2PC's performance remains consistently low since it fails to adapt to workloads in Figure~\ref{Fig.uniform.timeline}. 
Leap's aggressive migration disrupts transactions and prolongs jitter, evident in skew scenarios B, C, and D (as shown in Figure~\ref{Fig.shift.timeline}).
Clay can not eradicate all distributed transactions when facing cross-partitioned and skewed workloads (as shown in Figure~\ref{Fig.uniform.timeline} and C, D in Figure~\ref{Fig.shift.timeline}).
}

}
\maintext{
\begin{figure}[t]
\vspace{-3.5mm}
    \centering
 \begin{subfigure}{0.66\linewidth}
    \includegraphics[width=\linewidth]{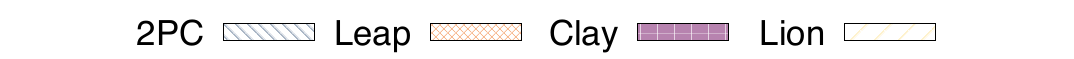}
\end{subfigure}

 \hfill
 \begin{subfigure}{0.48\linewidth}
    \includegraphics[width=\linewidth]{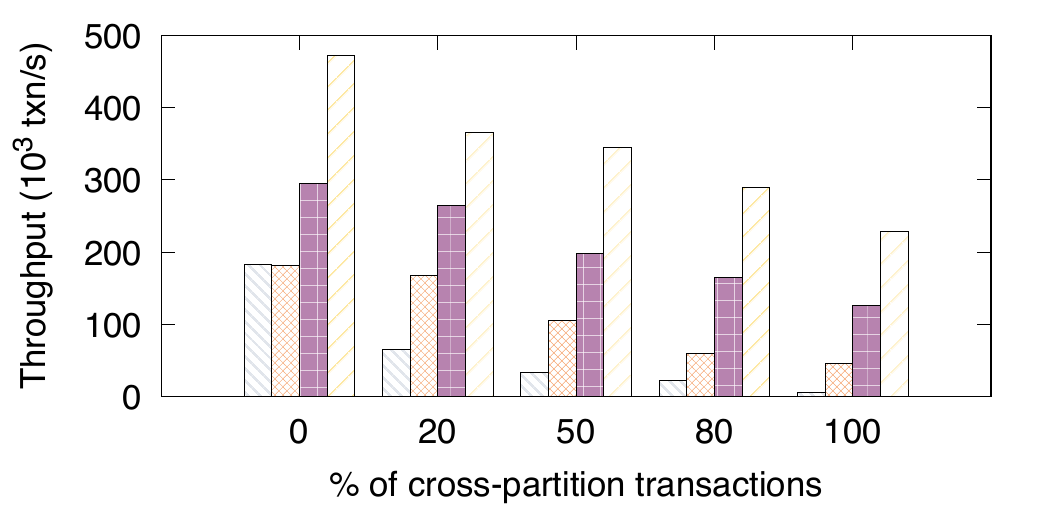}
    \vspace{-6mm}
    \caption{ Skewed workload on YCSB}
    \label{Fig.non.ycsb.skew}
\end{subfigure}
\begin{subfigure}{0.48\linewidth}
    \includegraphics[width=\linewidth]{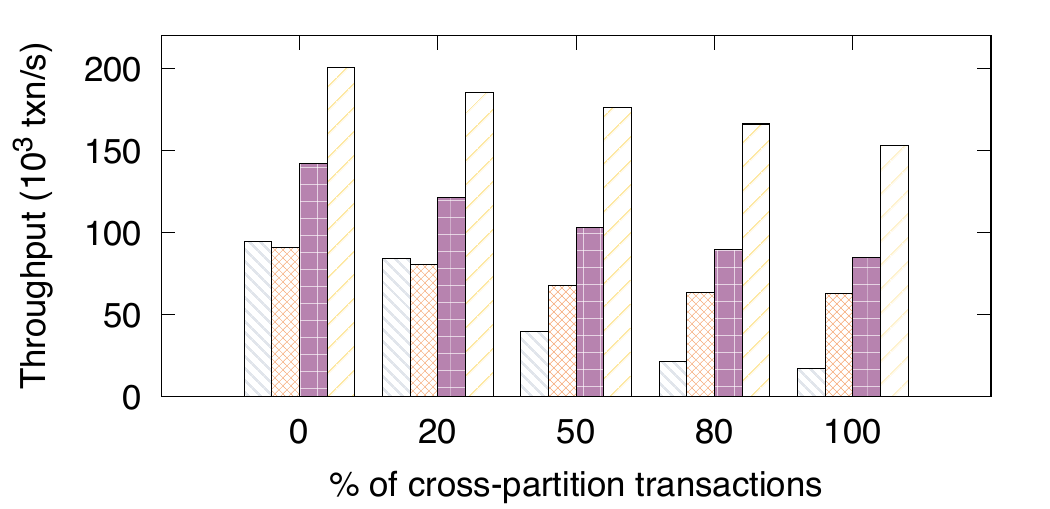}
    \vspace{-6mm}
    \caption{ Skewed workload on TPC-C}
    \label{Fig.non.tpcc.skew}
\end{subfigure}
\hfill
\vspace{-1mm}
  \caption{\textcolor{blue}{Impact of varying cross-partition ratios (non-batch)}}
  \label{fig.non-deterministic.experiment.throughput}
\vspace{-3mm}
\end{figure}

\begin{figure}[t]
    \centering
\begin{subfigure}{0.58\linewidth}
    \includegraphics[width=\linewidth]{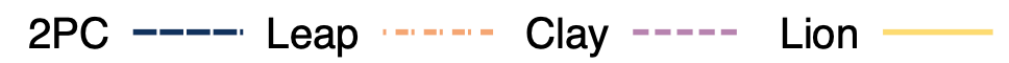}
\end{subfigure}
\begin{subfigure}{0.48\linewidth}
    \includegraphics[width=\linewidth]{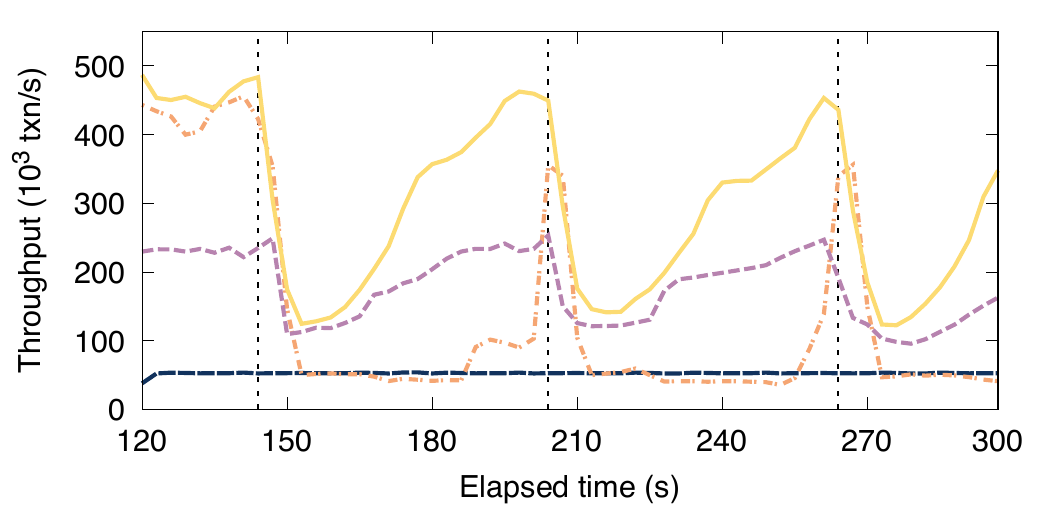}
    \vspace{-6mm}
    \caption{Varying hotspot interval}
    \label{Fig.uniform.timeline}
\end{subfigure}
\begin{subfigure}{0.48\linewidth}
    \includegraphics[width=\linewidth]{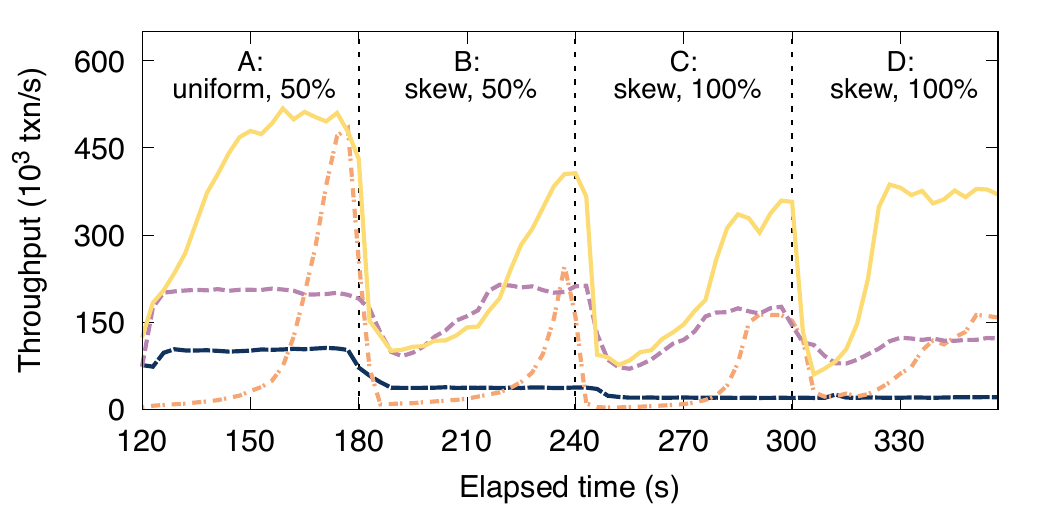}
    \vspace{-6mm}
    \caption{Varying hotspot position}
    \label{Fig.shift.timeline}
\end{subfigure}
\vspace{-1mm}    
  \caption{\textcolor{blue}{Impact of dynamic workloads (non-batch)}}
  \label{Fig.experiment.nondmtimeline}
\vspace{-6mm}
\end{figure}

}

\maintext{

\begin{figure}[t]
    \centering
    \vspace{-3.5mm}
 \begin{subfigure}{0.78\linewidth}
    \includegraphics[width=\linewidth]{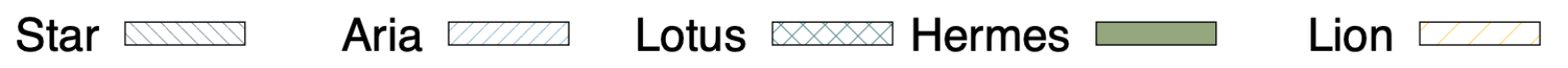}
    \vspace{-4mm}
\end{subfigure}
 \hfill
\begin{subfigure}{0.48\linewidth}
    \includegraphics[width=\linewidth]{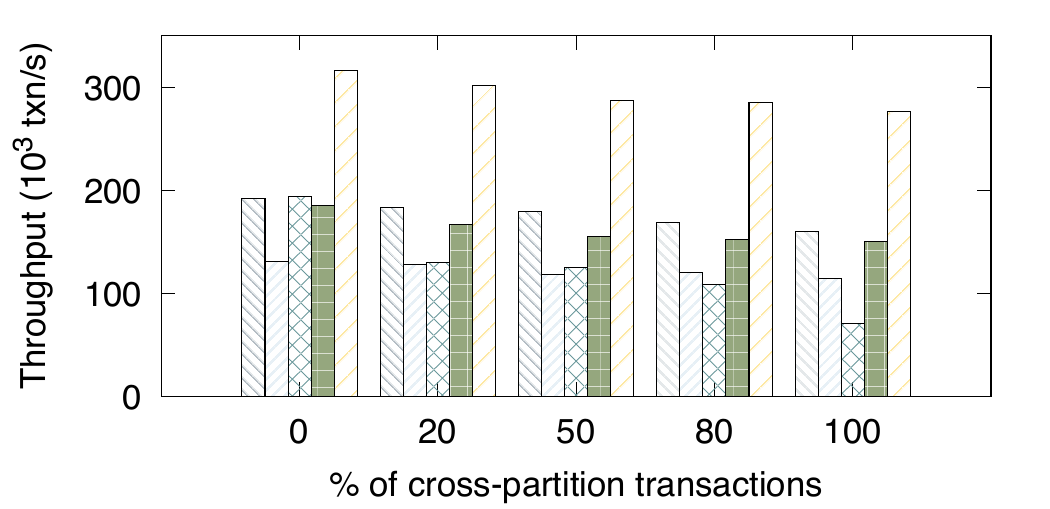}
    \vspace{-6mm}
    \caption{ Skewed workload on YCSB}
    \label{Fig.dm.ycsb.skew}
\end{subfigure}
\begin{subfigure}{0.48\linewidth}
    \includegraphics[width=\linewidth]{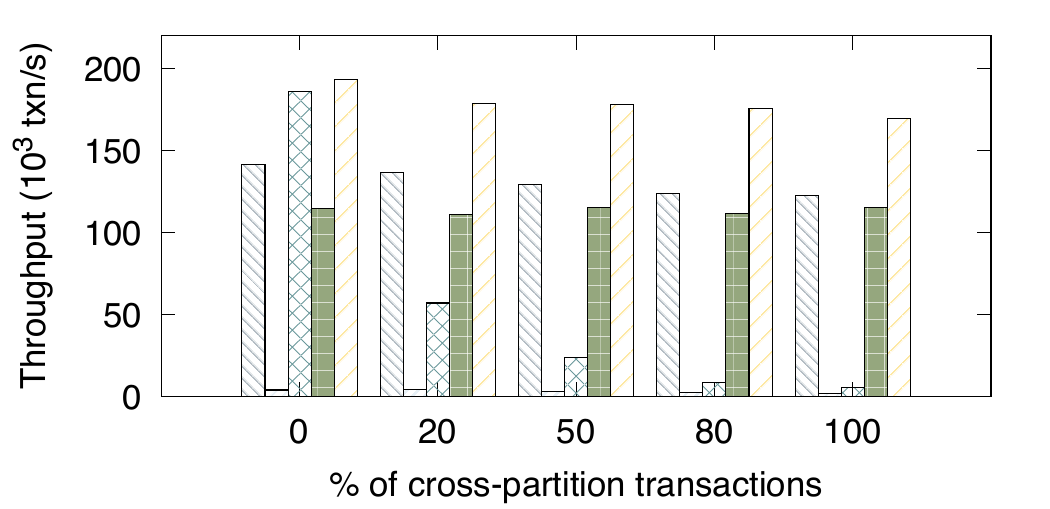}
    \vspace{-6mm}
    \caption{ Skewed workload on TPC-C}
    \label{Fig.dm.tpcc.skew}
\end{subfigure}
  \vspace{-1mm}
  \caption{\textcolor{blue}{Impact of varying cross-partition ratios (batch)}}
  \label{Fig.dm.experiment.throughput}
  \vspace{-3mm}
\end{figure}

\begin{figure}[t]
    \centering
\begin{subfigure}{0.78\linewidth}
    \includegraphics[width=\linewidth]{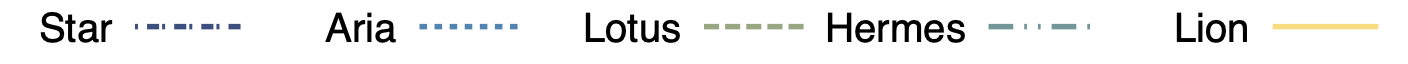}
\end{subfigure}
\begin{subfigure}{0.48\linewidth}
    \includegraphics[width=\linewidth]{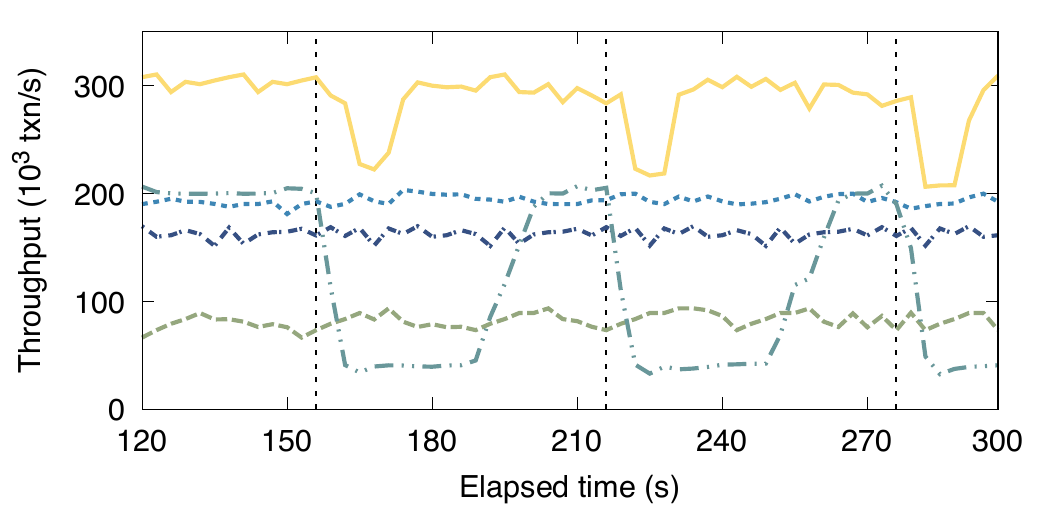}
    \vspace{-6mm}
    \caption{Varying hotspot interval}
    \label{Fig.ycsb.timeline.dm}
\end{subfigure}
\begin{subfigure}{0.48\linewidth}
    \includegraphics[width=\linewidth]{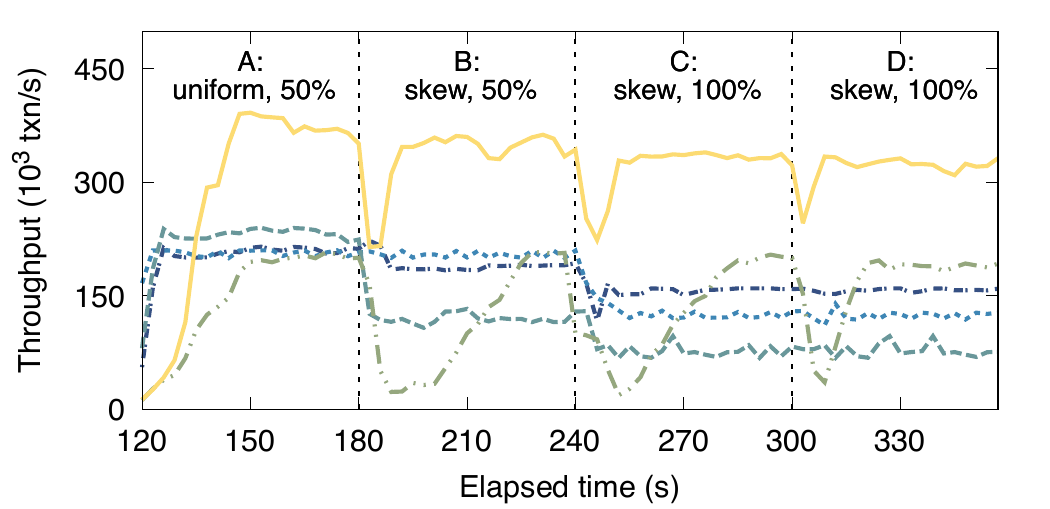}
    \vspace{-6mm}
    \caption{Varying hotspot position}
    \label{Fig.ycsb.shift.timeline.dm}
\end{subfigure}
\vspace{-1mm}
  \caption{\textcolor{blue}{Impact of dynamic workloads (batch)}}
  \label{Fig.experiment.timeline}
  \vspace{-6mm}
\end{figure}
}

\extended{
\begin{figure}[t]
\vspace{-3mm}
    \centering
 \begin{subfigure}{0.88\linewidth}
    \includegraphics[width=\linewidth]{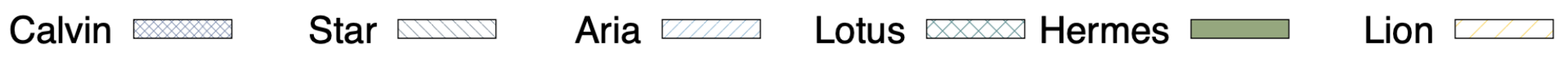}
\end{subfigure}
 \hfill
\begin{subfigure}{0.48\linewidth}
    \includegraphics[width=\linewidth]{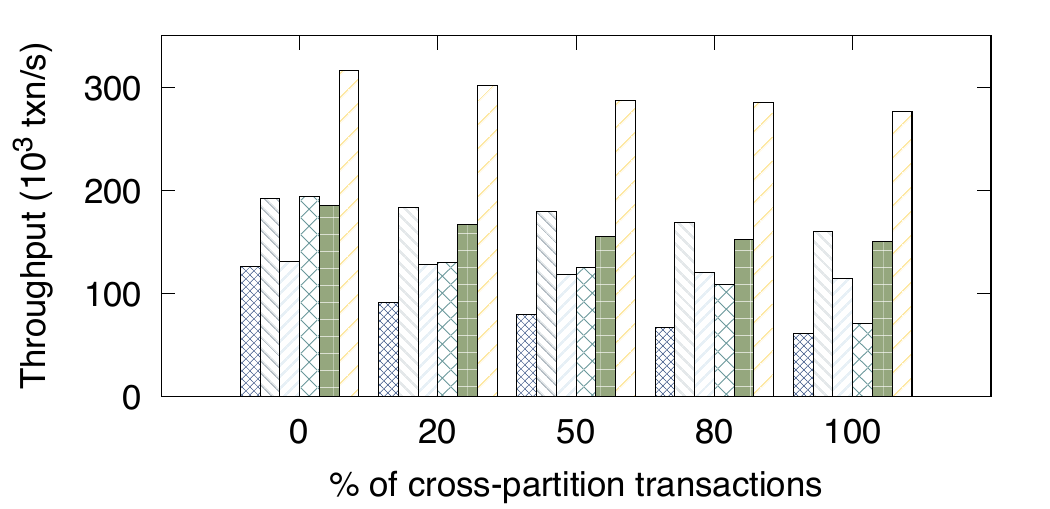}
    \vspace{-6mm}
    \caption{ Skewed workload on YCSB}
    \label{Fig.dm.ycsb.skew}
\end{subfigure}
\begin{subfigure}{0.48\linewidth}
    \includegraphics[width=\linewidth]{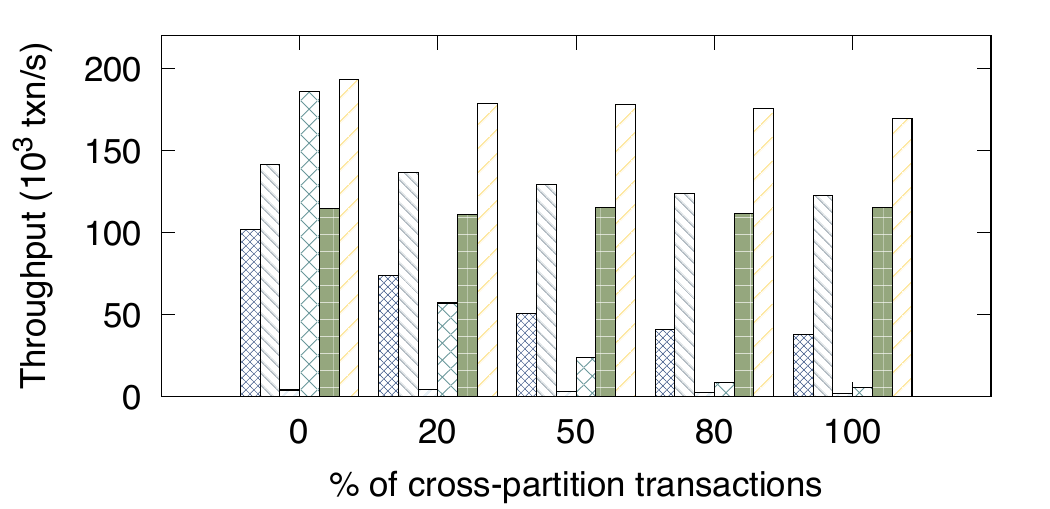}
    \vspace{-6mm}
    \caption{ Skewed workload on TPC-C}
    \label{Fig.dm.tpcc.skew}
\end{subfigure}
  \vspace{-1mm}
  \caption{Impact of varying cross-partition ratios (batch)}
  \label{Fig.dm.experiment.throughput}
  \vspace{-3mm}
\end{figure}
}

\extended{
\begin{figure}[t]
    \centering
\begin{subfigure}{0.78\linewidth}
    \includegraphics[width=\linewidth]{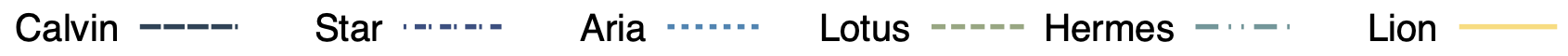}
\end{subfigure}
\begin{subfigure}{0.48\linewidth}
    \includegraphics[width=\linewidth]{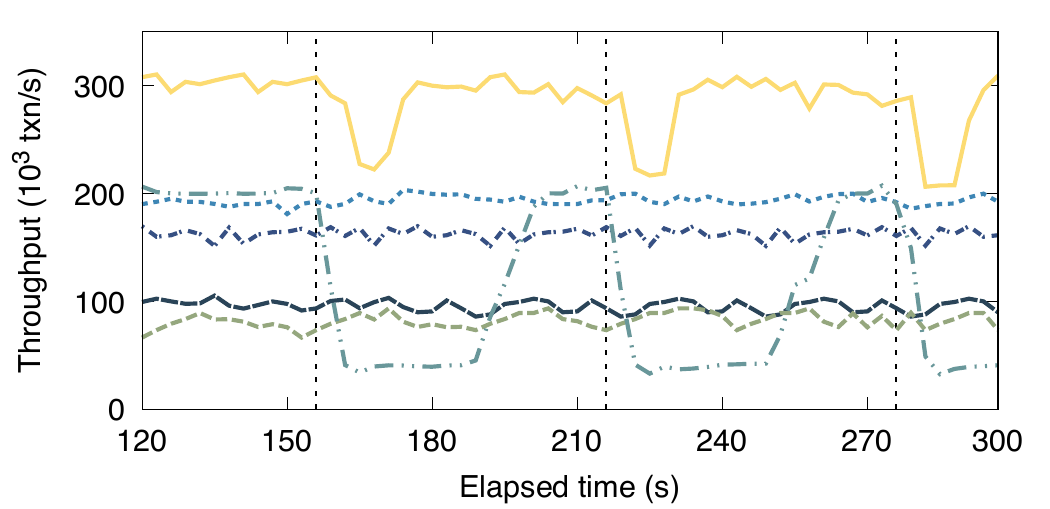}
    \vspace{-6mm}
    \caption{Varying hotspot interval}
    \label{Fig.ycsb.timeline.dm}
\end{subfigure}
\begin{subfigure}{0.48\linewidth}
    \includegraphics[width=\linewidth]{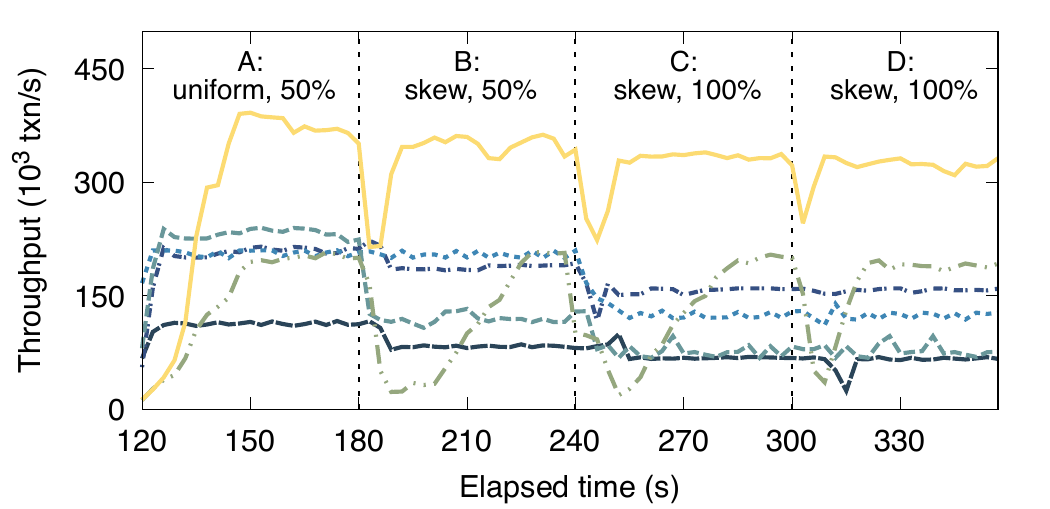}
    \vspace{-6mm}
    \caption{Varying hotspot position}
    \label{Fig.ycsb.shift.timeline.dm}
\end{subfigure}
\vspace{-1mm}
  \caption{Impact of dynamic workloads (batch)}
  \label{Fig.experiment.timeline}
  \vspace{-6mm}
\end{figure}
}

\subsection{Comparison with Batch Execution Approaches}
\label{sec.exp.comparison.deter}
We next compare \lion with other batch execution approaches, including Star, Lotus, and other deterministic methods. 
We deploy a single-threaded lock manager for all deterministic methods to grant locks to multiple executors following the deterministic order.  
To make a fair comparison, we implemented a batch-processing version of \lion which is introduced in Section~\ref{sec.batch.optimiaztion}. 
The batch size is set to be 10k transactions.

\subsubsection{Workloads with varying the cross-partition ratio}
We measure the throughput following the same way in \ref{sec.exp.comparison.non-deter.cross-ratio}.

\extended{
The results shown in Figure~\ref{Fig.dm.experiment.throughput} show that \lion has up to 1.7$\times$ higher than the next-best approach. 
\lion distributes unrelated co-located partitions evenly among nodes, which enables all workers in each node can independently process transactions with high parallelism. 
Calvin avoids the 2PC through deterministic execution but still suffers from the remote read caused by distributed transactions. 
As illustrated in Figure~\ref{Fig.dm.ycsb.skew} and Figure~\ref{Fig.dm.tpcc.skew}, the performance of Star and Hermes remains stable when varying the cross-ratio. 
Because they eliminate distributed transactions either through the full replication or the migration. 
But their throughput is limited by the bottleneck of a ``super node'' or a single lock manager.
Aria and Lotus perform well in low cross-ratio scenarios due to their specific epoch-based transaction processing. 
However, their performance decreases significantly as the cross-ratio increases, because they require a costly commit protocol for distributed transactions and lack optimizations for load balancing.
Moreover, their epoch-based schema exacerbates contention. For instance, Lotus maintains locks until the end of an epoch, leading to transaction aborts and re-executions, as evident in Figure~\ref{Fig.latency.ycsb.latency}.
}

\maintext{
\textcolor{blue}{
The  \marginpar[\textcolor{blue}{R2.O2}]{\textcolor{blue}{R2.O2}}results shown in Figure~\ref{Fig.dm.experiment.throughput} show that \lion has up to 1.7$\times$ higher than the next-best approach. 
\lion distributes unrelated co-located partitions evenly among nodes, which enables all workers in each node can independently process transactions with high parallelism. 
As illustrated in Figure~\ref{Fig.dm.ycsb.skew} and Figure~\ref{Fig.dm.tpcc.skew}, the performance of Star and Hermes remains stable when varying the cross-ratio. 
Because they eliminate distributed transactions either through the full replication or the migration. 
But their throughput is limited by the bottleneck of a ``super node'' or a single lock manager.
Aria and Lotus perform well in low cross-ratio scenarios due to their specific epoch-based transaction processing. 
However, their performance decreases significantly as the cross-ratio increases, because they require a costly commit protocol for distributed transactions and lack optimizations for load balancing.
Moreover, their epoch-based schema exacerbates contention. For instance, Lotus maintains locks until the end of an epoch, leading to transaction aborts and re-executions, as evident in Figure~\ref{Fig.latency.ycsb.latency}.
}
}

\extended{
\begin{figure}[t]
    \centering
\vspace{-4mm}
\begin{subfigure}{0.46\linewidth}
    \vspace{1mm}
    \includegraphics[width=\linewidth]{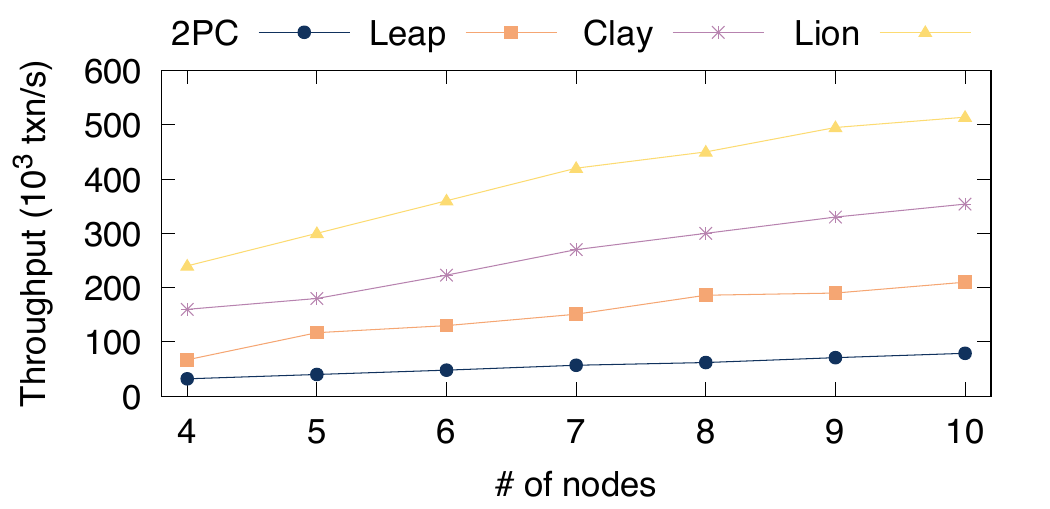}
    \vspace{-6mm}
    \caption{Standard approaches}
    \label{Fig.exp.scale.standard}
\end{subfigure}
\begin{subfigure}{0.5\linewidth}
    \includegraphics[width=\linewidth]{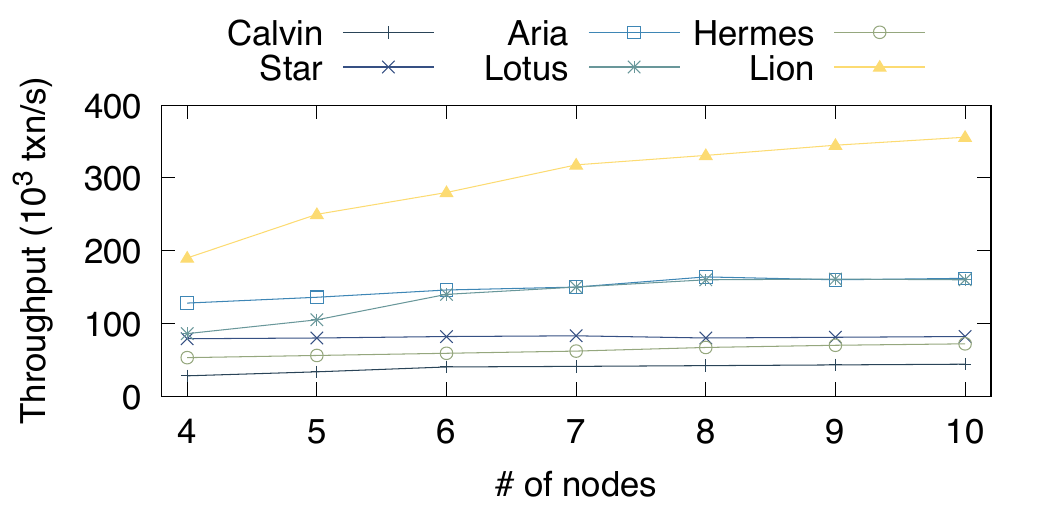}
    \vspace{-6mm}
    \caption{Batch-based approaches}
    \label{Fig.exp.scale.deter}
\end{subfigure}
\vspace{-1mm}
  \caption{Scalability}
  \label{Fig.exp.scale}
  \vspace{-4mm}
\end{figure}
}

\extended{
\begin{figure}[t]
    \centering

\begin{subfigure}{0.48\linewidth}
    \includegraphics[width=\linewidth]{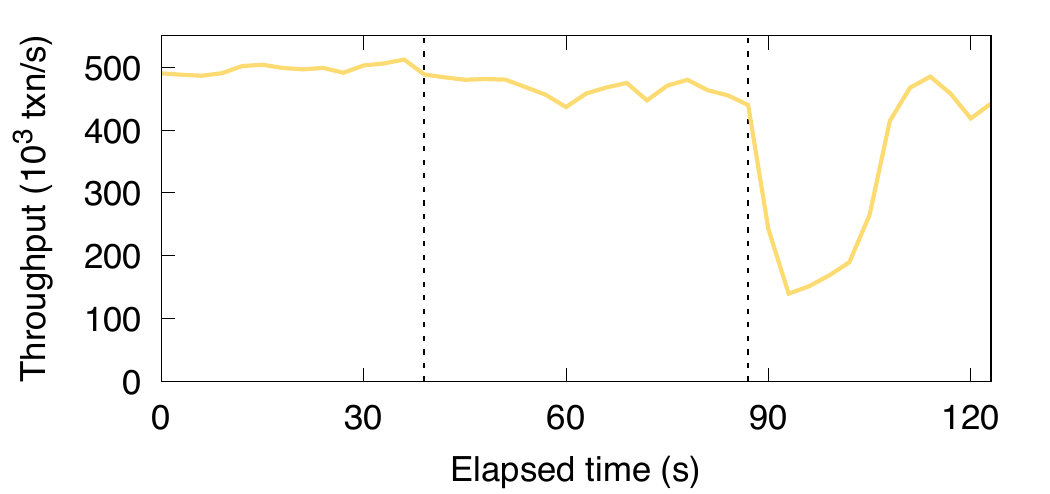}
    \vspace{-6mm}
    \caption{Throughput}
    \label{Fig.ycsb.migration_process.throughput}
\end{subfigure}
\begin{subfigure}{0.48\linewidth}
    \vspace{-0.5mm}
    \includegraphics[width=\linewidth]{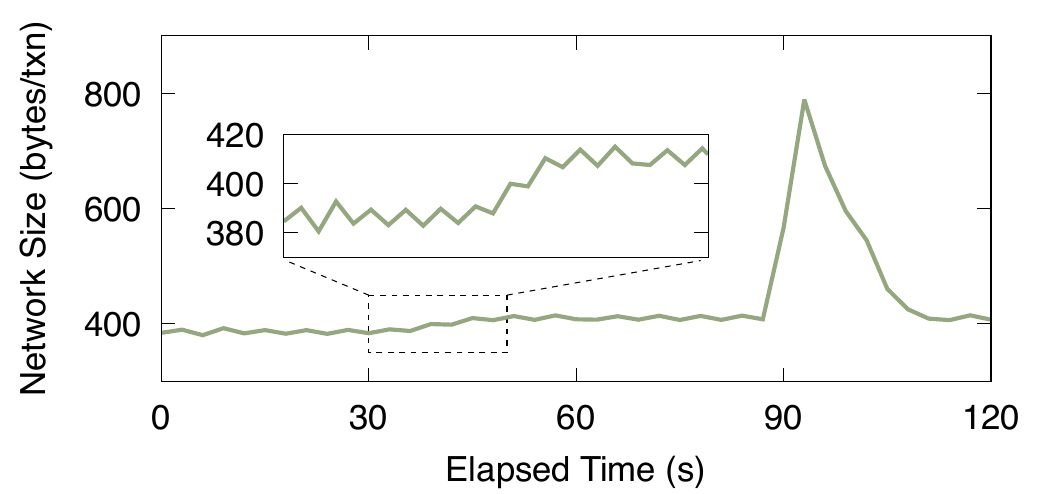}
    \vspace{-6mm}
    \caption{Network cost}
    \label{Fig.ycsb.migration_process.network_cost}
\end{subfigure}
\vspace{-1mm}
  \caption{The analysis of migration process}
  \label{Fig.experiment.migration_process.timeline}
  \vspace{-6mm}
\end{figure}

}

\extended{
\subsubsection{Dynamic workloads with a changing hotspot}
}
\maintext{
\subsubsection{\textcolor{blue}{Dynamic workloads with a changing hotspot}}
}
\label{sec.exp.comparison.deter.dynamicworkload}
\extended{
Following the workload introduced in Section~\ref{sec.exp.comparison.non-deter.dynamicworkload}, we assess \lion's performance using deterministic and batch-based approaches. 
Notably, \lion can adapt to the new workload with 4$\times$ faster than its non-batch version, experiencing a mere 26.6\% throughput degradation in Figure~\ref{Fig.ycsb.timeline.dm}. That can be attributed to its batch optimization with asynchronous remastering. 
%
Hermes shows its superior performance against other baselines. Because it eliminates distributed transactions through migration in a deterministic pre-defined order for each replica group. 
However, it experiences a severe performance jitter when adapting to new workloads due to its deterministic migration.  
Subsequent transactions will be unable to proceed until the preceding distributed transactions complete migration, which leads to a staggering 62.9\% even lower performance compared to Lotus(as shown in Figure~\ref{Fig.ycsb.timeline.dm} and the B-C switching boundary in Figure~\ref{Fig.ycsb.shift.timeline.dm}).
Others suffer from poor performance since they fail to adapt to the workload. 
Additionally, they overlook the load-balancing problem when facing skew workloads (as depicted in scenarios B, C, and D in Figure~\ref{Fig.ycsb.shift.timeline.dm}).
}
\maintext{
\textcolor{blue}{
Following 
\marginpar[\textcolor{blue}{R2.O3, R3.D7}]{\textcolor{blue}{R2.O3, R3.D7}}
the workload introduced in Section~\ref{sec.exp.comparison.non-deter.dynamicworkload}, we assess \lion's performance using deterministic and batch-based approaches. 
Notably, \lion can adapt to the new workload with 4$\times$ faster than its non-batch version, experiencing a mere 26.6\% throughput degradation in Figure~\ref{Fig.ycsb.timeline.dm}. That can be attributed to its batch optimization with asynchronous remastering. 
%
Hermes shows its superior performance against other baselines. Because it eliminates distributed transactions through migration in a deterministic pre-defined order for each replica group. 
However, it experiences a severe performance jitter when adapting to new workloads due to its deterministic migration.  
Subsequent transactions will be unable to proceed until the preceding distributed transactions complete migration, which leads to a staggering 62.9\% even lower performance compared to Lotus(as shown in Figure~\ref{Fig.ycsb.timeline.dm} and the B-C switching boundary in Figure~\ref{Fig.ycsb.shift.timeline.dm}).
Others suffer from poor performance since they fail to adapt to the workload. 
Additionally, they overlook the load-balancing problem when facing skew workloads (as depicted in scenarios B, C, and D in Figure~\ref{Fig.ycsb.shift.timeline.dm}).
}
}

\subsection{Scalability}
We now study the scalability of each approach, varying the number of executor nodes from 4 to 10 under the same workload (100\% cross-partition with uniform access pattern) detailed in Sections~\ref{sec.exp.comparison.deter} and \ref{sec.exp.comparison.deter}. 
Observing the results as presented in Figure~\ref{Fig.exp.scale}, \lion achieves approximately 2$\times$ higher throughput with 10 nodes compared to the scenario involving 4 nodes and has up to 76.4\% better scalability than other approaches.
This superiority stems from \lion's replica rearrangement strategy, considering distributed transaction elimination and load balancing. This approach ensures optimized performance across each node, and this cumulative advantage becomes more pronounced with an increase in the number of nodes. 
We note that the throughput of all non-deterministic approaches scales almost linearly as the number of nodes increases. However, Star demonstrates poor scalability due to a bottleneck arising from the super node. Deterministic approaches reach their limit post an increase in the number of nodes to 7, mainly resulting from the predefined order established by the sequencer and lock manager, which becomes less conducive in larger node clusters.

\maintext{
\begin{figure}[t]
    \centering
\vspace{-4mm}
\begin{subfigure}{0.46\linewidth}
    \vspace{1mm}
    \includegraphics[width=\linewidth]{figures/exp/ycsb/single_nondm/single_nondm_dist_scale.pdf}
    \vspace{-6mm}
    \caption{Standard approaches}
    \label{Fig.exp.scale.standard}
\end{subfigure}
\begin{subfigure}{0.5\linewidth}
    \includegraphics[width=\linewidth]{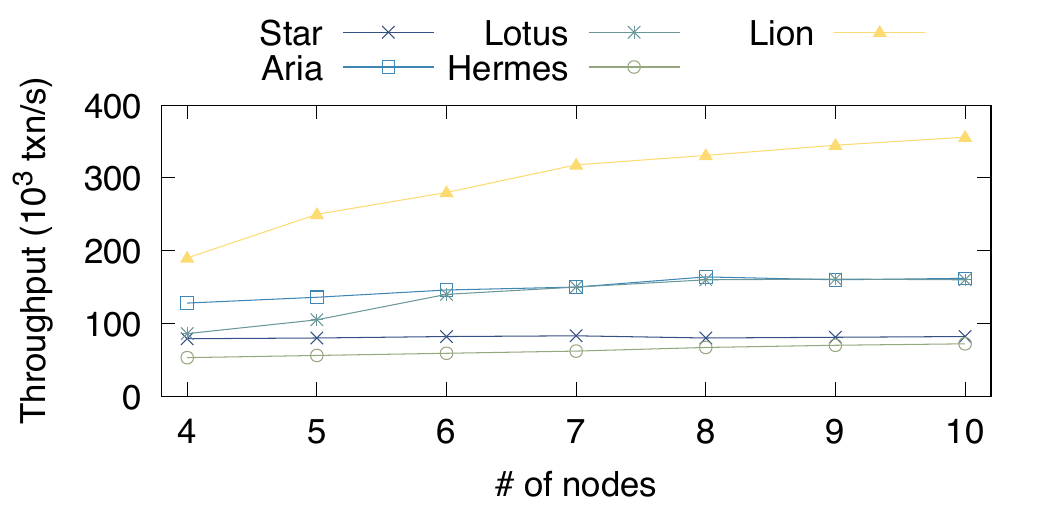}
    \vspace{-6mm}
    \caption{Batch-based approaches}
    \label{Fig.exp.scale.deter}
\end{subfigure}
\vspace{-1mm}
  \caption{Scalability}
  \label{Fig.exp.scale}
  \vspace{-4mm}
\end{figure}
}
\maintext{
\begin{figure}[t]
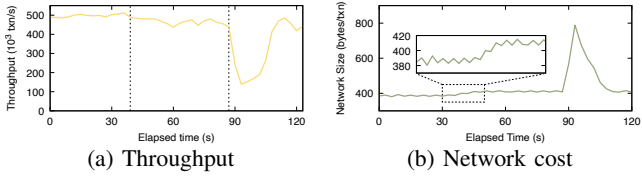

    \centering

\begin{subfigure}{0.48\linewidth}
    \includegraphics[width=\linewidth]{figures/exp/ycsb/lion/single_migration_cost.pdf}
    \vspace{-6mm}
    \caption{Throughput}
    \label{Fig.ycsb.migration_process.throughput}
\end{subfigure}
\begin{subfigure}{0.48\linewidth}
    \vspace{-0.5mm}
    \includegraphics[width=\linewidth]{figures/exp/ycsb/lion/single_nondm_network_cost.pdf}
    \vspace{-6mm}
    \caption{Network cost}
    \label{Fig.ycsb.migration_process.network_cost}
\end{subfigure}
\vspace{-1mm}
  \caption{The analysis of migration process}
  \label{Fig.experiment.migration_process.timeline}
  \vspace{-6mm}
\end{figure}
}

%

\subsection{Migration and remastering analysis}
\label{exp.overhead}
We now evaluate \lion's adaptability to new workloads in depth.
Our evaluation begins by analyzing the overhead of our proposed adaptive replica provision mechanism, employing the workload detailed in Section~\ref{sec.exp.comparison.non-deter.dynamicworkload}, with anticipated changes at the 90s, shown as the right dashed line in Figure~\ref{Fig.ycsb.migration_process.throughput}.
The planner, detecting an impending workload shift, initiates rearranging the replica arrangement by adding replications at the 30s, highlighted by the dashed line in the left segment of Figure~\ref{Fig.ycsb.migration_process.throughput}. 
This introduces additional synchronization costs to ensure consistency for these newly-added replicas, which elevates the network cost per transaction execution from 380 to nearly 420 bytes, as illustrated in Figure~\ref{Fig.ycsb.migration_process.network_cost}. 
However, with our replication arrangement strategy and group commit optimization, we cap the throughput decrease within 5\%.
A notable surge in network cost emerges around the 90s, peaking at 700 bytes per transaction (Figure~\ref{Fig.ycsb.migration_process.network_cost}) due to remastering requests for single-node conversion.
\extended{
With the pre-replication mechanism, \lion mitigates expensive data transmissions and demonstrates better adaptability to dynamic workloads, resulting in a 4x increase in efficiency as shown in Figure~\ref{Fig.ycsb.migration_process.pre-replication}.
Furthermore, Figure~\ref{Fig.ycsb.remaster_process.batch_op} showcases that batch processing, coupled with asynchronous remastering, experiences minimal impact from latency induced by the remastering process.
}

\subsection{Latency breakdown}
We finally analyze the latency of \lion and the breakdown of individual phases in comparison to deterministic approaches. Figure~\ref{Fig.latency.ycsb.latency} and Figure~\ref{Fig.latency.ycsb.breakdown} illustrate that \lion exhibits latency at the 95th percentile which is 48\% lower than Hermes. 
This stable latency is attributed to the group commit optimization and it dedicates 35\% of the time to replication synchronization while other deterministic approaches execute the same transaction batch on each replica to maintain consistency.
\extended{
Calvin consistently presents high latency across all percentiles due to the necessity of remote reads during distributed transaction execution, consuming over 90\% of the execution time.
}
\extended{
Aria employs an optimistic write reservation technique, fostering highly parallel execution without coordination. To reduce the abort ratio, it designs a reordering mechanism that costs an additional 20\% latency. 
Lotus introduces nearly zero scheduling time due to its epoch-based execution strategy with asynchronous commit and replication.
However, Aria and Lotus lead to occasional transaction aborts during successive batch processing rounds, resulting in high latency, especially at the 95th percentile.
}
\maintext{
\textcolor{blue}{
Aria 
\marginpar[\textcolor{blue}{R2.O2}]{\textcolor{blue}{R2.O2}}
employs an optimistic write reservation technique, fostering highly parallel execution without coordination. To reduce the abort ratio, it designs a reordering mechanism that costs an additional 20\% latency. 
Lotus introduces nearly zero scheduling time due to its epoch-based execution strategy with asynchronous commit and replication.
However, Aria and Lotus lead to occasional transaction aborts during successive batch processing rounds, resulting in high latency, especially at the 95th percentile.
}
}
Hermes demonstrates adaptability to workloads through partition migration, yet it still requires 19\% of the time for scheduling by a single lock manager, affecting executor concurrency.

\section{Related Works}
\label{sec.related}

\extended{
\begin{figure}[t]
    \centering
\vspace{-3mm}
\begin{subfigure}{0.48\linewidth}
    \includegraphics[width=\linewidth]{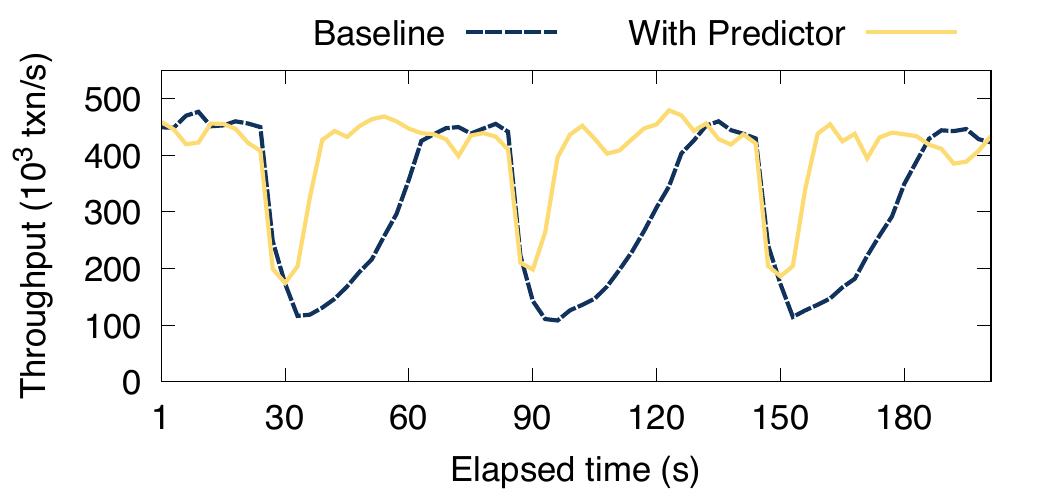}
    \vspace{-6mm}
    \caption{Impact of pre-replication}
    \label{Fig.ycsb.migration_process.pre-replication}
\end{subfigure}
\begin{subfigure}{0.48\linewidth}
    \includegraphics[width=\linewidth]{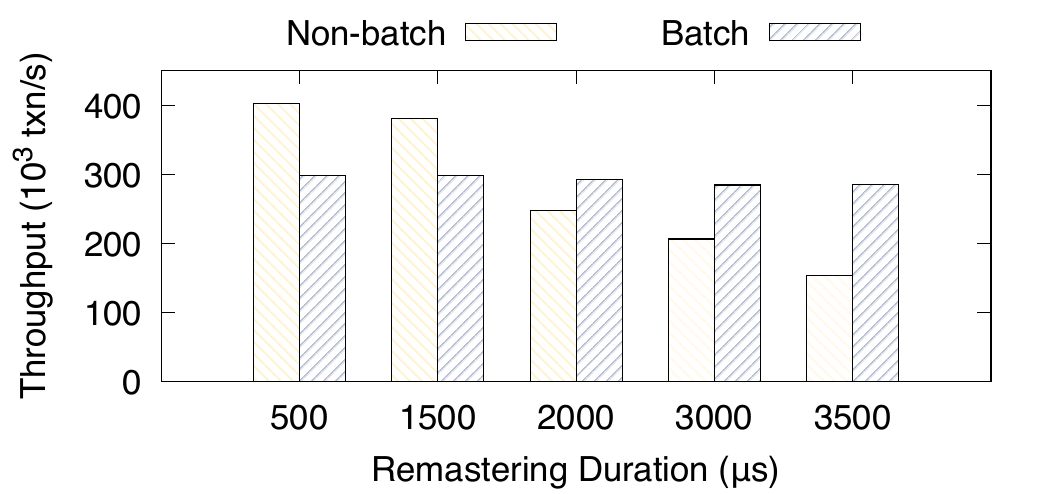}
    \vspace{-5mm}
    \caption{Impact of batch optimization}
    \label{Fig.ycsb.remaster_process.batch_op}
\end{subfigure}
\vspace{-1mm}
  \caption{The analysis of optimization}
  \label{Fig.experiment.migration_process.timeline}
  \vspace{-4mm}
\end{figure}
}

\label{sec:relatedwork}

\maintext{
\begin{figure}[t]
\vspace{-5mm}
    \centering
\begin{subfigure}{0.4\linewidth}
    \vspace{1.5mm}
    \includegraphics[width=\linewidth]{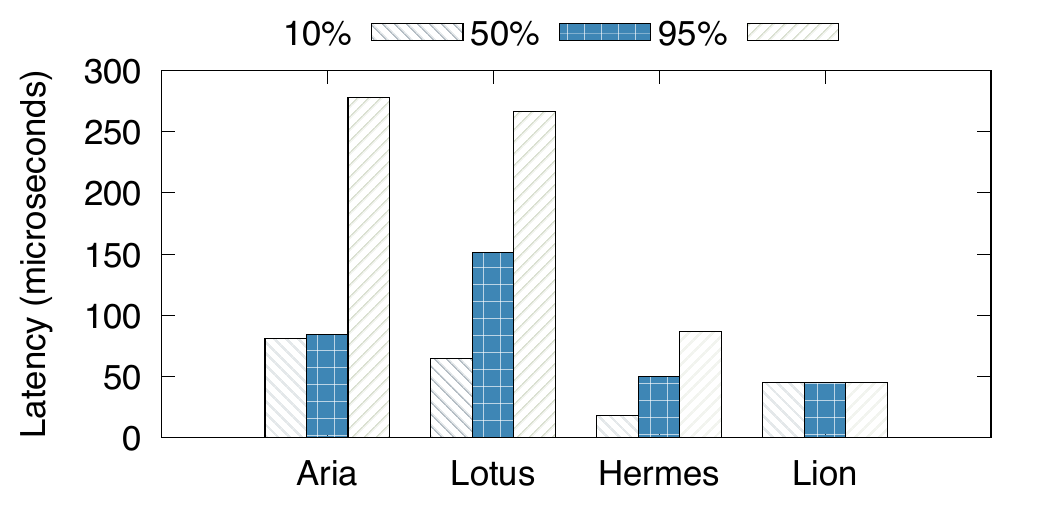}
    \vspace{-6mm}
    \caption{Latency}
    \label{Fig.latency.ycsb.latency}
\end{subfigure}
\begin{subfigure}{0.5\linewidth}
    \includegraphics[width=\linewidth]{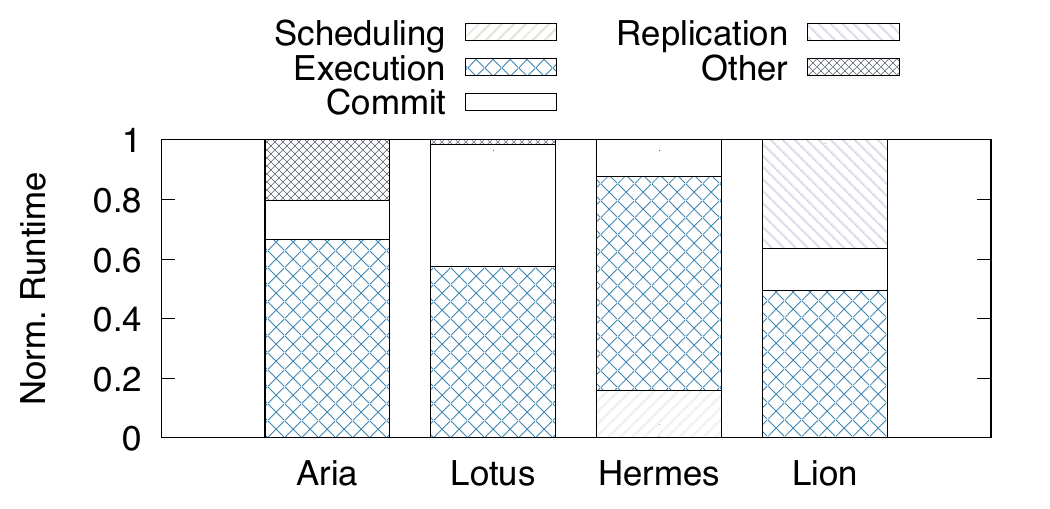}
    \vspace{-5mm}
    \caption{Breakdown}
    \label{Fig.latency.ycsb.breakdown}
\end{subfigure}
\vspace{-1mm}
  \caption{The analysis of average latency}
  \label{Fig.latency.experiment.timeline}
  \vspace{-6mm}
\end{figure}

}
\maintext{
Recent research focuses on optimizing distributed transaction performance by reducing network overheads through approaches like unifying 2PC and replica synchronization, which minimizes coordination costs but may increase consensus overhead~\cite{DBLP:conf/srds/StamosC90, DBLP:conf/sosp/ZhangSSKP15, DBLP:conf/eurosys/KraskaPFMF13, DBLP:journals/pvldb/MaiyyaNAA19, DBLP:journals/pvldb/ZhangLZXLXHYD23}. 
Others explore deterministic execution~\cite{DBLP:journals/pvldb/FaleiroA15, DBLP:conf/sigmod/ThomsonDWRSA12, DBLP:journals/pvldb/LuYCM20, DBLP:conf/sigmod/LinTLCW21, DBLP:conf/edbt/QadahGS20, DBLP:conf/sosp/QinBG21} and batch-based processing~\cite{DBLP:journals/pvldb/0010Y0M21,DBLP:journals/pvldb/ZhouYGS22,DBLP:conf/icde/LaiFZMPL023}, decomposing transactions to eliminate cross-node coordination, yet this may limit transaction interactivity and requires prior knowledge.
Additionally, leveraging new hardware like RDMA offers insights into mitigating network bottlenecks and re-implementing concurrency control algorithms for compatibility, but often requires special hardware support and system modifications~\cite{DBLP:journals/corr/ZamanianBKH16, DBLP:journals/pvldb/BinnigCGKZ16, DBLP:journals/usenix-login/KaliaKA16, DBLP:conf/osdi/WeiD0C18, DBLP:conf/sigmod/YoonCM18, DBLP:journals/tocs/ChenCWSCWZG17}.
\lion utilizes partition-based replication to reduce the occurrence of distributed transactions, which avoids the constraints of interactive transactions and hardware dependence. 
\textcolor{blue}{
Due to space constraints, a more detailed analysis of related work can be found in the extended version~\cite{lionExtended}.
}
}
\extended{
Substantial efforts have been devoted to optimizing distributed transaction performance by reducing network overheads has gained increasing attention in recent years.
For these works~\cite{DBLP:conf/srds/StamosC90, DBLP:conf/sosp/ZhangSSKP15, DBLP:conf/eurosys/KraskaPFMF13, DBLP:journals/pvldb/MaiyyaNAA19, DBLP:journals/pvldb/ZhangLZXLXHYD23}
, they focus on minimizing network round trips by
unifying 2PC and replica synchronization in a single framework.
In these approaches, the coordinator simultaneously engages both primary and secondary replicas for voting on whether to commit a transaction.
Consequently, they reduce network round trips by removing the need for sequential cross-node coordination and replica synchronization in 2PC-Paxos.
However, they commit a transaction only after achieving a majority consensus among all involved replicas.
This increased consensus across numerous nodes could potentially intensify the cross-node coordination cost.

As opposed to optimizing the number of network round-trips, quite a few studies~\cite{DBLP:journals/pvldb/FaleiroA15, DBLP:conf/sigmod/ThomsonDWRSA12, DBLP:journals/pvldb/LuYCM20, DBLP:conf/sigmod/LinTLCW21, DBLP:conf/edbt/QadahGS20, DBLP:conf/sosp/QinBG21} explores deterministic execution, where each transaction is decomposed into multiple sub-transactions that are executed individually on different nodes. 
Because the equivalent serializable schedule of sub-transactions in each individual node follows the same pre-determined order, expensive coordination among nodes, e.g., 2PC, can be eliminated.
Furthermore, lots of works are exploring the benefits of the group-based processing schema~\cite{DBLP:journals/pvldb/0010Y0M21,DBLP:journals/pvldb/ZhouYGS22,DBLP:conf/icde/LaiFZMPL023}. However, these methods require prior knowledge in advance and cannot handle interactive transactions.


Another line of research utilizes modern hardware such as RDMA for enhancing distributed transaction processing efficiency. Several studies~\cite{DBLP:journals/corr/ZamanianBKH16, DBLP:journals/pvldb/BinnigCGKZ16, DBLP:journals/usenix-login/KaliaKA16} offer valuable insights and architectural guidelines to mitigate network bottlenecks. Further, other works~\cite{DBLP:conf/osdi/WeiD0C18, DBLP:conf/sigmod/YoonCM18, DBLP:journals/tocs/ChenCWSCWZG17} re-implement concurrency control algorithms to be RDMA-compatible, employing efficient one-sided and two-sided verbs. However, these approaches often necessitate special network hardware support and substantial system modifications to leverage this technology.
}

\extended{
\begin{figure}[t]
\vspace{-5mm}
    \centering
\begin{subfigure}{0.42\linewidth}
    \vspace{1.5mm}
    \includegraphics[width=\linewidth]{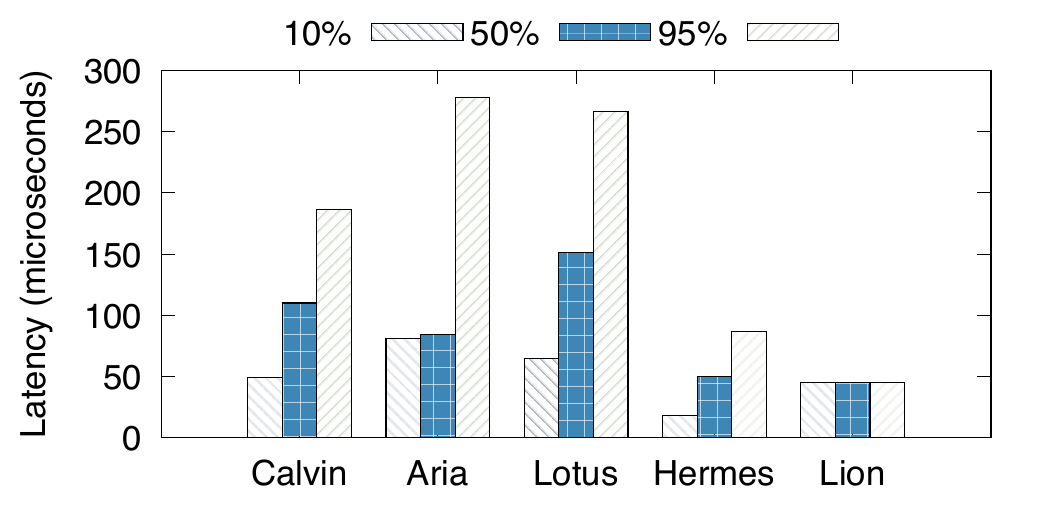}
    \vspace{-6mm}
    \caption{Latency}
    \label{Fig.latency.ycsb.latency}
\end{subfigure}
\begin{subfigure}{0.5\linewidth}
    \includegraphics[width=\linewidth]{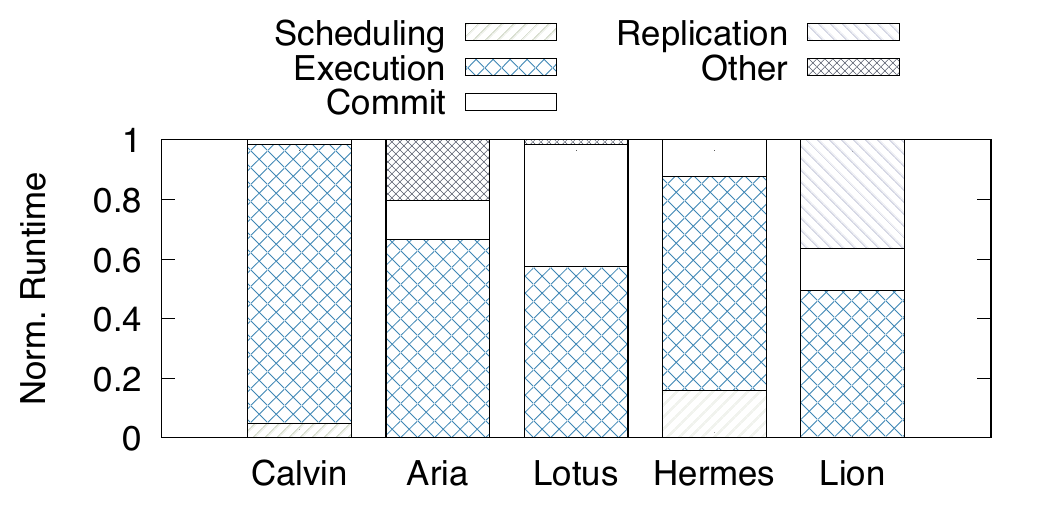}
    \vspace{-7mm}
    \caption{Breakdown}
    \label{Fig.latency.ycsb.breakdown}
\end{subfigure}
\vspace{-1mm}
  \caption{The analysis of average latency}
  \label{Fig.latency.experiment.timeline}
  \vspace{-6mm}
\end{figure}

}


\section{Conclusion}
\label{sec:conclusion}

In this paper, we present {\lion}, a general transaction processing protocol that minimizes distributed transactions by employing replication in today's distributed databases.
With an adaptive replica provision mechanism, {\lion} asynchronously adjusts the replica placement based on the workload, ensuring most transactions can execute on a single node containing all the necessary data replicas.
Further, we introduce a workload prediction technique to ensure the replica adjustment can be proactive, which maintains sustainable performance even as the workload dynamically changes.
We conduct extensive experiments to compare {\lion} against various transaction processing protocols.
The results show that {\lion} achieves up to 2.7x higher throughput and 76.4\% better scalability against these state-of-the-art approaches.

\balance

\bibliographystyle{IEEEtran}
\bibliography{sample}

\begin{thebibliography}{10}
\providecommand{\url}[1]{#1}
\csname url@samestyle\endcsname
\providecommand{\newblock}{\relax}
\providecommand{\bibinfo}[2]{#2}
\providecommand{\BIBentrySTDinterwordspacing}{\spaceskip=0pt\relax}
\providecommand{\BIBentryALTinterwordstretchfactor}{4}
\providecommand{\BIBentryALTinterwordspacing}{\spaceskip=\fontdimen2\font plus
\BIBentryALTinterwordstretchfactor\fontdimen3\font minus
  \fontdimen4\font\relax}
\providecommand{\BIBforeignlanguage}[2]{{%
\expandafter\ifx\csname l@#1\endcsname\relax
\typeout{** WARNING: IEEEtran.bst: No hyphenation pattern has been}%
\typeout{** loaded for the language `#1'. Using the pattern for}%
\typeout{** the default language instead.}%
\else
\language=\csname l@#1\endcsname
\fi
#2}}
\providecommand{\BIBdecl}{\relax}
\BIBdecl

\bibitem{DBLP:conf/osdi/CorbettDEFFFGGHHHKKLLMMNQRRSSTWW12}
J.~C. Corbett, J.~Dean, M.~Epstein, A.~Fikes, C.~Frost, J.~J. Furman,
  S.~Ghemawat, A.~Gubarev, C.~Heiser, P.~Hochschild, W.~C. Hsieh, S.~Kanthak,
  E.~Kogan, H.~Li, A.~Lloyd, S.~Melnik, D.~Mwaura, D.~Nagle, S.~Quinlan,
  R.~Rao, L.~Rolig, Y.~Saito, M.~Szymaniak, C.~Taylor, R.~Wang, and
  D.~Woodford, ``Spanner: Google's globally-distributed database,'' in
  \emph{{OSDI}}.\hskip 1em plus 0.5em minus 0.4em\relax {USENIX} Association,
  2012, pp. 251--264.

\bibitem{DBLP:conf/sigmod/TaftSMVLGNWBPBR20}
R.~Taft, I.~Sharif, A.~Matei, N.~VanBenschoten, J.~Lewis, T.~Grieger, K.~Niemi,
  A.~Woods, A.~Birzin, R.~Poss, P.~Bardea, A.~Ranade, B.~Darnell, B.~Gruneir,
  J.~Jaffray, L.~Zhang, and P.~Mattis, ``Cockroachdb: The resilient
  geo-distributed {SQL} database,'' in \emph{{SIGMOD} Conference}.\hskip 1em
  plus 0.5em minus 0.4em\relax {ACM}, 2020, pp. 1493--1509.

\bibitem{DBLP:journals/pvldb/HuangLCFMXSTZHW20}
D.~Huang, Q.~Liu, Q.~Cui, Z.~Fang, X.~Ma, F.~Xu, L.~Shen, L.~Tang, Y.~Zhou,
  M.~Huang, W.~Wei, C.~Liu, J.~Zhang, J.~Li, X.~Wu, L.~Song, R.~Sun, S.~Yu,
  L.~Zhao, N.~Cameron, L.~Pei, and X.~Tang, ``Tidb: {A} raft-based {HTAP}
  database,'' \emph{Proc. {VLDB} Endow.}, vol.~13, no.~12, pp. 3072--3084,
  2020.

\bibitem{DBLP:journals/pvldb/HardingAPS17}
R.~Harding, D.~V. Aken, A.~Pavlo, and M.~Stonebraker, ``An evaluation of
  distributed concurrency control,'' \emph{Proc. {VLDB} Endow.}, vol.~10,
  no.~5, pp. 553--564, 2017.

\bibitem{DBLP:journals/pvldb/CurinoZJM10}
C.~Curino, Y.~Zhang, E.~P.~C. Jones, and S.~Madden, ``Schism: a workload-driven
  approach to database replication and partitioning,'' \emph{Proc. {VLDB}
  Endow.}, vol.~3, no.~1, pp. 48--57, 2010.

\bibitem{DBLP:conf/edbt/QuamarKD13}
A.~Quamar, K.~A. Kumar, and A.~Deshpande, ``{SWORD:} scalable workload-aware
  data placement for transactional workloads,'' in \emph{{EDBT}}.\hskip 1em
  plus 0.5em minus 0.4em\relax {ACM}, 2013, pp. 430--441.

\bibitem{DBLP:journals/pvldb/SerafiniTEPAS16}
M.~Serafini, R.~Taft, A.~J. Elmore, A.~Pavlo, A.~Aboulnaga, and M.~Stonebraker,
  ``Clay: Fine-grained adaptive partitioning for general database schemas,''
  \emph{Proc. {VLDB} Endow.}, vol.~10, no.~4, pp. 445--456, 2016.

\bibitem{DBLP:conf/sigmod/LinC0OTW16}
Q.~Lin, P.~Chang, G.~Chen, B.~C. Ooi, K.~Tan, and Z.~Wang, ``Towards a non-2pc
  transaction management in distributed database systems,'' in \emph{{SIGMOD}
  Conference}.\hskip 1em plus 0.5em minus 0.4em\relax {ACM}, 2016, pp.
  1659--1674.

\bibitem{DBLP:journals/pvldb/AbebeGD20}
M.~Abebe, B.~Glasbergen, and K.~Daudjee, ``Morphosys: Automatic physical design
  metamorphosis for distributed database systems,'' \emph{Proc. {VLDB} Endow.},
  vol.~13, no.~13, pp. 3573--3587, 2020.

\bibitem{DBLP:journals/pvldb/LuYM19}
Y.~Lu, X.~Yu, and S.~Madden, ``{STAR:} scaling transactions through asymmetric
  replication,'' \emph{Proc. {VLDB} Endow.}, vol.~12, no.~11, pp. 1316--1329,
  2019.

\bibitem{DBLP:conf/icde/AbebeGD20}
M.~Abebe, B.~Glasbergen, and K.~Daudjee, ``Dynamast: Adaptive dynamic mastering
  for replicated systems,'' in \emph{{ICDE}}.\hskip 1em plus 0.5em minus
  0.4em\relax {IEEE}, 2020, pp. 1381--1392.

\bibitem{DBLP:conf/sigmod/LinTLCW21}
Y.~Lin, C.~Tsai, T.~Lin, Y.~Chang, and S.~Wu, ``Don't look back, look into the
  future: Prescient data partitioning and migration for deterministic database
  systems,'' in \emph{{SIGMOD} Conference}.\hskip 1em plus 0.5em minus
  0.4em\relax {ACM}, 2021, pp. 1156--1168.

\bibitem{DBLP:conf/usenix/WeiSCC17}
X.~Wei, S.~Shen, R.~Chen, and H.~Chen, ``Replication-driven live
  reconfiguration for fast distributed transaction processing,'' in
  \emph{{USENIX} Annual Technical Conference}.\hskip 1em plus 0.5em minus
  0.4em\relax {USENIX} Association, 2017, pp. 335--347.

\bibitem{DBLP:conf/usenix/OngaroO14}
D.~Ongaro and J.~K. Ousterhout, ``In search of an understandable consensus
  algorithm,'' in \emph{{USENIX} Annual Technical Conference}.\hskip 1em plus
  0.5em minus 0.4em\relax {USENIX} Association, 2014, pp. 305--319.

\bibitem{DBLP:conf/opodis/Lamport02}
L.~Lamport, ``Paxos made simple, fast, and byzantine,'' in \emph{{OPODIS}},
  ser. Studia Informatica Universalis, vol.~3.\hskip 1em plus 0.5em minus
  0.4em\relax Suger, Saint-Denis, rue Catulienne, France, 2002, pp. 7--9.

\bibitem{DBLP:conf/sigmod/ElmoreATPAA15}
A.~J. Elmore, V.~Arora, R.~Taft, A.~Pavlo, D.~Agrawal, and A.~E. Abbadi,
  ``Squall: Fine-grained live reconfiguration for partitioned main memory
  databases,'' in \emph{{SIGMOD} Conference}.\hskip 1em plus 0.5em minus
  0.4em\relax {ACM}, 2015, pp. 299--313.

\bibitem{DBLP:conf/sigmod/ElmoreDAA11}
A.~J. Elmore, S.~Das, D.~Agrawal, and A.~E. Abbadi, ``Zephyr: live migration in
  shared nothing databases for elastic cloud platforms,'' in \emph{{SIGMOD}
  Conference}.\hskip 1em plus 0.5em minus 0.4em\relax {ACM}, 2011, pp.
  301--312.

\bibitem{DBLP:conf/sigmod/CubukcuEPSS21}
U.~Cubukcu, O.~Erdogan, S.~Pathak, S.~Sannakkayala, and M.~Slot, ``Citus:
  Distributed postgresql for data-intensive applications,'' in \emph{{SIGMOD}
  Conference}.\hskip 1em plus 0.5em minus 0.4em\relax {ACM}, 2021, pp.
  2490--2502.

\bibitem{DBLP:journals/pvldb/DasNAA11}
S.~Das, S.~Nishimura, D.~Agrawal, and A.~E. Abbadi, ``Albatross: Lightweight
  elasticity in shared storage databases for the cloud using live data
  migration,'' \emph{Proc. {VLDB} Endow.}, vol.~4, no.~8, pp. 494--505, 2011.

\bibitem{DBLP:journals/tpds/ShenWCCZ22}
S.~Shen, X.~Wei, R.~Chen, H.~Chen, and B.~Zang, ``Drtm+b: Replication-driven
  live reconfiguration for fast and general distributed transaction
  processing,'' \emph{{IEEE} Trans. Parallel Distributed Syst.}, vol.~33,
  no.~10, pp. 2628--2643, 2022.

\bibitem{DBLP:journals/pvldb/YangYHZYYCZSXYL22}
\BIBentryALTinterwordspacing
Z.~Yang, C.~Yang, F.~Han, M.~Zhuang, B.~Yang, Z.~Yang, X.~Cheng, Y.~Zhao,
  W.~Shi, H.~Xi, H.~Yu, B.~Liu, Y.~Pan, B.~Yin, J.~Chen, and Q.~Xu,
  ``Oceanbase: {A} 707 million tpmc distributed relational database system,''
  \emph{Proc. {VLDB} Endow.}, vol.~15, no.~12, pp. 3385--3397, 2022. [Online].
  Available: \url{https://www.vldb.org/pvldb/vol15/p3385-xu.pdf}
\BIBentrySTDinterwordspacing

\bibitem{DBLP:journals/corr/abs-2206-07278}
\BIBentryALTinterwordspacing
M.~Wu, X.~Yi, H.~Yu, Y.~Liu, and Y.~Wang, ``Nebula graph: An open source
  distributed graph database,'' \emph{CoRR}, vol. abs/2206.07278, 2022.
  [Online]. Available: \url{https://doi.org/10.48550/arXiv.2206.07278}
\BIBentrySTDinterwordspacing

\bibitem{DBLP:phd/us/Ongaro14}
D.~Ongaro, ``Consensus: bridging theory and practice,'' Ph.D. dissertation,
  Stanford University, {USA}, 2014.

\bibitem{DBLP:conf/sigmod/MaAHMPG18}
L.~Ma, D.~V. Aken, A.~Hefny, G.~Mezerhane, A.~Pavlo, and G.~J. Gordon,
  ``Query-based workload forecasting for self-driving database management
  systems,'' in \emph{{SIGMOD} Conference}.\hskip 1em plus 0.5em minus
  0.4em\relax {ACM}, 2018, pp. 631--645.

\bibitem{DBLP:conf/icde/GaoHZ00C23}
Y.~Gao, X.~Huang, X.~Zhou, X.~Gao, G.~Li, and G.~Chen, ``Dbaugur: An
  adversarial-based trend forecasting system for diversified workloads,'' in
  \emph{{ICDE}}.\hskip 1em plus 0.5em minus 0.4em\relax {IEEE}, 2023, pp.
  27--39.

\bibitem{DBLP:journals/toms/Vitter85}
J.~S. Vitter, ``Random sampling with a reservoir,'' \emph{{ACM} Trans. Math.
  Softw.}, vol.~11, no.~1, pp. 37--57, 1985.

\bibitem{DBLP:conf/uai/BootsGG13}
B.~Boots, G.~J. Gordon, and A.~Gretton, ``Hilbert space embeddings of
  predictive state representations,'' in \emph{{UAI}}.\hskip 1em plus 0.5em
  minus 0.4em\relax {AUAI} Press, 2013.

\bibitem{DBLP:journals/pvldb/LeeZLHTLZ21}
R.~Lee, M.~Zhou, C.~Li, S.~Hu, J.~Teng, D.~Li, and X.~Zhang, ``The art of
  balance: {A} rateupdb experience of building a {CPU/GPU} hybrid database
  product,'' \emph{Proc. {VLDB} Endow.}, vol.~14, no.~12, pp. 2999--3013, 2021.

\bibitem{vaswani2017attention}
A.~Vaswani, N.~Shazeer, N.~Parmar, J.~Uszkoreit, L.~Jones, A.~N. Gomez,
  {\L}.~Kaiser, and I.~Polosukhin, ``Attention is all you need,''
  \emph{Advances in neural information processing systems}, vol.~30, 2017.

\bibitem{devlin2018bert}
J.~Devlin, M.-W. Chang, K.~Lee, and K.~Toutanova, ``Bert: Pre-training of deep
  bidirectional transformers for language understanding,'' \emph{arXiv preprint
  arXiv:1810.04805}, 2018.

\bibitem{DBLP:journals/pvldb/0010Y0M21}
Y.~Lu, X.~Yu, L.~Cao, and S.~Madden, ``Epoch-based commit and replication in
  distributed {OLTP} databases,'' \emph{Proc. {VLDB} Endow.}, vol.~14, no.~5,
  pp. 743--756, 2021.

\bibitem{TidbDocs}
PingCAP, ``Replica management in tidb: Pd control user guide,''
  \url{https://docs.pingcap.com/tidb/stable/pd-control}.

\bibitem{oceanbaseDocs}
Oceanbase, ``Modify locality,''
  \url{https://en.oceanbase.com/docs/common-oceanbase-database-10000000001105992}.

\bibitem{DBLP:conf/cloud/CooperSTRS10}
B.~F. Cooper, A.~Silberstein, E.~Tam, R.~Ramakrishnan, and R.~Sears,
  ``Benchmarking cloud serving systems with {YCSB},'' in \emph{SoCC}.\hskip 1em
  plus 0.5em minus 0.4em\relax {ACM}, 2010, pp. 143--154.

\bibitem{DBLP:conf/tpctc/NambiarWMTLCM11}
R.~O. Nambiar, N.~Wakou, A.~Masland, P.~Thawley, M.~Lanken, F.~Carman, and
  M.~Majdalany, ``Shaping the landscape of industry standard benchmarks:
  Contributions of the transaction processing performance council {(TPC)},'' in
  \emph{{TPCTC}}, ser. Lecture Notes in Computer Science, vol. 7144.\hskip 1em
  plus 0.5em minus 0.4em\relax Springer, 2011, pp. 1--9.

\bibitem{DBLP:conf/sosp/VandiverBLM07}
B.~Vandiver, H.~Balakrishnan, B.~Liskov, and S.~Madden, ``Tolerating byzantine
  faults in transaction processing systems using commit barrier scheduling,''
  in \emph{{SOSP}}.\hskip 1em plus 0.5em minus 0.4em\relax {ACM}, 2007, pp.
  59--72.

\bibitem{DBLP:journals/pvldb/ZhouYGS22}
X.~Zhou, X.~Yu, G.~Graefe, and M.~Stonebraker, ``Lotus: Scalable
  multi-partition transactions on single-threaded partitioned databases,''
  \emph{Proc. {VLDB} Endow.}, vol.~15, no.~11, pp. 2939--2952, 2022.

\bibitem{DBLP:conf/srds/StamosC90}
J.~W. Stamos and F.~Cristian, ``A low-cost atomic commit protocol,'' in
  \emph{{SRDS}}.\hskip 1em plus 0.5em minus 0.4em\relax {IEEE} Computer
  Society, 1990, pp. 66--75.

\bibitem{DBLP:conf/sosp/ZhangSSKP15}
I.~Zhang, N.~K. Sharma, A.~Szekeres, A.~Krishnamurthy, and D.~R.~K. Ports,
  ``Building consistent transactions with inconsistent replication,'' in
  \emph{{SOSP}}.\hskip 1em plus 0.5em minus 0.4em\relax {ACM}, 2015, pp.
  263--278.

\bibitem{DBLP:conf/eurosys/KraskaPFMF13}
T.~Kraska, G.~Pang, M.~J. Franklin, S.~Madden, and A.~D. Fekete, ``{MDCC:}
  multi-data center consistency,'' in \emph{EuroSys}.\hskip 1em plus 0.5em
  minus 0.4em\relax {ACM}, 2013, pp. 113--126.

\bibitem{DBLP:journals/pvldb/MaiyyaNAA19}
S.~Maiyya, F.~Nawab, D.~Agrawal, and A.~E. Abbadi, ``Unifying consensus and
  atomic commitment for effective cloud data management,'' \emph{Proc. {VLDB}
  Endow.}, vol.~12, no.~5, pp. 611--623, 2019.

\bibitem{DBLP:journals/pvldb/ZhangLZXLXHYD23}
Q.~Zhang, J.~Li, H.~Zhao, Q.~Xu, W.~Lu, J.~Xiao, F.~Han, C.~Yang, and X.~Du,
  ``Efficient distributed transaction processing in heterogeneous networks,''
  \emph{Proc. {VLDB} Endow.}, vol.~16, no.~6, pp. 1372--1385, 2023.

\bibitem{DBLP:journals/pvldb/FaleiroA15}
J.~M. Faleiro and D.~J. Abadi, ``Rethinking serializable multiversion
  concurrency control,'' \emph{Proc. {VLDB} Endow.}, vol.~8, no.~11, pp.
  1190--1201, 2015.

\bibitem{DBLP:conf/sigmod/ThomsonDWRSA12}
A.~Thomson, T.~Diamond, S.~Weng, K.~Ren, P.~Shao, and D.~J. Abadi, ``Calvin:
  fast distributed transactions for partitioned database systems,'' in
  \emph{{SIGMOD} Conference}.\hskip 1em plus 0.5em minus 0.4em\relax {ACM},
  2012, pp. 1--12.

\bibitem{DBLP:journals/pvldb/LuYCM20}
Y.~Lu, X.~Yu, L.~Cao, and S.~Madden, ``Aria: {A} fast and practical
  deterministic {OLTP} database,'' \emph{Proc. {VLDB} Endow.}, vol.~13, no.~11,
  pp. 2047--2060, 2020.

\bibitem{DBLP:conf/edbt/QadahGS20}
T.~Qadah, S.~Gupta, and M.~Sadoghi, ``Q-store: Distributed, multi-partition
  transactions via queue-oriented execution and communication,'' in
  \emph{{EDBT}}.\hskip 1em plus 0.5em minus 0.4em\relax OpenProceedings.org,
  2020, pp. 73--84.

\bibitem{DBLP:conf/sosp/QinBG21}
D.~Qin, A.~D. Brown, and A.~Goel, ``Caracal: Contention management with
  deterministic concurrency control,'' in \emph{{SOSP}}.\hskip 1em plus 0.5em
  minus 0.4em\relax {ACM}, 2021, pp. 180--194.

\bibitem{DBLP:conf/icde/LaiFZMPL023}
Z.~Lai, H.~Fan, W.~Zhou, Z.~Ma, X.~Peng, F.~Li, and E.~Lo, ``Knock out 2pc with
  practicality intact: a high-performance and general distributed transaction
  protocol,'' in \emph{{ICDE}}.\hskip 1em plus 0.5em minus 0.4em\relax {IEEE},
  2023, pp. 2317--2331.

\bibitem{DBLP:journals/corr/ZamanianBKH16}
E.~Zamanian, C.~Binnig, T.~Kraska, and T.~Harris, ``The end of a myth:
  Distributed transactions can scale,'' \emph{CoRR}, vol. abs/1607.00655, 2016.

\bibitem{DBLP:journals/pvldb/BinnigCGKZ16}
C.~Binnig, A.~Crotty, A.~Galakatos, T.~Kraska, and E.~Zamanian, ``The end of
  slow networks: It's time for a redesign,'' \emph{Proc. {VLDB} Endow.},
  vol.~9, no.~7, pp. 528--539, 2016.

\bibitem{DBLP:journals/usenix-login/KaliaKA16}
A.~Kalia, M.~Kaminsky, and D.~G. Andersen, ``Design guidelines for high
  performance {RDMA} systems,'' \emph{login Usenix Mag.}, vol.~41, no.~3, 2016.

\bibitem{DBLP:conf/osdi/WeiD0C18}
X.~Wei, Z.~Dong, R.~Chen, and H.~Chen, ``Deconstructing rdma-enabled
  distributed transactions: Hybrid is better!'' in \emph{{OSDI}}.\hskip 1em
  plus 0.5em minus 0.4em\relax {USENIX} Association, 2018, pp. 233--251.

\bibitem{DBLP:conf/sigmod/YoonCM18}
D.~Y. Yoon, M.~Chowdhury, and B.~Mozafari, ``Distributed lock management with
  {RDMA:} decentralization without starvation,'' in \emph{{SIGMOD}
  Conference}.\hskip 1em plus 0.5em minus 0.4em\relax {ACM}, 2018, pp.
  1571--1586.

\bibitem{DBLP:journals/tocs/ChenCWSCWZG17}
H.~Chen, R.~Chen, X.~Wei, J.~Shi, Y.~Chen, Z.~Wang, B.~Zang, and H.~Guan,
  ``Fast in-memory transaction processing using {RDMA} and {HTM},'' \emph{{ACM}
  Trans. Comput. Syst.}, vol.~35, no.~1, pp. 3:1--3:37, 2017.

\end{thebibliography}


\end{document}